\documentclass[
reprint,
superscriptaddress,
nofootinbib,
aps,
prd,
]{revtex4-1}

\usepackage{graphicx}
\usepackage{dcolumn}
\usepackage{bm}
\usepackage{xcolor}
\usepackage[colorlinks=true,allcolors=blue,linkcolor=blue,citecolor=blue]{hyperref}
\usepackage[mathlines]{lineno}
\usepackage{epsfig}
\usepackage{amsmath, amsbsy}
\usepackage{amsthm}
\usepackage{amssymb}
\usepackage{comment}
\usepackage{hhline}
\usepackage{makecell}
\usepackage{multirow}
\usepackage{enumitem}
\usepackage{tabularx}

\newcommand{\be}{\begin{equation}}
\newcommand{\ee}{\end{equation}}
\newcommand{\ba}{\begin{eqnarray}}
\newcommand{\ea}{\end{eqnarray}}

\newcommand{\apjl}{ApJ Letters}
\newcommand{\mnras}{Mon. Not. R. Astron. Soc.}
\newcommand{\aap}{A\&A}
\newcommand{\physrep}{Physics Reports}
\newcommand{\jcap}{J. Cosmol. Astropart. Phys.}

\setlist[description]{font=\normalfont}

\begin{document}


\title{Lensing reconstruction in post-Born cosmic microwave background weak lensing}

\author{Dominic Beck}
\email{Corresponding author. dbeck@apc.in2p3.fr}
\affiliation{AstroParticule et Cosmologie, Universit\'{e} Paris Diderot, CNRS/IN2P3,CEA/Irfu, Obs de Paris, Sorbonne Paris Cit\'{e}, France}

\author{Giulio Fabbian}
\email{Corresponding author. giulio.fabbian@ias.u-psud.fr}
\affiliation{Institut d'Astrophysique Spatiale, CNRS (UMR 8617), Univ. Paris-Sud, Universit\'{e} Paris-Saclay, b\^{a}timent 121, 91405 Orsay, France}

\author{Josquin Errard}
\affiliation{AstroParticule et Cosmologie, Universit\'{e} Paris Diderot, CNRS/IN2P3,CEA/Irfu, Obs de Paris, Sorbonne Paris Cit\'{e}, France}


\date{\today}

\begin{abstract}
The study of the Cosmic Microwave Background (CMB) lensing potential has established itself by now as a robust way of probing the physics of large-scale structure growth. The most common estimators of the lensing potential are derived under the assumption of Gaussianity of the matter distribution and in the Born approximation of the photon diffusion. In this paper we study the performance of quadratic estimators when applied to realistic sky maps extracted from multiple-lens ray tracing techniques in cosmological $N$-body simulations. These are expected to model accurately the effects due to the non-Gaussianity of the matter distribution induced by its nonlinearity and the deviation from the Born approximation. We show that both these effects on their own lead to reconstruction biases, but these tend to partially cancel each other when both these effects are considered together. We forecast the impact of these biases on the estimation of cosmological parameters for future high-sensitivity CMB experiments like CMB-S4. We find that the cold dark matter density, $\Omega_\textrm{cdm}$, the optical depth to reionization $\tau$, the amplitude of primordial inflationary perturbations, $A_s$ and the sum of neutrino masses, $M_\nu$, could be biased at the 1-2$\sigma$ level, if no external data set is used. We also observe a reduction of the bias if external data like baryon acoustic oscillations (BAO) is included.
\end{abstract}

\maketitle
%
\section{\label{sec:intro} Introduction}
The cosmic microwave background (CMB) photons detected today have interacted with the matter distribution in the universe throughout their journey from the last scattering surface towards us. Such interactions result in the generation of the so-called secondary anisotropies, i.e. fluctuations generated after the epoch of matter-radiation decoupling (see e.g \cite{aghanim2008} for a review). These can be either due to scattering between CMB photons and free electrons, such as inverse Compton or velocity-induced scatterings (the thermal and kinetic Sunyaev-Zel'dovich effect and the Ostriker-Vishniac effect) or to interactions of the photons with gravitational potential wells (e.g. the integrated Sachs-Wolfe \cite{isw} and Rees-Sciama \cite{rees-sciama} effects). Within this last class of secondary anisotropies the weak gravitational lensing of CMB anisotropies in temperature and polarization is one of the key signals exploited by current and future experiments to obtain constraints on cosmological models.
\\

CMB lensing is sourced by the growth of all matter located between $z=0$ and the last-scattering surface ($z\approx 1100$). It contains thus valuable information on the parameters affecting the formation of the large-scale structures (LSS) of the universe such as the sum of neutrino masses ($M_\nu$) and the properties of the dark energy (see \cite{lewis2006} for a review). 
\\

The effect of lensing on the CMB manifests itself in a scale-dependent smoothing of the acoustic oscillations in the angular power spectrum of temperature and $E$-mode polarization as well as in an increase of power in the damping tails. The first evidence of this effect was reported by the ACBAR experiment \citep{reichardt2009} and measured with high significance by SPT \cite{keisler2011}. In addition, lensing induces correlations in the harmonic coefficients of CMB anisotropies that can be used to reconstruct the distribution of the line-of-sight integrated gravitational potential that lensed the CMB, i.e. the so-called lensing potential. The first attempts to measure the latter in CMB data from WMAP using cross-correlation techniques with external LSS tracers were performed by \cite{smith2007,hirata2008} and the first significant direct detections were reported by the ACT and SPT Collaborations \cite{das2011, vanengelen2012}. Currently, the most precise measurement of the CMB lensing potential has been achieved by the Planck Collaboration, who measured this signal with a significance higher than $40 \sigma$ on nearly the full sky \cite{planck-lensing2016}. The effect of gravitational lensing on the CMB polarization anisotropies has been recently isolated by the current generation of ground-based CMB polarization experiments POLARBEAR \cite{pb-lens}, SPTpol \cite{story2015}, and ACTpol \cite{sherwin2017} using CMB data alone and in cross-correlation with LSS tracers \cite{pb-cib, sherwin2017,  hanson2013}. Additionally, limits on the CMB $B$-mode power on subdegree scales have been obtained~\cite{pb-bb2014, pb-bb2017, keisler2015}.
The $B$-mode signal of CMB polarization on these scales is largely sourced by the lensing distortion of the primordial $E$-mode polarization. Hence, achieving high-sensitivity measurements of the lensed CMB polarization is a crucial step to increase the precision of the CMB lensing potential reconstruction. With decreasing noise levels, higher angular resolutions and larger areas observed by future experiments (e.g. CMB-S4 \cite{cmbs4}), the accuracy of reconstruction techniques and theoretical modeling of the measurements has to improve alike.\\

To date, CMB lensing potential reconstruction analyses commonly rely on the assumption of Gaussianity of the unlensed CMB temperature field and of the lensing potential itself. The lensing potential, however, becomes non-Gaussian due to nonlinear structure formation mainly at late times. Although the level of non-Gaussianity is expected to be small due to the large number of potential wells that deflect CMB photons, the impact of this effect has to be quantified in light of future high-precision measurements \cite{merkel2011, namikawa2016, pratten2016}. \\

Moreover, the Born approximation (i.e. the evaluation of the deflections of the photons with respect to the original unperturbed line of sight), usually employed for modeling CMB lensing, does not account accurately for all features of the actual deflection process (e.g. the correlation between subsequent lensing events) neglecting therefore some of the sources of non-Gaussian statistics in the lensing potential. Earlier attempts to model CMB lensing including the effects of nonlinear structure formation were presented in \cite{carbone2008, carbone2009, merkel2011}.
Recent works investigated the effect of the relaxation of the Born approximation on lensed CMB power spectra and CMB lensing power spectra, from both an analytical and a numerical point of view \cite{hagstotz2015, pratten2016, lewis2016, fabbian2018}.\footnote{We warn the reader that the findings of \cite{hagstotz2015} disagrees with later studies of \cite{pratten2016, fabbian2018}, probably due to numerical error in the evaluation of their analytical expressions. We refer the reader to the discussion in \cite{pratten2016} for more details.} Similar analytical studies were previously performed also in the context of the weak lensing shear power spectrum \cite{cooray2002, shapiro2006, krause2010, petri2017}. While the most recent studies showed that the main post-Born effects are observed on the higher-order statistics of the CMB lensing potential rather than on its power spectrum, the impact of such effects on lensing reconstruction has not yet been evaluated. Recent theoretical works further suggested that the presence of non-Gaussianities in the CMB lensing potential could lead to percent level biases in the reconstructed CMB lensing potential power spectrum if they are left unaccounted for \cite{bohm2016}. This could in turn lead to a biased estimation of cosmological parameters. \\

In this paper we evaluate the impact of the non-Gaussian statistics of the CMB lensing potential on the commonly employed quadratic estimator techniques for the CMB lensing reconstruction. As these effects are often too complex to model analytically, we use the simulations of \cite{fabbian2018} that include both the nonlinear evolution of LSS and post-Born effects to model and investigate this problem numerically. The paper is organized in the following way. In Sec.~\ref{sec:theory} we review the theoretical aspects of weak lensing in the Born and post-Born regimes, and in Sec.~\ref{sec:quest} we review the properties of the statistical estimators to extract this effect in the CMB. In Sec.~\ref{sec:sims} we review the details of the modeling implemented in the simulations used in this work. In Sec.~\ref{sec:results} we show the results of our numerical experiments as their impact on the lensing potential power spectrum and in Sec.~\ref{sec:params} we describe the impact of our findings on the estimation of several cosmological parameters with a particular focus on the total mass of neutrinos, $M_\nu$, which is one of the main science targets of future CMB experiments. Finally, conclusions are made in Sec.~\ref{sec:conclusions}.

\section{Gravitational lensing formalism}\label{sec:theory}
In the weak lensing formalism the effect of deflections of light rays coming from a source plane is described by the lens equation. This maps the final position $(\boldsymbol\beta, \chi)$ of a
ray to the angular position of its source $\boldsymbol\theta$, i.e.,
\be
\beta_i (\boldsymbol\theta, \chi) = \theta_i -
\frac{2}{c^2}\int_0^{\chi}\frac{D_A(\chi -
  \chi')}{D_A(\chi)D_A(\chi')}
\Psi,_{\beta_i}\left(\boldsymbol\beta(\boldsymbol\theta,
\chi'), \chi' \right) {\rm d}\chi', 
\label{eq:beta}
\ee
where $\chi$ is the conformal time, $\Psi(\boldsymbol\beta, \chi)$ is a gravitational potential located on the photon path, $\Psi(\boldsymbol\beta, \chi),_{\beta_i}$ their angular derivatives\footnote{The derivatives in the small angle limit should be computed using a coordinate
system orthogonal to the current light ray's direction of travel. Numerical tests have shown that using angular derivatives causes a negligible
error (see e.g. \cite{becker2013} and references therein).} and $D_A(\chi)$ is the comoving angular diameter distance.
The linearized mapping between an image at the source plane and the lensed image at a given lens plane is described by the lensing magnification matrix (or lensing Jacobian). This can be computed as the simple derivative of the equation above.\footnote{We note that the following formula can be extended to the full-sky case by promoting the partial derivatives to covariant derivatives.}\\

\begin{widetext}
\be
A_{ij}(\boldsymbol\theta,\chi) \equiv \frac{\partial
  \beta_i(\boldsymbol\theta,\chi)}{\partial \theta_j}  
 = \delta^K_{ij} - \frac{2}{c^2}\int_0^{\chi}\frac{D_A(\chi -
  \chi')}{D_A(\chi)D_A(\chi')}
\Psi,_{\beta_i\beta_k}\left(\boldsymbol\beta(\boldsymbol\theta,
\chi'), \chi' \right) A_{kj}(\boldsymbol\theta,\chi'){\rm d}\chi',
\label{eq:distortionmatrix}
\ee
\newpage
\end{widetext}

In the weak lensing regime the magnification matrix is usually decomposed into 

\ba
A_{ij}  &\approx&
\begin{pmatrix}
  1 - \kappa -\gamma_1 & -\gamma_2 + \omega \\
  -\gamma_2 - \omega & 1-\kappa+\gamma_1
\end{pmatrix},
\ea
where $\kappa$ is the lensing convergence, $\gamma_{1,2}$ are the components of the lensing shear, and $\omega$ is the lensing rotation angle \citep{bartelmann-shneider}. The components of the magnification matrix are not independent and are connected through a series of consistency relations \cite{stebbins96, hu2000}.\\*

In the leading-order computations of the lensing effect, the photon path is approximated by the unperturbed photon geodesic $\mathbf{x}(\chi) \approx \boldsymbol\theta \chi$, such that the line integral of the Newtonian potential $\Psi$ simplifies to
\be
\beta_i (\boldsymbol\theta, \chi) = \theta_i -
\frac{2}{c^2}\int_0^{\chi}\frac{D_A(\chi -
  \chi')}{D_A(\chi)D_A(\chi')}
\Psi,_{\beta_i}\left(\boldsymbol\theta, \chi' \right) {\rm d}\chi'.
\label{eq:lenseq}
\ee
At linear order in $\Psi$, the overall deflection of a photon $\boldsymbol\alpha$ is then given by 

\ba
\boldsymbol\alpha(\boldsymbol\theta) &=& \frac{2}{D_A(\chi_*)} \int_0^{\chi_*} d \chi \frac{D_A(\chi_* - \chi)}{D_A(\chi)} \nabla\Psi(\boldsymbol\theta, \chi),
\label{eq:lenspot}
\ea
where $\chi_*$ is the distance to the source plane. In the case of CMB lensing, for instance, it is the distance to the last scattering surface. The lens equation is usually rewritten in terms of the lensing potential $\phi$, which is connected to the total photon deflection, as
\be
\beta(\boldsymbol\theta,\chi^*) = \boldsymbol\theta - \boldsymbol\alpha(\boldsymbol\theta) \equiv\boldsymbol\theta - \nabla\phi(\boldsymbol\theta).
\label{eq:remapping}
\ee
We note that the lensing potential and the lensing convergence can be connected in the weak lensing regime through the relations \footnote{Despite being derived in the Born approximation, these relations hold in the post-Born regime at sub-percent accuracy as discussed in \cite{fabbian2018}.}
\ba
\kappa &=& -\frac{1}{2}\nabla^2\phi,\label{eq:conskappa}\\
C^{\kappa\kappa}_{L} &=& \frac{\left[L(L+1)\right]}{4}^2C^{\phi\phi}_{L}.
\ea
If we want to evaluate the lens equation at higher order, i.e. beyond the Born approximation (post-Born), we have to account for the fact that photons do not travel along the unperturbed background geodesics. Higher-order corrections are typically introduced perturbatively in Eq.~\eqref{eq:beta} by Taylor expanding the potential $\Psi$ around the unperturbed geodesic position.\\

The distinct additional couplings that arise reflect the change in the shape of a light ray bundle by one lensing event affecting the amount of lensing generated by a later lensing event (lens-lens coupling) as well as by changing gravitational potentials in the direction in which the ray path is bent. We refer the reader to \cite{cooray2002, pratten2016, krause2010, marozzi2016} for further details.
Post-Born corrections affect the angular power spectrum of CMB lensing observables in a minor way. In particular, the amplitude of $C_{L}^{\kappa\kappa}$ is suppressed on scales $L\lesssim 1000$ by roughly 0.2\% due to lens-lens coupling and enhanced above the cosmic variance uncertainties at $L\gtrsim 1000$, mimicking thus an additional nonlinear large-scale structure growth \cite{pratten2016}. Higher-order correlations of the $\kappa$ field, such as the bispectrum, are, however, more affected and we will discuss these effects in the coming sections. \\

A characteristic signature of post-Born corrections is the appearance of curl-like modes in the overall lensing deflection angle \cite{hirata2003-curl, pratten2016}, such that
\be
\beta(\boldsymbol\theta) = \boldsymbol\theta - \nabla\phi(\boldsymbol\theta)  - \nabla\times\Omega(\boldsymbol\theta).\label{eq:full-remapping}
\ee
Here we define $(\nabla\times\Omega)_i \equiv \epsilon_{ij}\partial_j\Omega$,where $\epsilon_{ij}$ is the  Levi-Civita symbol in two dimensions and $\Omega$ is a pseudo-scalar field. In analogy to the case of $\kappa$ and $\psi$, $\Omega$ is related to the lensing rotation $\omega$ as 
\ba
\omega &=& -\frac{1}{2}\nabla^2\Omega,\label{eq:consomega}\\
C_{L}^{\omega\omega} &=& \frac{[L(L+1)]^2}{4}C_{L}^{\Omega\Omega}.\label{eq:kappa-lensing}
\ea

\section{\label{sec:quest} CMB lensing reconstruction with quadratic estimators}
\subsection{Formalism}
Weak lensing by the large-scale structure of the Universe
remaps the primary CMB anisotropies according to Eq.~\eqref{eq:remapping} such that its observed lensed Stokes parameter $X$ along the $\boldsymbol\theta$ direction is given by
\begin{equation}
X(\boldsymbol\theta) = \tilde{X}(\boldsymbol\theta -\nabla\phi) \qquad X\in(T, Q, U)
\end{equation}
where $\tilde{X}$ is the primordial unlensed CMB and $\phi$ the lensing potential. In the harmonic domain the remapping operation acts as a convolution that mixes power in different multipoles and therefore correlates modes across a band determined by the power in the lensing potential \cite{fabbian2013}. The lensing potential itself can be extracted statistically using the observed CMB, assuming that the underlying, unlensed CMB is on average homogeneous and isotropic. This operation commonly involves the use of the so-called quadratic estimator \cite{hu-okamoto,okamoto-hu}, which relies on the lensing information in the two-point correlation of the CMB fields. Although higher-order correlations will become more important in reconstructing the CMB lensing potential to exploit the full power of future data sets to a more optimal precision \cite{hirata2003, carron2017, millea2017}, the quadratic estimator is proven to be a very robust tool thanks to its well understood biases and capability of quick forward modeling of instrumental and systematic effects. Furthermore we note that, to date, none of the proposed alternative estimators can dispense easily with the assumption of Gaussianity of the CMB lensing potential. In the following, we will use an implementation of the quadratic estimator, the publicly available code {\sc lensquest} \footnote{\url{http://github.com/doicbek/lensquest}}. 
In the quadratic estimator context we assume the primordial unlensed CMB to be a Gaussian field such that the harmonic coefficients for its temperature, $E$-mode and $B$-mode anisotropies, $a_{\ell m}^X$, $X \in {T,E,B}$, have a variance given by the four nonzero power spectra $C_\ell^\textrm{TT}$, $C_\ell^\textrm{TE}$, $C_\ell^\textrm{EE}$ and $C_\ell^\textrm{BB}$. Likewise, for the harmonic coefficients after lensing, $\tilde{a}_{\ell m}^X$, $X \in {T,E,B}$, we can write the variance of the harmonic coefficients, computed taking the ensemble average over primordial CMB and matter realizations, as $\left\langle \tilde{a}_{\ell m}^{X~ \dagger} \tilde{a}^Y_{\ell' m'} \right\rangle = \delta_{\ell\ell'}\delta_{mm'} \tilde{C}^{XY}_\ell$. \\

For a given realization of the lensing potential, this variance acquires non-diagonal terms due to the characteristic introduction of correlations in the harmonic coefficients due to gravitational lensing. This can be used to construct an estimator for the lensing potential \cite{hu-okamoto,okamoto-hu}. 
On the full sky, this estimator takes the form 
\be
\hat{\phi}_{LM}^{XY}=\sum_{\ell_1 m_1 \ell_2 m_2} \mathcal{K}^{XY}_{L M \ell_1 m_1 \ell_2 m_2} \hat{a}^X_{\ell_1 m_1} \hat{a}^Y_{\ell_2 m_2}, \label{eq:estimator}
\ee
where the convolution kernel is given by
\begin{align} \mathcal{K}^{XY}_{L M \ell_1 m_1 \ell_2 m_2} = \frac{A^{XY}_L}{L(L+1)} (-1)^M 
\begin{pmatrix}
L & \ell_1 & \ell_2 \\
-M & m_1 & m_2
\end{pmatrix}
g^{XY}_{L\ell_1\ell_2}.
\label{eq:quadestkernel} \end{align}
This kernel has cosmology and experiment-dependent weights, which read 
\begin{align} g^{XY}_{L\ell_1\ell_2} =
\frac{f^{XY *}_{L\ell_1\ell_2}}{2 \tilde{C}_{\ell_1}^{XXn} \tilde{C}_{\ell_1}^{YYn}} ~~ \text{or} ~~ g^{XY}_{L\ell_1\ell_2} = \frac{f^{XY *}_{L\ell_1\ell_2}}{\tilde{C}_{\ell_1}^{XXn} \tilde{C}_{\ell_1}^{YYn}}, \end{align}
if $X=Y$ or $X \neq Y$, respectively. These weights are chosen such that the variance of $\hat{\phi}_{LM}^{XY}$ is minimal \footnote{Correlation between T and E is neglected in the estimator weights, causing the estimator for $XY=TE$ to be slightly suboptimal in favor of computational time.} and we adopt the measured power spectra including the instrumental noise power spectrum $N^{XY}_\ell$, i.e. $\tilde{C}_{\ell}^{XYn}= \tilde{C}^{XY}_\ell + N_\ell^{XY}$.\\

The response functions $f^{XY}_{L\ell_1\ell_2}$ used in this work are those of \cite{okamoto-hu}, with the distinction of using the lensed CMB power spectra $\tilde{C}_\ell$ to mitigate the biases of $\mathcal{O}\left({C_L^{\phi\phi}}^2\right)$ that arise in the lensing potential power spectrum calculation using this estimator \cite{hanson2011}.
We note that this choice of weights might still be suboptimal and lead to biased results from very small-scale CMB temperature signal \cite{peloton2017}. This bias could be mitigated replacing the temperature autopower spectrum with the lensed temperature-gradient power spectrum  $C_{\ell}^{\tilde{T}\nabla\tilde{T}}$, appearing in the nonperturbative response function calculation \cite{lewis2011}. Because in the following we will compare lensed CMB realizations among each other and do not compare to a specific model, this has a negligible effect on our results. \\

The normalization vector $A_L^{XY}$ in Eq.~\eqref{eq:estimator} is given by
\be
A^{XY}_L=L(L+1)(2L+1) \left( \sum_{\ell_1 \ell_2} g^{XY*}_{L \ell_1 \ell_2} f^{XY}_{L l_1 l_2} \right)^{-1}
\ee
and ensures that the quadratic estimator is unbiased. The three CMB anisotropy fields allow for six separate estimators of $\phi$. The estimator for $XY=BB$ has a vanishing signal-to-noise in cosmological scenarios where gravitational wave perturbations are negligible compared to scalar perturbations, effectively reducing the number of estimators to 5. We will thus ignore it in the following without introducing an appreciable loss in the overall sensitivity. \\

The power spectrum of the quadratic estimate of the lensing potential is then a contraction of the CMB four-point function, which includes three terms up to first order in $\phi$
\begin{multline}
\left\langle \frac{1}{2L+1} \sum_M \left(\hat{\phi}_{LM}^{AB}\right)^\dagger \hat{\phi}_{LM}^{CD} \right\rangle \approx \\ \approx  N_L^{(0),ABCD} +C_L^{\phi\phi}  + N_L^{(1),ABCD}.
\label{powspec}
\end{multline}
The biases $N_L^{(0),~ABCD}$ and $N_L^{(1),~ABCD}$ arise from disconnected Gaussian two-point contractions of the CMB fields and --- in the case of the latter --- of the lensing potential up to first order in $C_L^{\phi\phi}$ \cite{kesden2003}. An analytic expression for the zero-order bias, $N_L^{(0)~ABCD}$, can be found in \cite{okamoto-hu}.
In practice, the computation of the realization-dependent zero-order bias \cite{namikawa2013,namikawa2014} with the help of Monte Carlo simulations is preferred to the evaluation of the analytic formula, since it accounts for small mismatches in the two-point statistic between simulation and data. Analytic expressions for $N_L^{(1)}$ can be found in \cite{kesden2003, anderes2013} and an analog method to compute it using Monte Carlo simulations in \cite{story2015}.\\
\newpage
The different estimators for $\phi$ can be combined into an optimal minimum-variance estimator as
\be
\hat{\phi}^\text{mv}_{LM}\equiv\sum_{XY} w^{AB} \hat{\phi}_{LM}^{AB},
\ee
with weights
\be
w^{AB}_L = N^{mv}_L \sum_{CD} \left( \mathbf{N}^{-1}_L \right)^{ABCD}
\ee
and minimum variance lensing noise
\be
N^\text{mv}_L=\left(  \sum_{ABCD} \left( \mathbf{N}^{-1}_L \right)^{ABCD} \right)^{-1}.
\ee

\subsection{Effect of non-Gaussianities on quadratic estimators}
The formalism derived in the previous section assumes that all the non-Gaussianity in the CMB is entirely due to the lensing effect and that the lensing potential is a Gaussian field. However, this is just an approximation and if the lensing potential (or equivalently the lensing convergence) has nonzero higher-order correlations, there are additional terms involving four-point functions of lensed CMB fields that create distinct biases. This problem was first studied in \cite{bohm2016} in the context of assessing the impact of the nonlinear evolution of the matter distribution in the lensing reconstruction. In this work the authors derived expressions for the bias induced by a nonzero bispectrum in the lensing potential caused by the nonlinear gravitational collapse that is of order $\mathcal{O}\left(\left(C_L^{\phi\phi}\right)^{3/2}\right)$ and is referred to as $N_L^{(3/2)}$. The TT reconstruction channel was found to be the most sensitive on angular scales $\ell\lesssim 1000$ considered in their work and could reach the level of $2.5\%$ for low noise and large sky coverage experiments. This level of bias is significant in light of the expected future experimental sensitivity. Understanding the amplitude and nature of higher-order biases and their effect on our ability of constraining the cosmology is therefore crucial.\\

Modeling these effects analytically becomes cumbersome very quickly. Therefore, we decide to adopt a numerical approach and assess the impact of these biases through accurate and realistic numerical simulations. In order to tackle the problem in its full complexity we decide to use simulations that include not only the nonlinear evolution of matter studied in \cite{bohm2016} but also non-Gaussianity induced by post-Born effects. Analytical predictions of the shape and amplitude of these non-Gaussian correlations have recently been computed in \cite{pratten2016, lewis2016, marozzi2016pol}.  
\section{Modeling CMB lensing at higher order}\label{sec:sims} 
To test the bias in the lensing reconstruction induced by non-Gaussian evolution of the matter and post-Born effect we need to simulate the lensing of CMB anisotropies including both these effects. For this purpose we use the simulation method and results of \cite{fabbian2018} (hereafter FCC18). This work produced a collection of lensing observables $\kappa, \omega, \phi, \Omega$, derived from a $\Lambda$CDM $N$-body simulation of the DEMNUni suite \cite{carbone2016, castorina2015} in the Born approximation and using multiple-lens ray tracing techniques. The $N$-body simulation employed in FCC18 used $2048^3$ dark matter particles and a box size of 2 Gpc/$h$ from $z=99$ to $z=0$. This redshift range cover allows to reproduce the CMB lensing kernel $D_A(\chi^* -\chi)/D_A(\chi)D_A(\chi^*)$ with sub-percent precision. The mass resolution of the simulation at $z = 0$ is $M_{\rm CDM} = 8.27 \times 10^{10} M_{\odot} / h$ and the gravitational softening length is set to $\epsilon_s = 20$ kpc$/h$ corresponding to 0.04 times the mean linear inter-particle separation.
Below, we briefly summarize the specificities of the light-cone construction and ray tracing algorithm adopted in these simulations as well as further tests complementary to the one presented in FCC18 and specifically performed for this work. We refer the reader to FCC18 for a more detailed discussion.

\subsection{Ray tracing algorithm for CMB lensing}

Starting from a series of snapshots in time of an $N$-body simulation, the algorithm adopted in FCC18 reconstructs the full-sky past light cone of the observer from redshift $z=0$ to the maximum redshift covered by the simulation $z_\textrm{max}$ (in our case $z_\textrm{max}=99$). Because the universe volume simulated in the $N$-body is finite, we replicate the box volume in space to fill the whole observable volume between $0\leq z \leq z_\textrm{max}$. To avoid repeating the same structures along the line of sight and to recover (at least partially) structures on scales comparable to the box size, the algorithm employs a specific randomization procedure for the particle positions as in \cite{carbone2008, calabrese2015}. The light cone is then sliced into spherical shells of thickness $\Delta\chi=150\ {\rm Mpc}/h$. The particles inside each of these volumes are then projected onto spherical planes of surface mass density, as described in \cite{fosalba2008}. The algorithm then converts the surface mass density planes into convergence fields. With this discrete version of the light cone at hand, it is convenient to discretize the geodesic and lens equation of Eqs.~\eqref{eq:beta} and \eqref{eq:distortionmatrix} \cite{jain2000, hamana2001, schneider1992}
\be
\boldsymbol\beta(\boldsymbol\theta,\chi) = \boldsymbol\theta -
\sum_{k=0}^{N-1}\frac{D_A(\chi -
  \chi_k)}{D_A(\chi)}\boldsymbol\alpha^{(k)}(\boldsymbol\beta^{(k)}).
\label{eq:fulllenseq_dis}
\ee
Here $k$ is the shell index and we define the gradient of the two-dimensional (2D) projected gravitational potential as
\be
\boldsymbol\alpha^{(k)}(\boldsymbol\beta^{(k)}) = \frac{2}{c^2} \int_{\chi_k-\Delta \chi}^{\chi_k+\Delta \chi} d \chi \frac{\boldsymbol\nabla\Psi(\boldsymbol\beta^{(k)}, \chi)}{D_A(\chi_k)} .
\ee
$\boldsymbol\alpha^{(k)}$ is easily computed starting from the convergence field of each shell $(k)$ using a spin-1 spherical harmonic transform \cite{das2008, becker2013} in the $E$ and $B$ decomposition
\be
_1\alpha^{(k),E}_{\ell m}=\frac{2\kappa^{(k)}_{\ell m}}{\sqrt{\ell(\ell+1)}} \qquad  _1\alpha^{(k), B}_{\ell m}=0
\label{eq:alpha}.
\ee
The latest operation requires the computation of the spherical harmonic coefficients $\kappa^{(k)}_{\ell m}$ using a fast spherical harmonic transform up to a given cut-off in power $\ell_\textrm{max}$. The choice of $\ell_\textrm{max}$ for each different shell is optimized to ensure the total deflection is computed with sub-percent precision for scales $\ell\lesssim 8000$.The magnification matrix follows straightforwardly from Eq.~\eqref{eq:fulllenseq_dis} as
\be
A^{N}_{ij}(\boldsymbol\theta, \chi_N) = \delta^K_{ij} - \sum^{N-1}_{k=0} \frac{D_{k,N}}{D_N}
U^{(k)}_{ip}(\boldsymbol\beta^{(k)},\chi_k)
A^{(k)}_{pj}(\boldsymbol\theta,\chi_k),
\label{eq:distortionmatrix_dis}
\ee
where $N$ is the number of planes necessary to reach the source at comoving distance $\chi_N$ and $U_{ij}$ is the matrix of the second derivatives of the gravitational potential, $\partial\Psi/\partial{\beta_i}\partial{\beta_j}$. $U_{ij}$ can be computed easily as derivatives of the component of the spin-1 field $\boldsymbol\alpha^{(k)}$ (see Appendix A of FCC18). In Eq.~\eqref{eq:distortionmatrix_dis} we use the notation $D_{k,N} \equiv D_A(\chi_N - \chi_k)$ and $D_k \equiv D_A(\chi_k)$ for simplicity. 
Implementing Eq.~\eqref{eq:distortionmatrix_dis} in numerical simulations becomes quickly prohibitive for a large number of lens planes and large sky fraction. FCC18 adopted the multiple lens approach of \cite{hilbert2009}, who showed that the equation can be rewritten in a more efficient form that requires one to store in a memory for a given $k$th iteration just the position of the light rays at the two previous positions $\boldsymbol\beta^{(k-2)}$ and
$\boldsymbol\beta^{(k-1)}$,
\ba
&\boldsymbol\beta^{(k)}&  = \left(1 -
\frac{D_{k-1}}{D_k}\frac{D_{k-2,k}}{D_{k-2,k-1}}\right) \boldsymbol\beta^{(k-2)} \\
& + &\frac{D_{k-1}}{D_k}\frac{D_{k-2,k}}{D_{k-2,k-1}} \boldsymbol\beta^{(k-1)} 
 - \frac{D_{k-1,k}}{D_k} \boldsymbol\alpha^{(k-1)} (\boldsymbol\beta^{(k-1)}).\nonumber
\label{eq:hilbert-beta}
\ea
By differentiating with respect to $\boldsymbol\theta$ as in
Eq.~\eqref{eq:distortionmatrix}, we obtain
the recurrence relation for the magnification matrix
\ba
&A^{(k)}_{ij}& = \left( 1 -
\frac{D_{k-1}}{D_k}\frac{D_{k-2,k}}{D_{k-2,k-1}}\right) A^{(k-2)}_{ij} \\
&+& \frac{D_{k-1}}{D_k}\frac{D_{k-2,k}}{D_{k-2,k-1}} A^{(k-1)}_{ij}
 - \frac{D_{k-1,k}}{D_k} U^{(k-1)}_{ip} A^{(k-1)}_{pj}.\nonumber
\label{eq:hilbert-a}
\ea

This algorithm was originally developed in the context of galaxy lensing, but adapted to spherical geometry in \cite{das2008} and developed first in \cite{calabrese2015, fabbian2018} for CMB lensing. This approach is also convenient to derive the magnification matrix and lensing observables in the Born approximation that we will use later to isolate the contribution coming from post-Born effects. Assuming the background distortion, the first-order magnification matrix is 
\be
\label{eq:first_order_aij}
A^{(N), {\rm 1st}}_{ij}(\boldsymbol\theta, \chi_s) = \delta^K_{ij} - \sum^{N-1}_{k=0}
\frac{D_{k,N}}{D_N} U^{(k)}_{ij}(\boldsymbol\theta,\chi_k).
\ee
We note that the $U_{ij}$ matrix is symmetric because mixed derivatives commute and thus the rotation, $\omega$, is identically zero. 

\subsection{Impact of the LSS bispectrum}\label{sec:cmb-bispectrum}
FCC18 carried out an accurate characterization of the post-Born corrections on $\kappa$, $\omega$ and lensed CMB power spectra and compared extensively with their analytical predictions derived in \cite{pratten2016, lewis2016, marozzi2016, marozzi2016pol, lewis2017}. However, the analysis did not investigate in detail the impact of the nonlinear evolution of large-scale structures and how simulation properties match with analytical predictions of the higher-order statistics of the $\kappa$ field. Below we present additional validation tests performed to assess the reliability of these simulations in modeling higher-order statistics of post-Born corrections and nonlinear LSS evolution. We limit our analysis to the statistics of the $\kappa$ field and its cross-correlation with $\omega$. Higher-order statistics of the curl mode of the deflection field beyond the mixed $\kappa\kappa\omega$ bispectrum \cite{pratten2016}, which appear at higher order in the perturbative expansion, are lacking theoretical predictions. The measurement of the $\kappa\kappa\omega$ and $\kappa\omega\omega$ bispectrum in the simulations used in this work through its effect on lensed $B$-modes power spectrum was presented in FCC18, together with the measurement of the post-Born induced curl mode on lensed CMB power spectra. We refer the reader to that work for a more in-depth discussion and comparison with theoretical predictions. 

\subsubsection{Higher-order statistics of the CMB convergence}

To verify the accuracy of the simulations in reproducing the expected level of non-Gaussianity in $\kappa$, we compare its skewness as measured in the simulations with the values obtained by contracting the predicted theoretical bispectrum including LSS nonlinearity and post-Born corrections. The definition of skewness given a pixelized map of a scalar field, $X$, is
\begin{align}
S_3[X]&=\left\langle XXX \right\rangle = \frac{1}{N_\textrm{pix}}\sum_p^{N_\textrm{pix}} X_p^3,
\end{align}
where $p$ is the pixel index and $N_\textrm{pix}$ the total number of pixels in the map.
Following \cite{Srednicki,komatsu2001}, we compute the skewness in terms of the reduced bispectrum $b_{L_1 L_2 L_3}$ as
\begin{widetext}
\begin{align}
S_3[b_{L_1 L_2 L_3}]&=
\sum_{L_1 L_2 L_3}^{L_\textrm{max}}  \frac{(2 L_1+1)(2L_2+1)(2L_3+1)}{(4\pi)^2}
\begin{pmatrix}
L_1 & L_2 & L_3 \\
0 & 0 & 0
\end{pmatrix}^2
b_{L_1 L_2 L_3},
\end{align}
with corresponding variance dominated by the disconnected six-point function
\begin{align}
\sigma_{S_3}^2\simeq \frac{6}{4\pi} \sum_{L_1 L_2 L_3}^{L_\textrm{max}} \frac{(2 L_1+1)(2 L_2+1)(2 L_3+1)}{(4\pi)^2} 
\begin{pmatrix}
L_1 & L_2 & L_3 \\
0 & 0 & 0
\end{pmatrix}^2
C_{L_1}C_{L_2}C_{L_3}.
\end{align}
\end{widetext}

In particular, the skewness of the Born-approximated convergence, $\kappa^{F}$, obtained from the first-order magnification matrix, provides a measurement of the LSS-induced bispectrum. The bispectrum of the convergence computed using the multiple lens ray tracing algorithm, $\kappa^R$, receives a contribution from the LSS-induced bispectrum as well as from the post-Born corrections induced bispectrum. The difference of the skewness of $\kappa^R$ and $\kappa^F$ gives thus a direct measurement of the collapsed post-Born-induced bispectrum.\\*

\begin{figure}[!htbp]
\centering
\includegraphics[width=1.\linewidth]{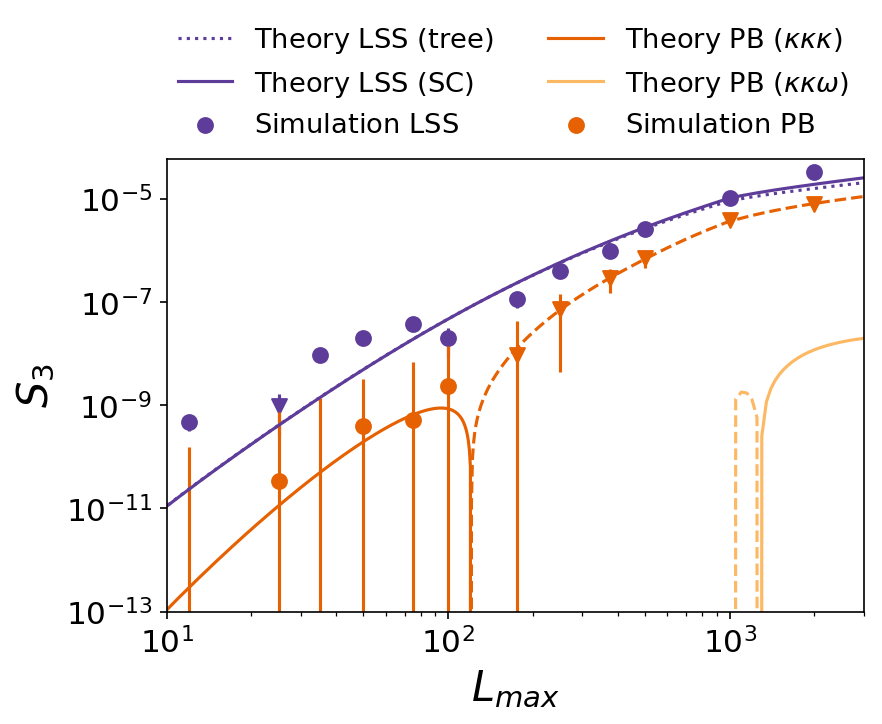}
\caption{Comparison of the skewness for different cutoff values of the convergence multipoles. The theory curves are computed using the tree-level expression of the LSS convergence bispectrum including the Scoccimarro \& Couchman fit of \cite{SC}, as well as post-Born corrections of the $b^{\kappa\kappa\kappa}$ and $b^{\kappa\kappa\omega(+)}$ bispectra of \cite{pratten2016}. Only the absolute values are shown; negative values are marked by a dashed line or triangular marker.}
\label{fig:theoconsistency}
\end{figure}

We use the formulas presented in \cite{Bernardeau} and \cite{pratten2016} to compute the bispectrum of $\kappa$ due to LSS nonlinearity (at tree level in density perturbations or adopting the nonlinear fitting formula from \cite{SC}) and post-Born effects, respectively. In Fig.~\ref{fig:theoconsistency} we show a comparison between the skewness measured in the low-pass-filtered simulations and their expected theoretical value as a function of the maximum multipole cutoff used in the calculations. 
We find a good agreement between simulation and theoretical expectations for the post-Born bispectrum part, confirming the findings of FCC18 on the level of the lensed CMB $B$-mode power spectrum. For this observable, the post-Born $\kappa\kappa\kappa$ bispectrum is the dominant correction while the contribution of the curl mode in terms of the $\kappa\kappa\omega$ bispectrum is negligibly small (see also \cite{lewis2016}). The LSS skewness agrees well with theoretical expectation on scales $75\lesssim L_\textrm{max}\lesssim 2000$ and starts deviating outside this range, yet still with reasonable agreement.
On the largest scales, the discrepancy might be due to the adoption of Limber approximation or by spurious numerical correlations induced by the box size replication during the light-cone construction or simply sample variance of the matter bispectrum.  In fact, \cite{ruggeri2018} measured the three-dimensional matter bispectrum from the same $N$-body simulation used for this work and found an excess of power at low values of $k\lesssim 0.1\ \textrm{Mpc}^{-1}h$ for both squeezed and equilateral configurations. These scales contribute significantly to the signal on angular scales $\ell\lesssim 100$ (see e.g. \cite{lewis2006}) and could be responsible of the excess of skewness observed when only such scales are included.  Although in FCC18 the replication procedure was shown to produce accurate results on the large scales of $C_{L}^{\kappa\kappa}$ and no significant spurious excess of power was observed, we tested the stability of our results on lensing reconstruction with respect to the minimum multipole employed in the analysis. We found negligible differences when excluding CMB angular scales $\ell\leq100$.\\*

\begin{figure*}[t!]
\centering
\includegraphics[width=.5\textwidth]{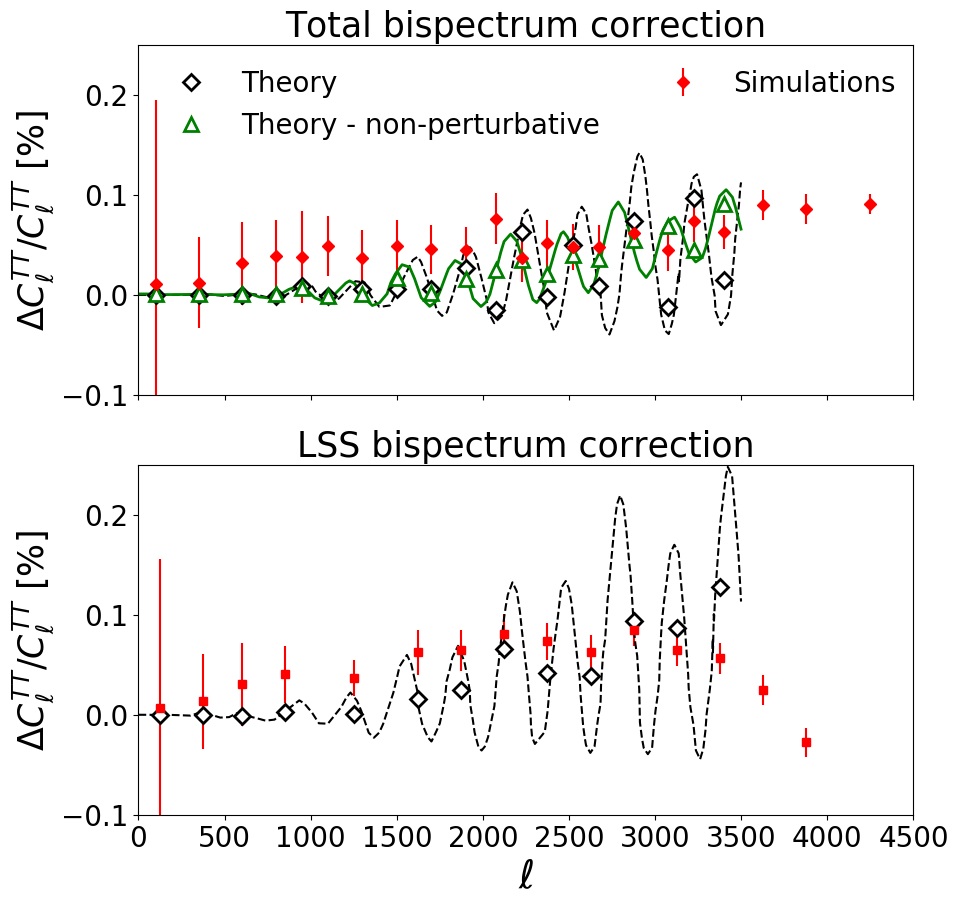}\includegraphics[width=.5\textwidth]{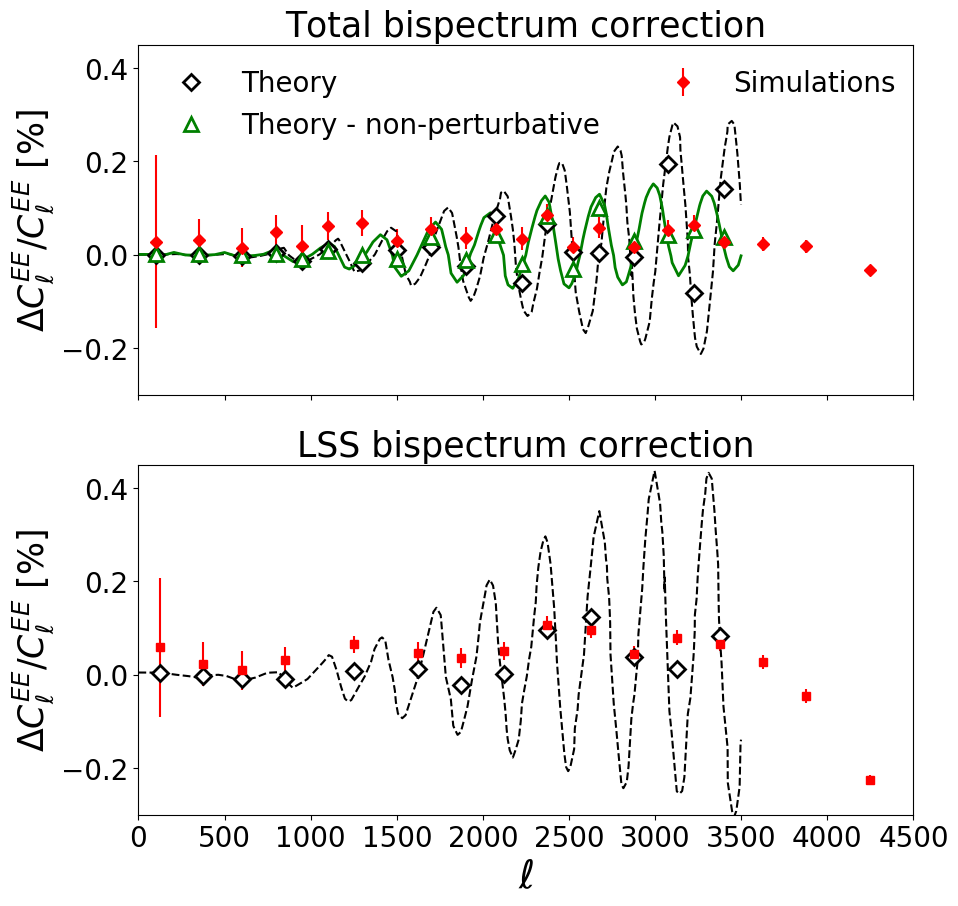}
\caption{Impact of CMB convergence bispectrum on lensed temperature (left) and $E$-mode (right) power spectra. The top panel shows the total correction accounting for the LSS and post-Born induced bispectrum, while the bottom panel shows the correction due to only nonlinear LSS evolution. The theoretical predictions of \cite{lewis2016} are shown in black and simulation results in red. The green curves show the values of the nonperturbative corrections computed in \cite{lewis2016} for the temperature and $E$-mode power spectra. Binned theoretical predictions are shown with empty markers. The error bars include only the uncertainty on the average over the Gaussian MC realizations and do not include the sample variance of the convergence bispectrum.}\label{fig:cl-bispectrum-corr}
\end{figure*}

\begin{figure}[b]
\centering
\includegraphics[width=.5\textwidth]{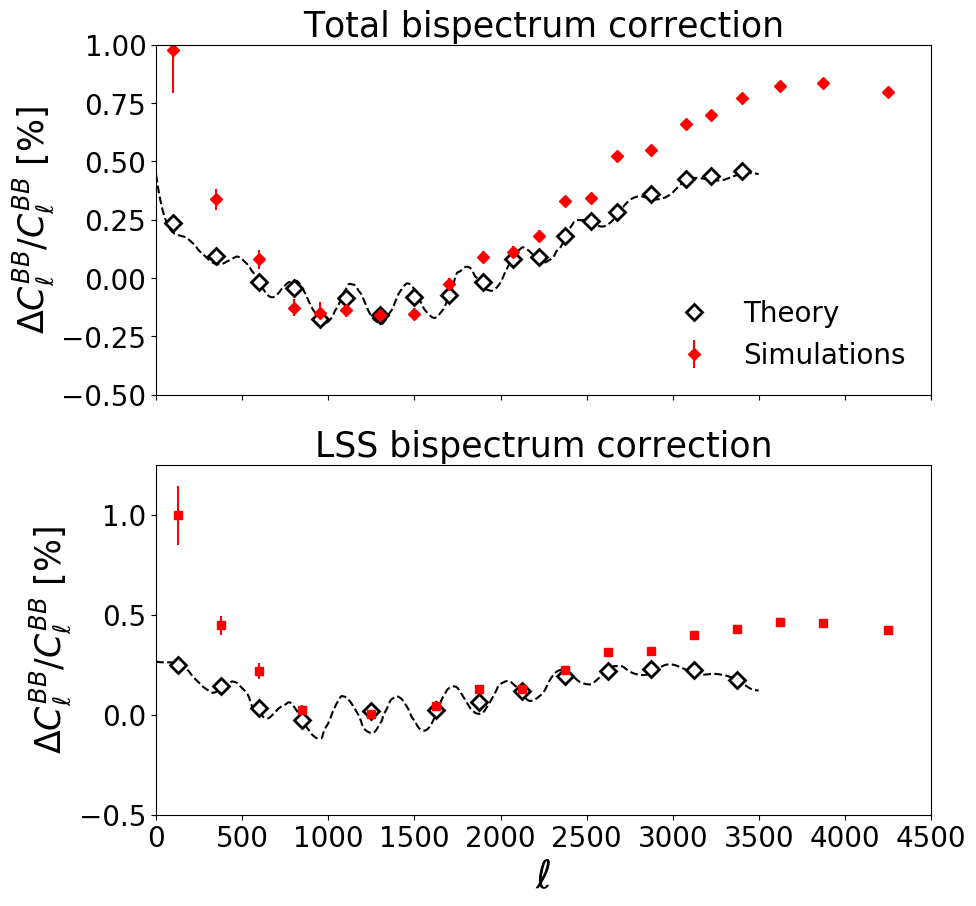}
\caption{Same as Fig.~\ref{fig:cl-bispectrum-corr} for the BB power spectrum.}
\label{fig:bb-bispectrum-corr}
\end{figure}

At angular scales $L_\textrm{max}\gtrsim 2000$ we expect to see discrepancies due to the limitation of the fitting formulas used to compute the theoretical expectation as well as theoretical uncertainties in the modeling of the nonlinear matter power spectrum used to compute the theoretical expectation of the skewness. In particular, at $L \approx2000$, the CMB convergence receives a non-negligible contribution from structures at scales $k\gtrsim 1~\textrm{Mpc}^{-1}h$ \cite{fabbian2018, lewis2006} and on these angular scales uncertainties on the matter power spectrum are already of the order of 15\% \cite{takahashi2012}. The use of nonlinear fitting formulas improves the agreement with simulation results with respect to the tree-level bispectrum. We note that we do not investigate possible improvement using alternative nonlinear bispectrum fitting formulas, as, for example, the one introduced in \cite{gilmarin2012}. The validity of these equations at high redshifts was not validated and the differences with respect to the Scoccimarro \& Couchman formulas formulas \cite{SC} were shown to be marginal and relevant only for a subset of the bispectrum configurations (see discussion in \cite{pratten2016}).

\subsubsection{LSS bispectrum effect on lensed CMB}

Non-Gaussianity in the lensing potential can affect the shape of the lensed CMB power spectra. The authors of \cite{lewis2016} (hereafter LP16) computed the effect on the CMB power spectrum induced by the bispectrum of the CMB convergence due to the nonlinear evolution of matter (hereafter LSS term) and the one due to post-Born corrections (hereafter PB term). FCC18 showed that the corrections computed by LP16 for the PB term match very well the results extracted from ray tracing simulations. As a validation test for this work, we focused on measuring the corrections to lensed CMB power spectra generated by the LSS term alone, as well as those due to the combination of LSS and PB terms. We then compared the results of the simulations with the theoretical prediction of LP16. To isolate the LSS term, we lens 100 Gaussian realizations of unlensed CMB maps with a deflection field extracted from the $\kappa^F$ map as performed in FCC18. From the average of the power spectra of these maps we subtract the average power spectrum of the 100 CMB realizations that were lensed with a deflection field computed from a Gaussian realization of the lensing convergence $\kappa^G$ with power spectrum $C_{L}^{\kappa^F\kappa^F}$. Similarly, to measure the total correction, we repeat the same procedure with $\kappa^R$ and $C_{L}^{\kappa^R\kappa^R}$ to produce the Gaussian realizations of the deflection field. In Figs.~\ref{fig:cl-bispectrum-corr} and \ref{fig:bb-bispectrum-corr} we show the results of this analysis together with a comparison with the prediction of LP16. \\

The theoretical predictions for both the total and LSS bispectrum (which is the dominant term) agree quite well with the simulation results on the relevant angular scales, especially the ones implementing the nonperturbative formalism for the TT and EE power spectrum as discussed in PL16 and FCC18. This approach accounts for the fact that even in the Gaussian approximation lensing is a $\mathcal{O}(1)$ effect at small scales and therefore treating the corrections due to non-Gaussianity as perturbations around an unlensed field leads to inaccurate results.\\*

Despite the overall good agreement, however, some differences can be observed. This is expected because, unlike the analytical approximations, simulations include the effects of non-Gaussianity nonperturbatively and the exact shape of the correction depends on the detailed shape of the bispectrum. In particular, simulation results show an excess of power on the $B$-mode power spectrum compared to analytical predictions. This is consistent since $B$-modes are more sensitive to small scale lenses and thus non-Gaussianities due to strongly nonlinear density fields are expected to give larger corrections where the perturbative expansion becomes less accurate. The discrepancies at scales $\ell\lesssim 100$ could conversely arise due to the excess of skewness discussed in the previous section, although we stress that a larger skewness does not seem to affect significantly the temperature and $E$-mode power spectrum, where the corrections are dominated by structures at $\ell\lesssim 300$. Nevertheless, we decide to perform dedicated robustness tests in the following section to assess the impact of this discrepancy as a potential systematic effect. 

\section{Results}\label{sec:results} 

\subsection{Numerical setup}
To measure the $N_L^{(3/2)}$ bias, we produce several sets of lensed CMB maps using the Lenspix code\footnote{We found consistent results when analyzing maps simulated with the LenS$^{2}$HAT code \cite{fabbian2013} which implements a different interpolation scheme to resample the unlensed CMB realization at the displaced ray position.}. These are later combined in different ways to isolate different contributions to this bias and to perform consistency and robustness tests. A subset of these simulations are briefly described in Sec.~\ref{sec:cmb-bispectrum}, and here we review the procedure in more detail.
First, we simulate 100 Gaussian realizations of the primordial CMB. Each of these simulations is then lensed using seven different simulated deflection fields $\boldsymbol\alpha^\textrm{eff} = \nabla\phi +\nabla\times\Omega$ and adopting the effective remapping for the CMB photons as in Eq.~\eqref{eq:full-remapping}. The $\phi$ and $\Omega$ potentials are obtained from the $\kappa$ and $\omega$ field of FCC18 using the consistency relations in Eqs.~\eqref{eq:conskappa} and \eqref{eq:consomega} in the harmonic domain. For this operation as well as in the synthesis of the unlensed CMB, we adopted a bandlimit parameter $\ell_\textrm{max}=6200$. According to the findings of \cite{fabbian2013}, this setup allows us to recover lensed CMB with a precision of $\mathcal{O}(10^{-3})$ on scales $\ell\lesssim 4000$ and $\mathcal{O}(10^{-2})$ at $\ell\approx 5000$. The full set of deflection fields used to lens the CMB are therefore as follows:

\begin{itemize}
\item $\kappa^G$\!. A Gaussian realization of convergence with power spectrum $C_L^{\kappa^F\kappa^F}$.
\item $\pm\kappa^{F}$\!. These simulations measure the bias including only the effects of the nonlinear LSS evolution.
\item $\pm\kappa^{R}$ alone. These simulations measure the bias due to LSS nonlinearity and PB effects in the convergence field.
\item $\pm\kappa^{R}$ and $\pm\omega^{R}$ ($\pm\kappa^{R\omega}$ hereafter). They include the full set of nonlinearity of LSS and PB corrections, including the so-called mixed bispectrum correlations $\kappa\kappa\omega$ and $\kappa\omega\omega$ (we refer the reader to \cite{pratten2016, fabbian2018} for further discussion). 
\end{itemize}

We denote the resulting lensed CMB simulations with a given deflection field by a superscript $G$, $\pm F$, $\pm R$ or $\pm R \omega$, respectively. For the results described in this paper we use
maps having an angular resolution of 52 arcsec in \textsc{Healpix} pixelization, corresponding to $N_\textrm{side}=4096$. On each of these sets we run the lensing reconstruction using a quadratic estimator and compare them to extract different sources of biases. 
Each simulation set is designed to contain a lensing potential with the same mean power spectrum $C_L^{\phi\phi}$.\footnote{$C_{L}^{\phi\phi}$ extracted from a $N$-body simulation has a potential bias at small angular scales due to the presence of shot noise due to the finite number of particles in the $N$-body simulation. According to the estimates of FCC18, the shot noise accounts for roughly $15\%$ of the amplitude of the power spectrum on the maximum multipole relevant for this analysis. Because in the following text we compare simulated quantities, all including the shot-noise term, the impact of the shot-noise term on the results is expected to be highly reduced.} Remaining relative deviations from the fiducial $C_L^{\phi\phi}$ due to post-Born corrections are below $0.2\%$ on the relevant scales considered in this paper. Hence, in the following, we assume $N_L^{(0)}$ and $N_L^{(1)}$ to be equal for all simulations. Under this assumption we can write
\begin{eqnarray}
\hat{C}_L^{\phi\phi}[\kappa]&=&\frac{1}{2L+1} \sum_M \hat{\phi}^{\dagger}_{LM} \hat{\phi}_{LM}  \\ 
&\approx& C_L^{\phi\phi} + N_L^{(0)} + N_L^{(1)} + N_L^{(3/2)}[\kappa] + \mathcal{O}\left( \phi^4, \Omega^2 \right),\nonumber
\end{eqnarray}
where only the $N_L^{(3/2)}$ bias depends on the specific statistic of the $\kappa$ field used to lens a specific simulation. We will test the validity of this assumption in Sec.~\ref{sec:consistency}. \\

In order to evaluate the bias in a specific experimental configuration we add Gaussian noise realizations with corresponding power spectrum $N_{\ell}=\sigma_n^2 B^2_\ell$, with white noise level, $\sigma_n$, and a circular Gaussian beam with FHWM size, $\theta$ \cite{knox1995},
\begin{equation}
B_\ell = \exp\left( \ell(\ell+1) \frac{\theta^2}{16\log 2} \right).
\end{equation} 

\subsection{Measurements of $N_L^{(3/2)}$ bias }\label{sec:n32-measurements}

To measure the $N_L^{(3/2)}$ biases from the simulations and distinguish the contributions to the biases originating from all the different contributions of the $\kappa$ bispectrum and correlations involving curl modes ($\kappa\kappa\omega+\kappa\omega\omega$, PB$\omega$ hereafter), we combine the reconstructed CMB lensing potential power spectrum on each set of lensed CMB realizations as follows:

\begin{description}
\item[LSS] $N_L^{(3/2)}=\left\langle \hat{C}_L^{\phi\phi}[\kappa^{F}]-\hat{C}_L^{\phi\phi}[\kappa^{G}]\right\rangle_\textrm{Lensed CMB}$
\item[PB] $N_L^{(3/2)}=\left\langle \hat{C}_L^{\phi\phi}[\kappa^{R}]-\hat{C}_L^{\phi\phi}[\kappa^{F}]\right\rangle_\textrm{Lensed CMB}$
\item[PB$\omega$] $N_L^{(3/2)}=\left\langle \hat{C}_L^{\phi\phi}[\kappa^{R\omega}]-\hat{C}_L^{\phi\phi}[\kappa^{R}]\right\rangle_\textrm{Lensed CMB}$
\item[Total] $N_L^{(3/2)}=\left\langle \hat{C}_L^{\phi\phi}[\kappa^{R\omega}]-\hat{C}_L^{\phi\phi}[\kappa^{G}]\right\rangle_\textrm{Lensed CMB}$,
\end{description}
\noindent
where we denote in squared brackets the corresponding set of CMB realizations used in the lensing reconstruction. The total bias is equal to the sum of the former three, well within the uncertainties shown later in the text.\\

We report the measurement of the $N_L^{(3/2)}$ as the average over the 100 lensed CMB simulations at our disposal for each deflection field configuration. The error bars shown in the following figures are computed from the dispersion of the lensed CMB simulations and represent the uncertainty on the mean of the simulations. Due to the fact that the realizations of primordial CMB are the same for all sets of simulations, we avoid realization-dependent biases (up to bispectrum terms) and cosmic variance noise. 
In the following we discuss the impact of $N_L^{(3/2)}$ bias in terms of the ratio between the bias and the lensing potential power spectrum measured in the FCC18 simulations. The reported signal-to-noise ratio (SNR) is computed as the ratio between $N_L^{(3/2)}$ and the error bar expected for a specific experimental configuration
\be
\sqrt{\frac{2}{2L+1} \frac{1}{f_\textrm{sky}\Delta L}}\left( N_L^{(0)} + N_L^{(1)}\right),
\ee
where we assume the observed sky fraction to be $f_\textrm{sky}=40\%$, to match the expected sky coverage of CMB-S4, and the bin size $\Delta L \approx 140$. For all configurations the minimal CMB multipole used is $\ell_\textrm{min}=2$. \\

\begin{figure*}[!htbp]
\includegraphics[width=.93\linewidth]{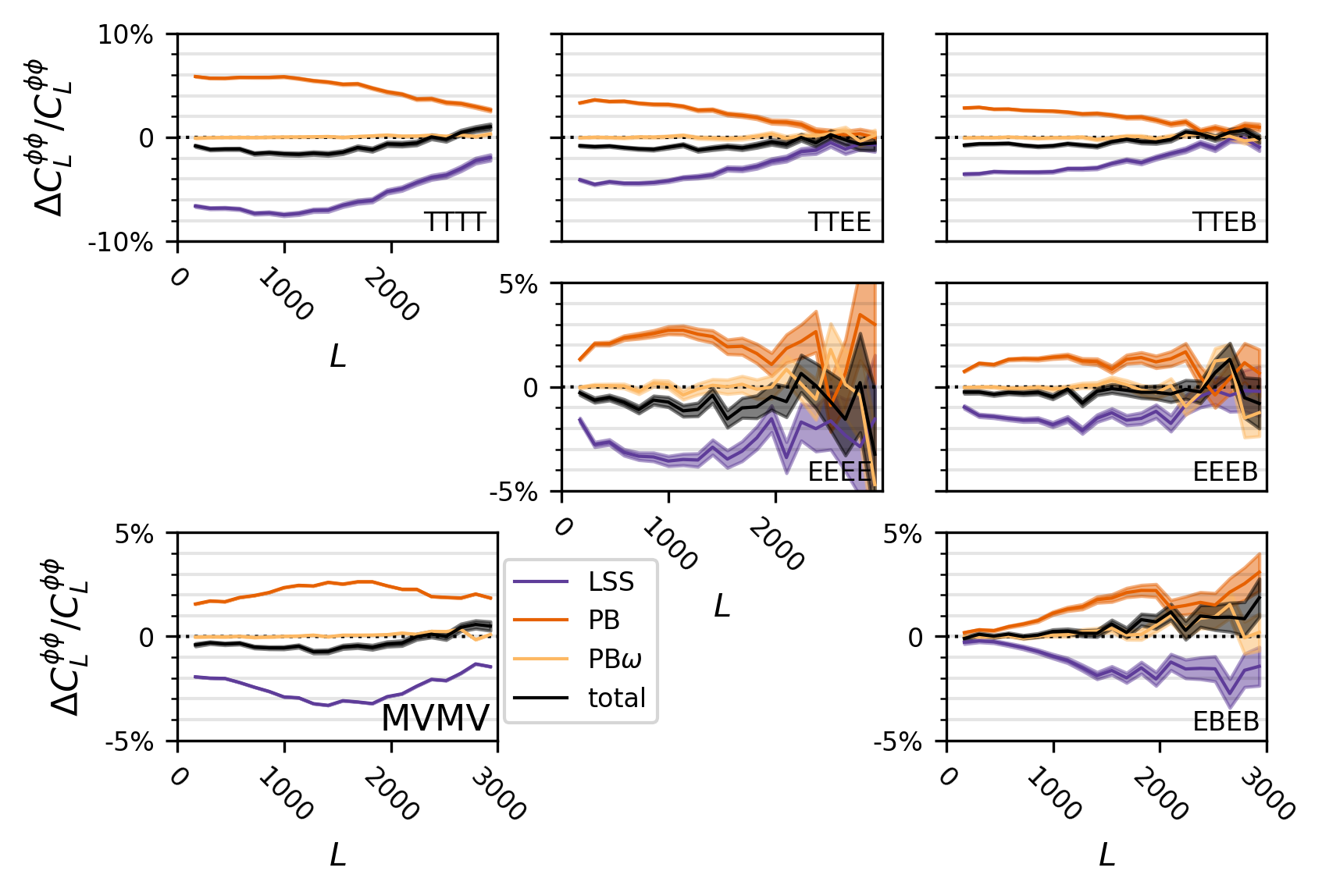}
\caption{Relative biases in the estimated lensing potential power spectrum induced by non-Gaussian statistics of the underlying lensing potential (black curves) as measured in the FCC18 simulations. This case included lensed CMB modes up to $\ell_\textrm{max}=4000$ and CMB-S4-like experimental configuration. We differentiate the effects caused by nonlinearities of large-scale structures (LSS, purple curve), post-Born lensing effects (PB, orange curve) as well as post-Born mixed bispectrum terms (PB$\omega$, yellow curve) accounting for higher-order correlation between the lensing gradient and curl potential. The shaded areas show the uncertainty on the bias computed from the dispersion of 100 lensed CMB simulations.}
\label{fig:4knoise}
\end{figure*}

In Fig.~\ref{fig:4knoise} we show the total $N_L^{(3/2)}$ bias for the minimum-variance quadratic estimator due to non-Gaussianity in the lensing deflection field, along with the breakdown of the contribution of each source of non-Gaussianity (LSS, PB, and PB$\omega$). These results are derived performing lensing reconstruction using a sharp cutoff in harmonic space that removed all the CMB harmonic coefficients having $\ell\geq\ell_\textrm{max}=4000$ and assuming an experiment with $1.4\ \mu K\textrm{-arcmin}$ white noise in polarization ($1\ \mu K\textrm{-arcmin}$ in temperature) and a $1\ \textrm{arcmin}$ beam size to match CMB-S4 configuration. 
We find that post-Born effects produce a positive bias in the lensing reconstruction, while LSS effects suppress power in the reconstructed potential. This leads to an important cancellation of the two effects and, in fact, the total $N_L^{(3/2)}$ bias becomes a subpercent effect. The amplitude of $N_L^{(3/2)}$, however, changes quite significantly depending on which combination of the quadratic estimator is used for the lensing reconstruction. 
At low multipoles the individual relative contributions to the biases induced by LSS and PB can reach up to 7\% in the autopower spectrum of the $TT$ estimator. Generally, the bias amplitude grows with the number of contributing temperature fields used in the estimator. For polarization-based estimators the overall bias can reach 2\% for both LSS and PB terms when considered separately. In our experimental setup, the polarization-based estimators provide the most important contribution to the minimum variance combination below $L\approx 1500$, while for larger multipoles the temperature reconstruction, which is more sensitive to small-scale lenses, starts to dominate in the minimum-variance combination. \\

\begin{figure}[!t]
\centering
\includegraphics[width=.49\textwidth]{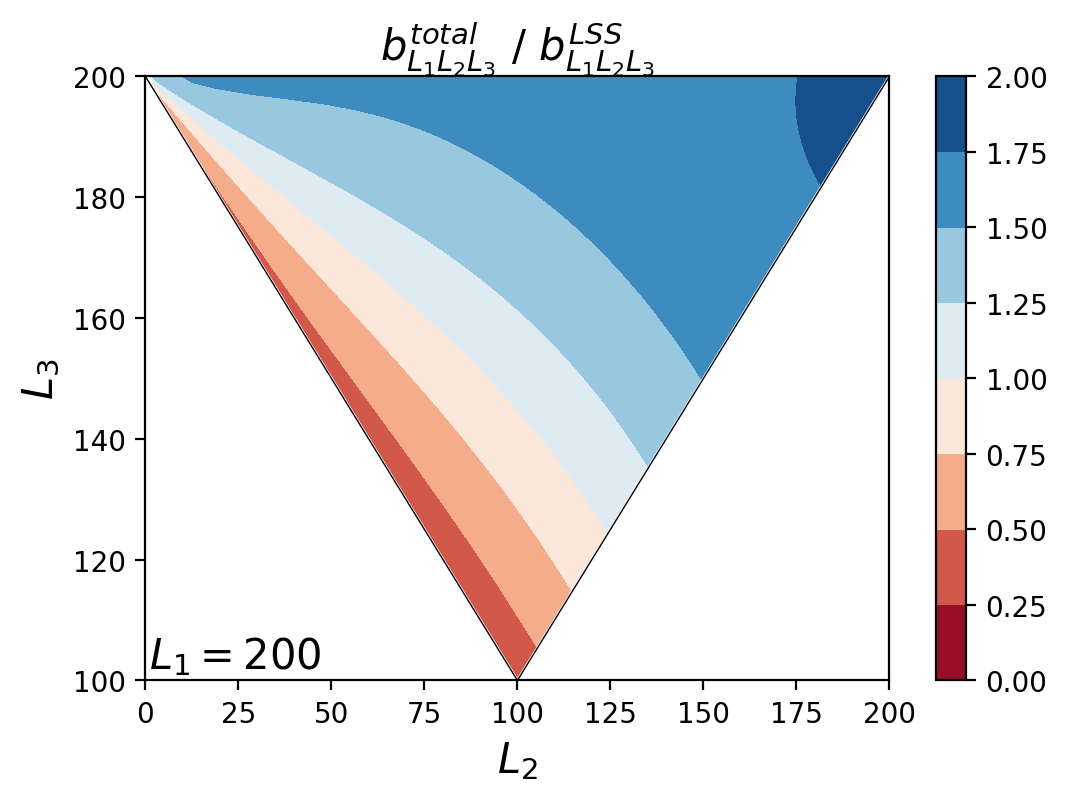}\\
\includegraphics[width=.49\textwidth]{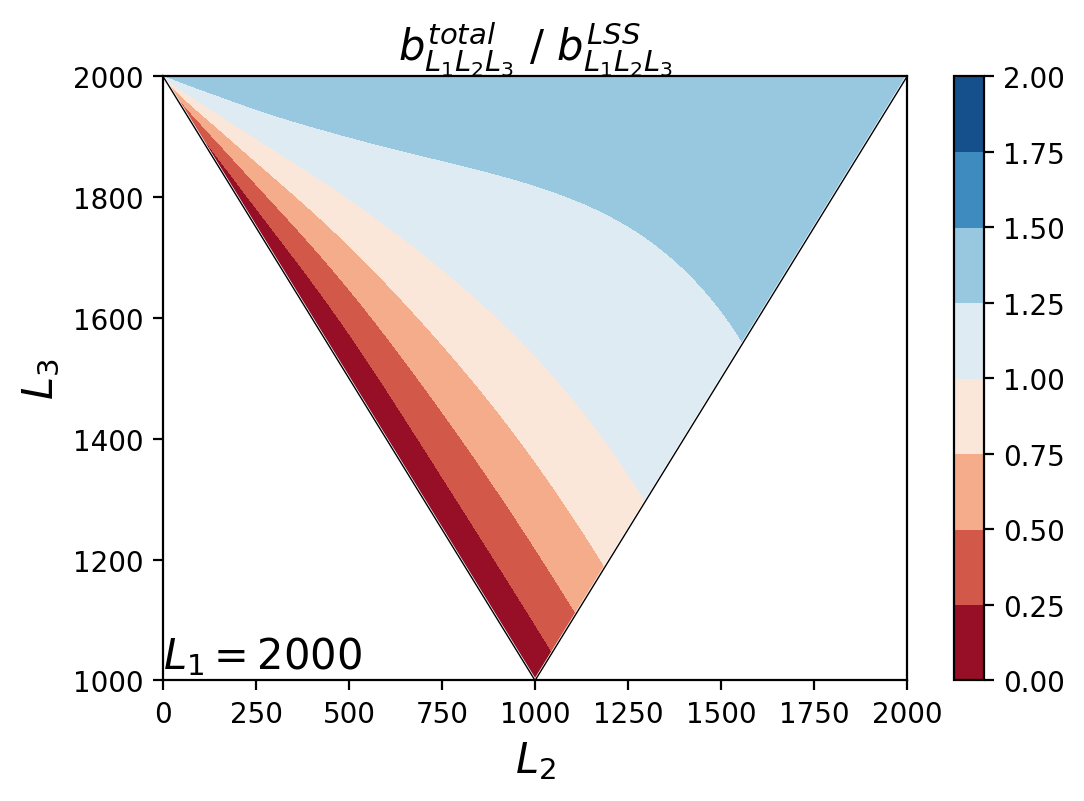}
\caption{The relative suppression or enhancement of the convergence bispectrum from large-scale structure nonlinearity (derived using the fitting formula of \cite{SC}) due to post-Born effects, for $L_1=200$ and $L_1=2000$.}
\label{fig:bispectratriangles}
\end{figure}

The cancellation effect observed between LSS and PB term can be understood noting that post-Born effects tend to reduce significantly the bispectrum amplitude on a large fraction of bispectrum configurations. The post-Born bispectrum has, in fact, mainly negative contributions while the LSS bispectrum due to nonlinearities has strictly positive contributions. This effect and its analytical modeling was discussed first in \cite{pratten2016}, and FCC18 observed it as a general reduction of the amplitude of higher-order moments on numerical simulations (see also the results in Fig.~\ref{fig:theoconsistency}). 
Fig.~\ref{fig:bispectratriangles} shows the ratio between the CMB convergence bispectrum including post-Born and LSS nonlinear evolution effects and the one including only the latter. The LSS bispectrum is strictly positive, since density perturbations grow faster if they are denser and, hence, large-scale overdensities correlate with small-scale lenses. 
One can observe a suppression of the bispectrum in the flattened configurations, when $L_1\approx L_2+L_3$, while for equilateral configurations, i.e. $L_1\approx L_2\approx L_3$, the bispectrum gets enhanced. Simple arguments can be made to understand why there is a sign difference in the bispectrum when all the convergence modes are aligned, i.e. in the flattened limit \cite{pratten2016}. In this case lens-lens deflection, i.e. the deflection of a light ray bundle off two consecutive lenses, dominates. In this case, the first lens induces a contraction of the light bundle area. This in turn causes the second lens to have a smaller effect than it would have without the first lens. This results in an anti-correlation between large and small scale convergence modes, leading to a negative sign of the bispectrum in the flattened limit. The positive contributions conversely represent a change in the deflection field along the direction in which the ray is deflected. A ray passing the edge of an overdensity could be deflected towards the center, where the potential gradients are larger. This generates more lensing than if the two contributions had been added independently and a positive correlation between angular scales.
The fact that the post-Born and LSS contributions roughly match in amplitude is coincidental and not anymore the case when the source plane is at low redshifts \cite{pratten2016}.\\\phantom{1}

We note, however, that due to the complex convolution of the bispectrum configurations in the quadratic estimator, the details of the cancellations happening on the $N_L^{(3/2)}$ bias are nontrivial and their analytical modeling for the different combinations of quadratic estimators is challenging. A more detailed discussion can be found in \cite{bohm2016, bohm2018}. The important cancellation effects between the LSS and PB terms observed for CMB lensing might not be as effective in the case of lensing of other diffuse background emissions that have a redshift kernel peaking at lower redshift, such as the cosmic infrared background or line intensity mapping data \cite{schaan2018}. With a shorter line of sight integration, the relative importance of the post-Born effect is in fact decreased and the LSS term for the $N_L^{(3/2)}$ bias will become the leading one, thus increasing the impact of $N_L^{(3/2)}$ on the reconstructed power spectrum.\\

\begin{figure}[!t]
\centering
\includegraphics[width=\linewidth]{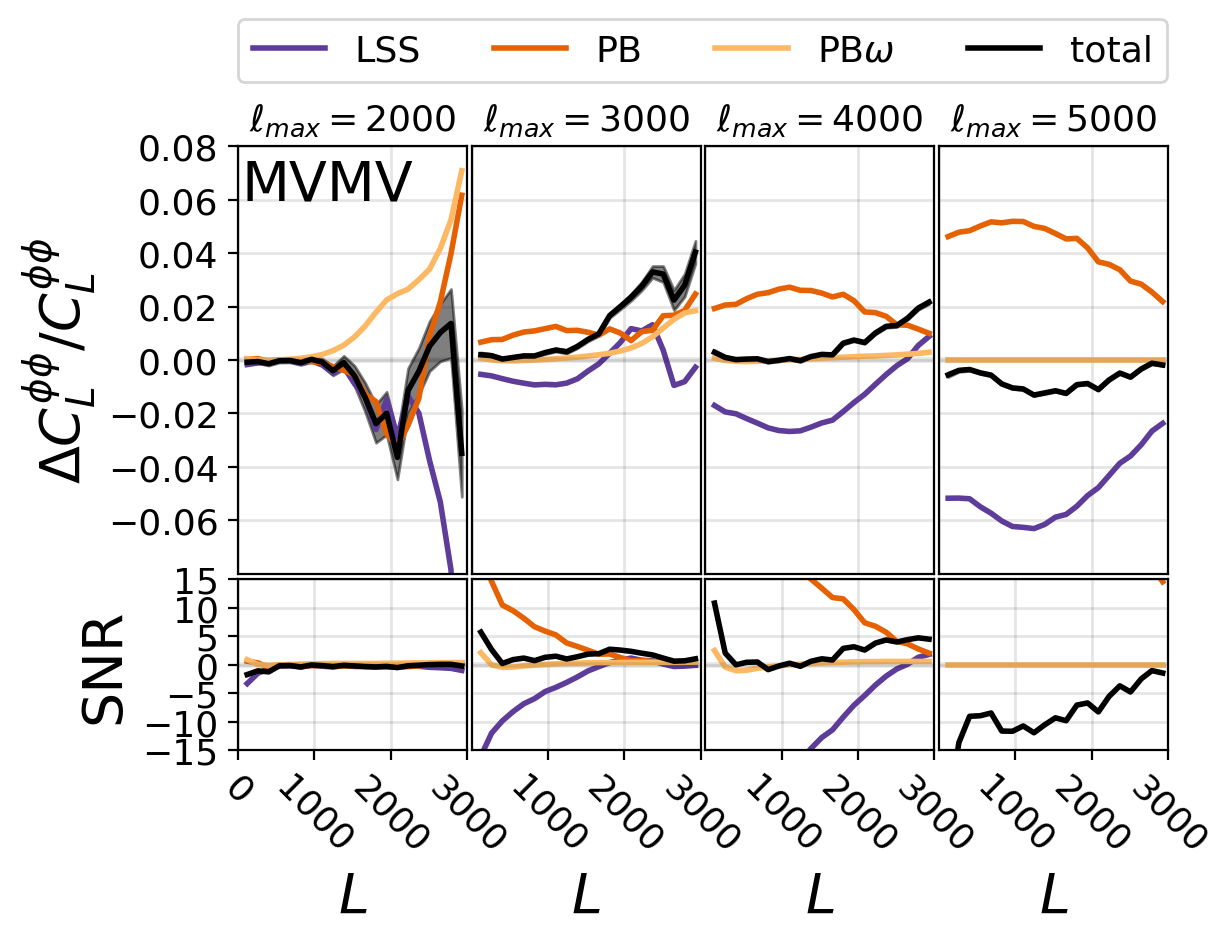}
\caption{Dependence of the $N_L^{(3/2)}$ bias for the minimum-variance lensing estimator on the maximum lensed CMB multipole used in the reconstruction algorithm in the limit of no instrumental noise. The shaded areas show the uncertainty on the bias, computed from the dispersion of the 100 simulations.}
\label{fig:lmaxcomp}
\end{figure}

The shape of the $N_L^{(3/2)}$ biases depends not only on the type of reconstruction channel used, but also on the range of multipoles included in the reconstruction. We perform the lensing reconstruction using different cutoff values $\ell_\textrm{max}$ for the harmonic coefficient used in the lensing reconstruction and show the value of $N_L^{(3/2)}$ for the minimum-variance estimator for a cosmic-variance limited experiment in Fig.~\ref{fig:lmaxcomp}. Because of the differences with analytical predictions discussed in Sec.~\ref{sec:cmb-bispectrum}, we test the stability of our results with respect to the choice of $\ell_\textrm{min}$ and verified that increasing the cutoff to $\ell_\textrm{min}=200$ did not affect our results. As expected, we can observe that the non-Gaussian effects become more prominent when we include progressively smaller angular scales in the lensing reconstruction. For $\ell_\textrm{max}=2000$ the bias is not detectable and its signal-to-noise ratio is smaller than one. In the case of $\ell_\textrm{max}=3000$, at small scales the LSS bias becomes positive, such that the total bias includes positive contributions from LSS and the post-Born gradient and curl fields, which causes the previously detected cancellation to fail. The total bias can therefore reach levels up to 4\%, although at multipoles with poor SNR. Including progressively smaller scales causes the LSS terms to increase in amplitude faster than the PB term, and as a result, the cancellation become less effective, causing the $N_L^{(3/2)}$ bias to grow. In this scenario the bias becomes very significant and its SNR could be larger than 10. We warn the reader that such an extreme case serves an illustrative purpose and should be taken with a grain of salt. In fact, the matter distribution on scales $k\geq 2\ \textrm{Mpc}^{-1}h$ affects significantly the CMB lensing signal at $\ell\simeq 5000$ and the simulations employed for this work have significant uncertainties on these scales due to the limited resolution of the $N$-body simulations used to model the deflection field and the absence of baryonic effects. These might become more important when analyzing non-Gaussian effects (see, e.g., \cite{castro2018, natarajan2014}).
Furthermore, one can observe that the cross-bispectrum contribution from the curl potential dominates at scales $\ell\lesssim 2000$ and gets subsequently downweighted in the reconstruction including larger multipoles. \\

\begin{figure}[!t]
\centering
\includegraphics[width=\linewidth]{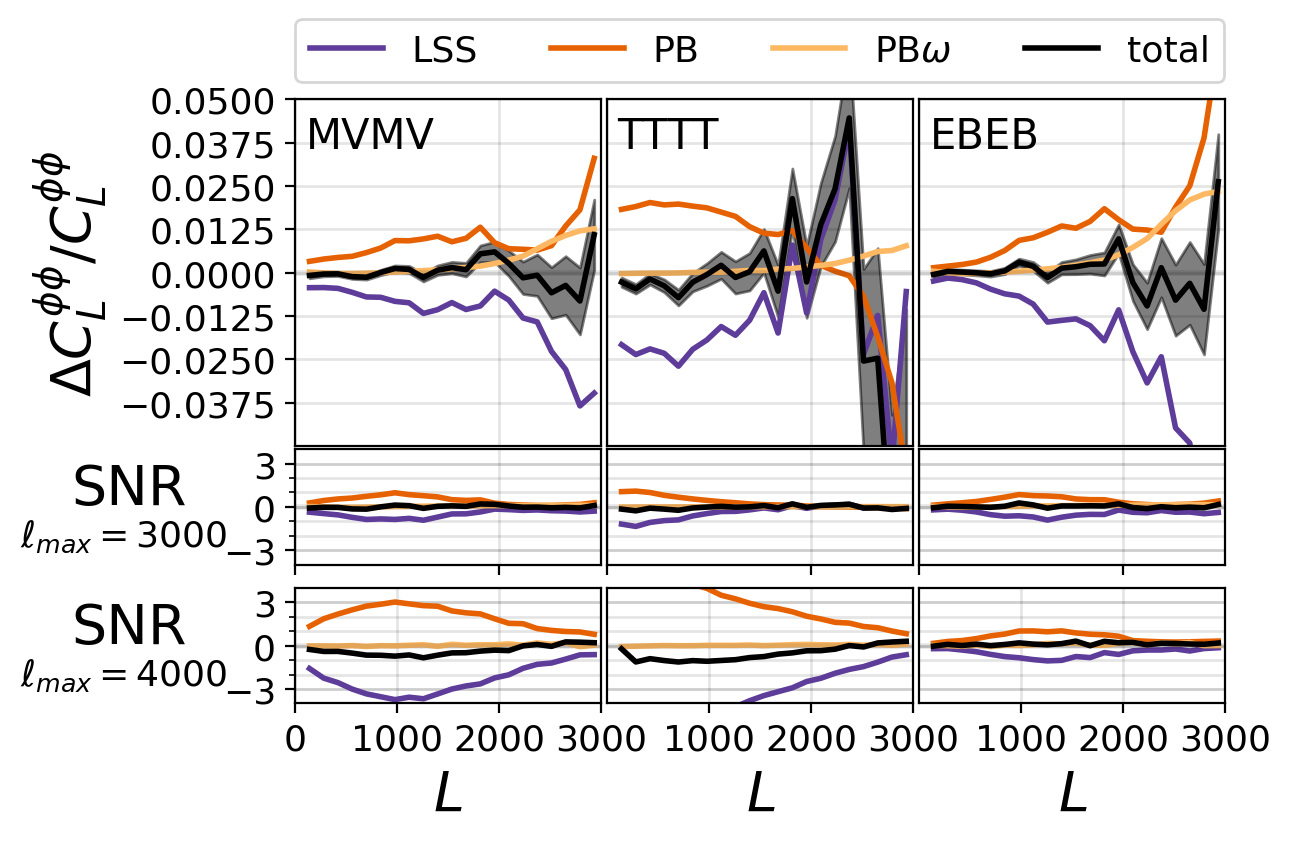}
\caption{The bias in the reconstructed minimum-variance, TTTT and EBEB lensing power spectrum for an instrument with $1.4~\mu K\textrm{-arcmin}$ white noise in polarization, $1\ \textrm{arcmin}$ beam, and including only lensed CMB multipoles up to $\ell_\textrm{max}=3000$ in the reconstruction. For comparison the signal-to-noise ratios of the biases in this case ($\ell_\textrm{max}=3000$) and the case with $\ell_\textrm{max}=4000$ (cf. Fig.~\ref{fig:4knoise}) are shown in the bottom two rows.}
\label{fig:spectra_3000_0101}
\end{figure}

\begin{figure}[!t]
\centering
\includegraphics[width=\linewidth]{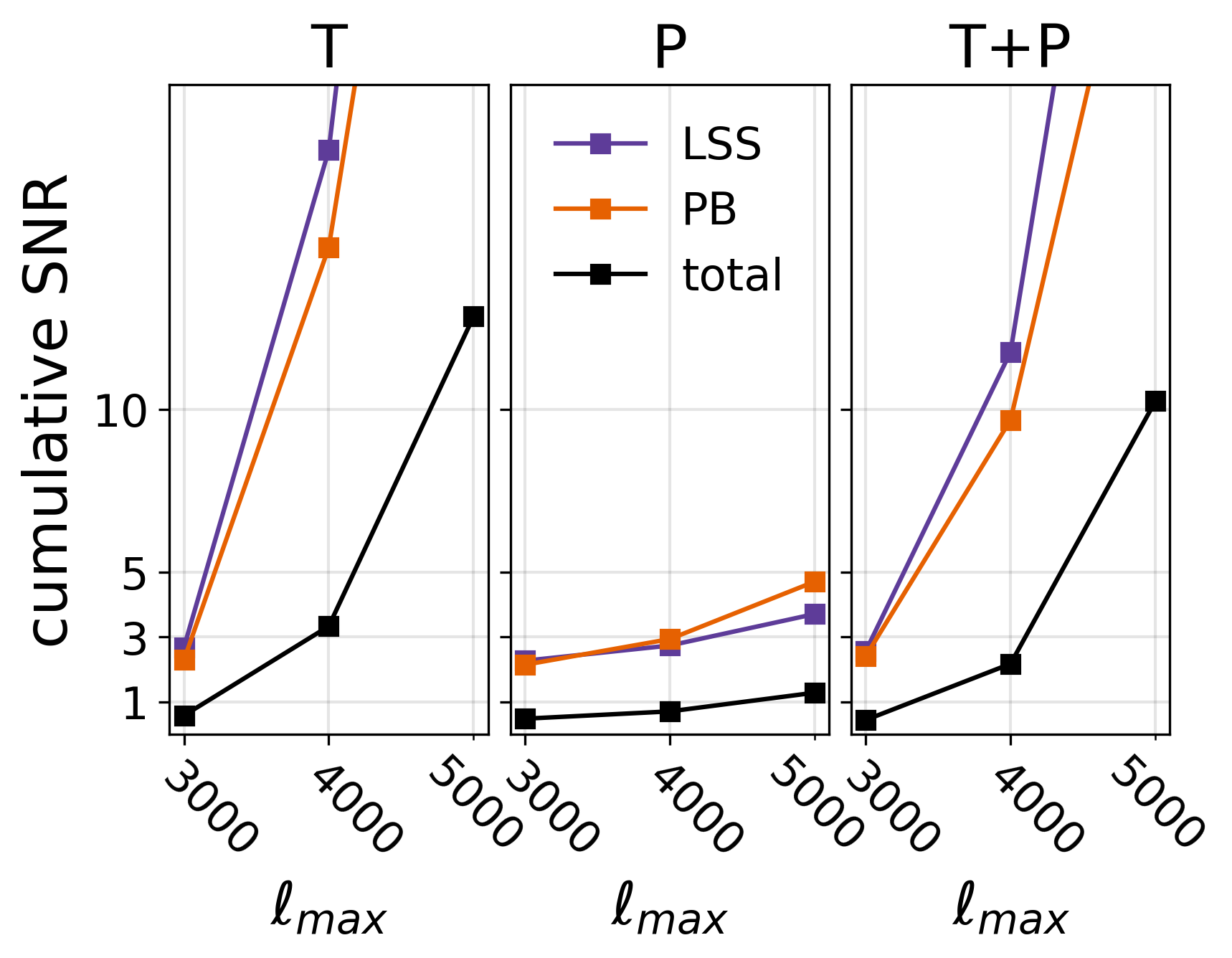}
\caption{Summary of cumulative signal-to-noise ratio of the bias for an instrument with $1.4~\mu K\textrm{-arcmin}$ white noise in polarization, $1\ \textrm{arcmin}$ beam, and different CMB multipole cutoffs $\ell_\textrm{max}$, comparing temperature (T), polarization (P), and minimum-variance (T+P) estimators.}
\label{fig:cumsnr}
\end{figure}

The changes in the weighting of the CMB harmonic coefficients used in the lensing reconstruction in the presence of experimental noise --- even with CMB-S4 sensitivity --- reduces the sharp features observed in the results of Fig.~\ref{fig:lmaxcomp} and the total $N_L^{(3/2)}$ gets suppressed compared to the cosmic-variance limit case. Reducing the cutoff in power for the reconstruction to $\ell_\textrm{max}=3000$ has a net effect of making the bias practically disappearing, despite that the individual LSS and PB effects can be of order of the error bar.\\
\vfill\null
In Fig.~\ref{fig:spectra_3000_0101} we show a comparison of the SNR obtained using these two cutoffs in CMB multipoles. As can be seen in this figure, we observe a rapid increase in the bias amplitudes between the two cases, in particular in the temperature reconstruction channels. Using polarization-only lensing reconstruction and comparing the results with temperature-only reconstruction can be appropriate tools to identify and potentially mitigate the $N_L^{(3/2)}$ biases.
Since the TTTT reconstruction is the most sensitive for the CMB-S4 experimental configurations for $L\gtrsim 1500$, dropping this reconstruction channel has an important effect in terms of the sensitivity of the reconstruction and thus, using a different cutoff in power for the temperature-based and polarization-based reconstruction might be an effective strategy to minimize the effect of $N_L^{(3/2)}$ biases while mitigating the loss of sensitivity. The contamination by unresolved extragalactic foreground residual might in any case prevent the use of multipoles $\ell\gg3000$ of temperature anisotropies. The significance to measure the bias in the lensing power spectrum, when combining all bins, is summarized in Fig.~\ref{fig:cumsnr} in terms of the cumulative signal-to-noise ratio for different CMB multipole cutoffs.\\

\begin{figure}[!t]
\centering
\includegraphics[width=\linewidth]{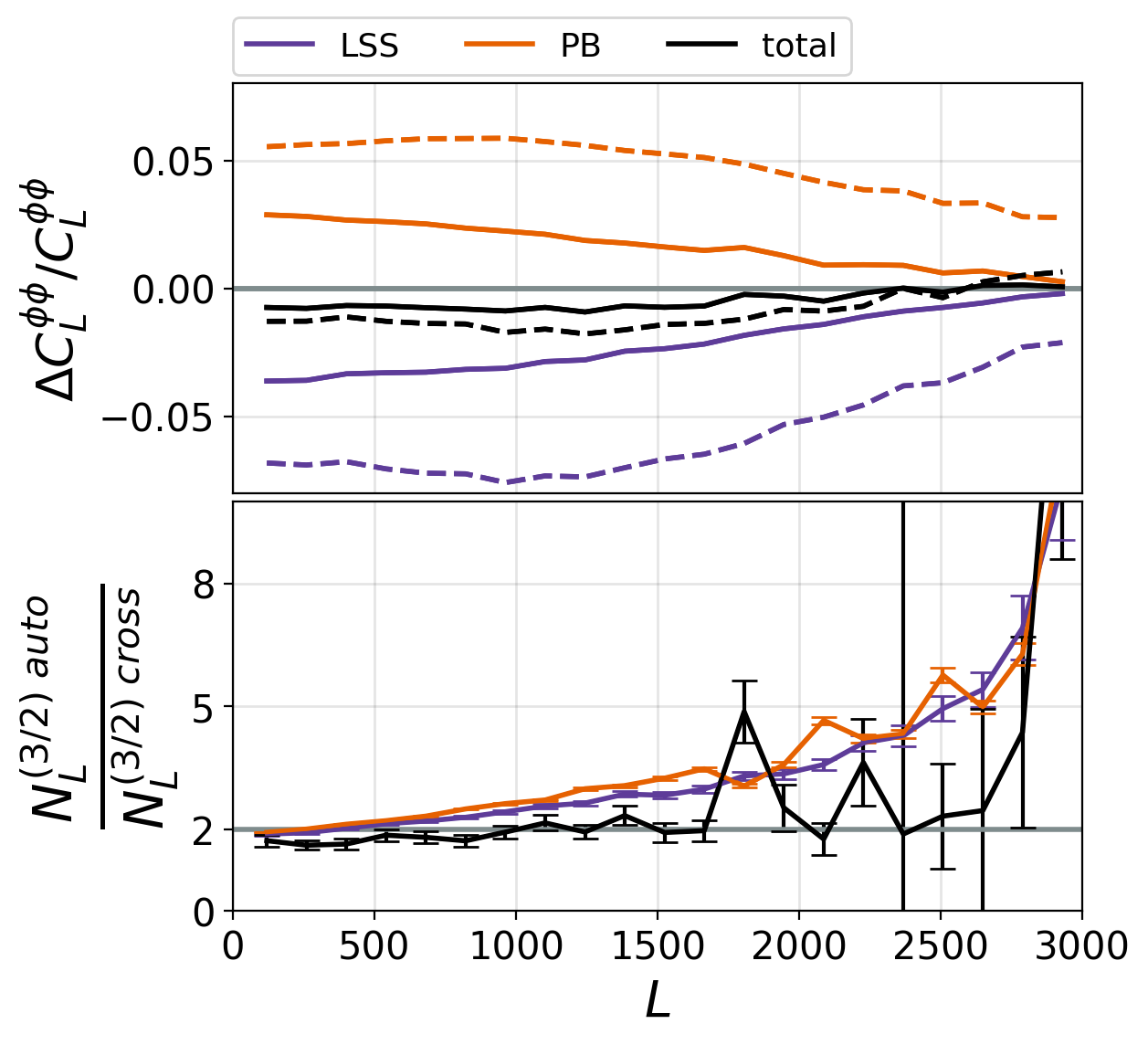}\\
\caption{Top: $N_L^{(3/2)}$ bias for the reconstructed CMB lensing potential autospectrum (dashed lines) and cross-correlation with the input CMB lensing potential of FCC18 simulations (solid lines) for a CMB-S4 experiment and a cutoff  in power $\ell_\textrm{max}=4000$ for the lensing reconstruction with the temperature estimator. Bottom: The ratio of the $N_L^{(3/2)}$ biases for the cross-correlation and autopower spectrum compared with the leading order predictions of \cite{bohm2016} (dashed black lines).}
\label{fig:crossspectra}
\end{figure}

Finally, we measure the effect of the $N_L^{(3/2)}$ in the cross-correlation power spectrum between the reconstructed lensing potential and an external large-scale structure tracer. The bias of the cross-spectrum, induced by a nonzero CMB lensing potential bispectrum, is mainly caused by the correlation of the external large-scale structure tracer with the second-order response of the reconstructed lensing potential to the true lensing potential. For the sake of simplicity we limit our analysis to the case of the cross-correlation with a perfect tracer of the CMB lensing potential, i.e., the lensing potential directly extracted from the FCC18 simulations. Since in the cross-correlation case the tracer is almost uncorrelated with the CMB, there are fewer contractions of the matter field that contribute to the $N_L^{(3/2)}$ bias, and thus we should see a reduction in the amplitude of $N_L^{(3/2)}$ by a factor of roughly 2 with respect to the bias on the autospectrum, in particular for the TTTT estimator \cite{bohm2016}. We verify that this prediction holds, as we show in Fig.~\ref{fig:crossspectra}. A similar level of suppression is observed also for other estimators and, in particular, for EBEB we saw a reduction of more than a factor of 4 for $L\gtrsim 2000$. This analysis might suggest that cosmological constraints based on cross-correlations of CMB lensing with an external tracer sufficiently correlated with the CMB lensing potential might be less biased if we cannot account for the $N_L^{(3/2)}$ bias in the autospectrum analysis. However, we stress that due to the distinctive impact of the post-Born term with respect to the LSS one in the case of CMB lensing, the overall variation in amplitude of the bias in cross-correlation might change significantly if a tracer of structures at lower redshift is considered. Nevertheless, these techniques might be affected by other type of biases, such as those due to the galaxy intrinsic alignements in the case of galaxy weak lensing \cite{merkel2017, larsen2016, troxel2014}. In addition, the tracers at lower redshift are in fact more sensitive to the non-Gaussianity due to matter nonlinearity and less sensitive to post-Born effects. Therefore we expect to observe an increase in the $N^{3/2}$ bias as the cancellation between LSS and post-Born becomes less effective in this case. We leave the investigation of this topic to future work.

\subsection{Consistency checks}\label{sec:consistency}

\begin{figure}[!h]
\centering
\includegraphics[width=1.\linewidth]{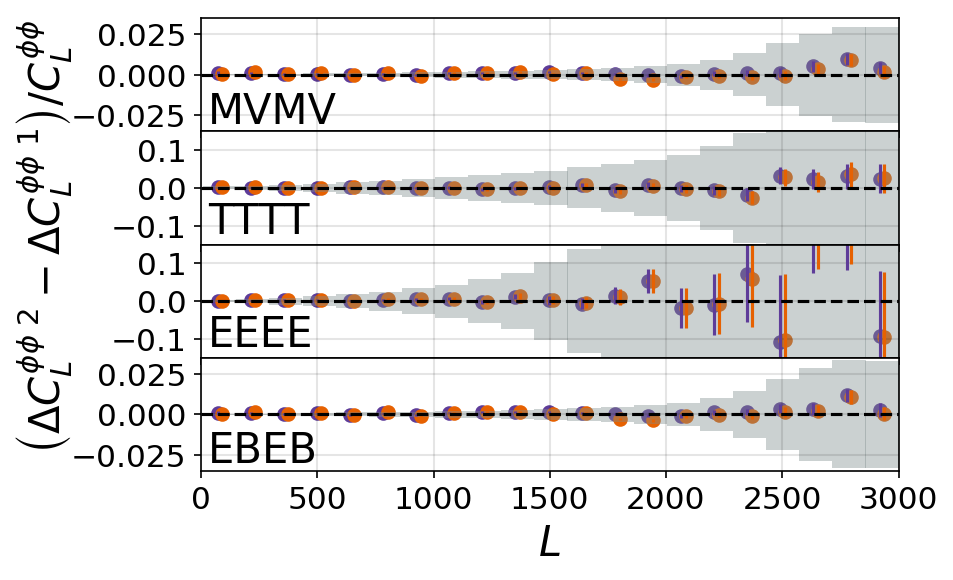}
\caption{The null spectra obtained taking the difference between $\Delta C_L^{\phi\phi, 1}$ and $\Delta C_L^{\phi\phi, 2}$ as defined in Eqs.~\eqref{eq:biastypei} and \eqref{eq:biastypeii} for the minimum-variance, TTTT, EEEE, and EBEB lensing reconstruction in the limit of no instrumental noise. The reconstructions on $\kappa^F$-lensed CMB fields are shown in purple (LSS only contribution), the same with $\kappa^R$ are shown in orange [LSS and PB (total) contributions]. The error bars show the uncertainties as measured from the scatter in the simulations while the shaded area show the expected statistical uncertainty in the respective bin.}
\label{fig:nullspectra}
\end{figure}

To ensure that the reported biases were not caused by a mismatch in the CMB and lensing potential power spectra and therefore are not residual $N_L^{(0)}$ and $N_L^{(1)}$ biases, we check the consistency of our measurements with an alternative method to extract the $N_L^{(3/2)}$ bias. 
In particular, we compare the spectra
\begin{align}
\Delta C_L^{\phi\phi, 1} [\kappa^X]&= \left\langle \hat{C}_L^{\phi\phi}[\kappa^X]-\hat{C}_L^{\phi\phi}[\kappa^G]\right\rangle_\text{100 sims} \label{eq:biastypei} \\
\Delta C_L^{\phi\phi, 2}[\kappa^X] &= \frac{1}{2}\left\langle \hat{C}_L^{\phi\phi}[\kappa^X]-\hat{C}_L^{\phi\phi}[-\kappa^X] \right\rangle_\text{100 sims}\label{eq:biastypeii},
\end{align}
where $X\in\left\{F, R\right\}$.  The averaging in Eq.~\eqref{eq:biastypei} is performed over the 100 realizations of lensed CMB derived with the set of simulations including a Gaussian convergence, and the averaging in Eq.~\eqref{eq:biastypeii} is computed over the 100 realizations of lensed CMB lensed with the non-Gaussian convergence $\kappa^{X}$. We have that
\begin{align}
\left\langle\hat{C}_L^{\phi\phi}[\kappa^X]\right\rangle \approx &N_L^{(0)}\left[C_\ell^{CMB}\right] + C_L^{\phi\phi} \\+&N_L^{(1)}\left[C_\ell^{CMB},C_L^{\phi\phi}\right] \notag\\+ &\text{sgn}(\kappa^X) N_L^{(3/2)}\left[C_\ell^{CMB},C_L^{\phi\phi},b_{L_1L_2L_3}^{\phi\phi\phi}\right],\notag
\end{align}
where we denote in squared brackets the functional dependencies of the biases for clarity. Hence both techniques in Eqs.~\eqref{eq:biastypei} and \eqref{eq:biastypeii} isolate in principle the $N_L^{(3/2)}$ bias. However, a mismatch of $N_L^{(0)}$ and $N_L^{(1)}$ between simulations lensed with $\kappa^F$, $\kappa^R$ and $\kappa^G$ or correlations at order higher than the bispectrum should manifest themselves in a discrepancy between the two spectra. We constructed null spectra and computed Welch's $t$-test statistics for both the $\kappa^F$ and the $\kappa^R$ sets of simulations to test separately LSS effects alone and LSS and PB together. In both cases we use the spectra from the three most relevant reconstruction channels (TTTT, EBEB, and the autopower spectrum of the minimum-variance estimator (MVMV)) binned in 21 bins within $L \in [30,3000]$. With this approach we test the hypothesis that the two curves are realizations of a common underlying distribution and quantify the validity of the assumptions used to isolate the biases above. The variances used in the tests are obtained from simulations. We show a subset of the null spectra $\Delta C_L^{\phi\phi, 2} -\Delta C_L^{\phi\phi, 1}$ in Fig.~\ref{fig:nullspectra}. The deviations from zero in the high signal-to-noise regions are subdominant, while small deviations at mostly large multipoles are well within the $1\sigma$ error bar. We furthermore obtained global $p$ values by averaging over the bins for each estimator and find no PTE lower than 5\%, as summarized in Table \ref{tab:ptes}. These results made us conclude that the simulation and reconstruction pipeline up to the lensing power spectrum step are internally consistent, increasing our confidence in the results shown in Sec.~\ref{sec:n32-measurements}.

\begin{table}
\centering
\caption{Global $p$ values of null spectra of noiseless configuration and $\ell_\textrm{max}=3000$.}
\label{tab:ptes}
\begin{ruledtabular}
\begin{tabular}{lcc}
  \textbf{$\mathbf{p}$ Value [\%]} & LSS & Total
    \\ 
\hline&&\\[-.8em]
MVMV &13.8 & 47.2 \\
TTTT &38.9 & 52.7 \\
TTTE &16.0 & 17.8 \\
TTEE &87.8 & 96.0 \\
TTTB &76.3 & 87.5 \\
TTEB &67.6 & 99.6 \\
TETE &43.6 & 47.6 \\
TEEE &5.3  & 9.2  \\
TETB &97.8 & 86.0 \\
TEEB &83.6 & 98.9 \\
EEEE &30.5 & 27.9 \\
EETB &21.7 & 20.7 \\
EEEB &46.1 & 49.0 \\
TBTB &45.2 & 20.4 \\
TBEB &60.1 & 73.4 \\
EBEB &17.3 & 51.5 \\
\end{tabular}
\end{ruledtabular}
\end{table}

\section{$N^{3/2}_L$ impact on cosmological parameter estimation}\label{sec:params}

Future sensitive measurements of the CMB lensing potential will provide important constraints on cosmological parameters. Therefore a biased reconstruction of the lensing potential power spectrum could affect their estimation. For example, we find that at the high sensitivities envisioned for CMB-S4 measurements the total $N_L^{(3/2)}$ bias could produce deviations of more than $3\sigma$ from the fiducial value of 1 when fitting the lensing amplitude parameter $A_\textrm{lens}$. In Table \ref{tab:Alens} we show the fitted $A_\textrm{lens}$ parameter for different CMB multipole cutoffs obtained by maximizing the simple one-parameter likelihood defined by
\be
-2 \ln \mathcal{L}=\sum_L (2L+1)f_\textrm{sky} \left(\ln \left(\frac{C_L}{D_L}\right) + \frac{D_L}{C_L} - 1\right),
\ee
where $C_L=A_\textrm{lens}\times C_L^\textrm{fid.}+N_\ell^{\phi\phi, \textrm{tot.}}$, $D_L=C_L^\textrm{fid.}+N_\ell^{\phi\phi, \textrm{tot.}}+N_L^{(3/2)}$, and $N_L^{\phi\phi, \textrm{tot.}}=N_L^{(0)}+N_L^{(1)}$.\\

\begin{table}[b]
\renewcommand{\arraystretch}{1.5}
\centering
\caption{Fitted $A_\textrm{lens}$ parameter of the biased reconstructed lensing power spectrum with a fiducial value of $A_\textrm{lens}=1$ for temperature-only (T), polarization-only (P), and minimum variance (T+P) lensing estimators and no noise in the CMB. Cases with significant bias are marked in bold. }\label{tab:Alens}
\begin{ruledtabular}
\begin{tabular}{lccc} 
    \textbf{Total bias} & {$\ell_\textrm{max}=3000$} & {$\ell_\textrm{max}=4000$} & {$\ell_\textrm{max}=5000$}\\ \hline
T      & $0.997 \pm 0.006$  & $\mathbf{0.988 \pm 0.003}$  & $\mathbf{0.973 \pm 0.002}$  \\ 
P      & $\mathbf{1.005 \pm 0.002}$  & $\mathbf{1.009 \pm 0.001}$  & $\mathbf{1.005 \pm 0.001}$  \\ 
T+P    & $\mathbf{1.004 \pm 0.002}$  & $\mathbf{1.004 \pm 0.001}$  & $\mathbf{0.992 \pm 0.001}$  \\ 
\end{tabular}
\end{ruledtabular}
\end{table}

Because of the nontrivial scale dependence of the $N^{3/2}_L$ bias, we expand our cosmological parameter estimation study to the exploration of a broader parameter space using Markov chain Monte Carlo (MCMC) techniques. The goal is to quantify the significance of possible biases in parameters like the total neutrino mass, $M_\nu$, or the amplitude of primordial inflationary perturbations, $A_s$, if $N^{3/2}_L$ is unaccounted for in the power-spectra modeling and cosmological parameters sampling. For this purpose we use the publicly available package \textsc{MontePython} \footnote{\url{http://baudren.github.io/montepython.html}} \cite{Brinckmann:2018cvx,Audren:2012wb} based on the Metropolis-Hastings sampling algorithm. 
In this analysis we consider the CMB and lensing likelihood for a set of parameters $\theta$ given the measured power spectra of CMB temperature, $E$-modes and lensing potential as Gaussian in the respective fields. Under these assumptions the likelihood function is given by (e.g., \cite{divalentino2018})
\be
-2\log\mathcal{L}(\theta|\hat{\mathbf{C}})=\sum_\ell  (2\ell+1) f_\textrm{sky} \left( \ln\frac{|\mathbf{C}_\ell|}{|\hat{\mathbf{C}}_\ell|} + \mathbf{C}_\ell^{-1}\hat{\mathbf{C}}_\ell-3 \right),
\label{eq:like}
\ee
where the covariance matrix for the fiducial model $\hat{\mathbf{C}}_\ell$ and the theoretical signal $\mathbf{C}_\ell$ are constructed as
\[
\mathbf{C}_\ell=
\begin{pmatrix}
C_\ell^{TT}+N_\ell^{TT} & C_\ell^{TE} & C_\ell^{T\phi} \\
C_\ell^{TE} & C_\ell^{EE}+N_\ell^{EE} & 0 \\
C_\ell^{T\phi} & 0 & C_\ell^{\phi\phi}+N_\ell^{\phi\phi, \textrm{tot.}} 
\end{pmatrix},
\]
where $N_\ell^\textrm{TT}$ and $N_\ell^\textrm{EE}$ are the white noise power spectra for the temperature and the $E$-modes and $N_\ell^{\phi\phi, \textrm{tot.}}=N_\ell^{(0)}+N_\ell^{(1)}$. All these quantities are computed assuming the fiducial cosmology with CMB-S4 sensitivities and considered to be independent of the cosmological parameters in order to simplify and speed up the sampling. In the evaluation of the fiducial $\hat{\mathbf{C}}_\ell$ we use the biased lensing potential power spectrum, $\tilde{C}_L^{\phi\phi}$, which includes the $N_L^{(3/2)}$ bias measured in the simulations and depends on the cosmological parameters of the fiducial model as
\begin{eqnarray}
\tilde{C}_L^{\phi\phi} &\equiv& C_L^{\phi\phi}[\theta^\textrm{fid.}]+N_L^{(0)}+N_L^{(1)}+\frac{C_L^{\phi\phi}[\theta^\textrm{fid.}]}{C_L^{\phi\phi\ \textrm{sims}}}N_L^{(3/2)}.~~~~~
\end{eqnarray}

This definition allows one to mitigate the impact of the shot-noise term and the difference in the modeling of the nonlinear evolution between the simulation results and the Boltzmann solvers which typically employ the Halofit fitting formulas \cite{takahashi2012}. This enables us to have a consistent modeling of nonlinearity between the fiducial and the fitted model, reducing the chance to obtain spurious results in the fitting that are driven by the differences in the CMB lensing potential power spectrum modeling. We note, however, that the uncertainties in the modeling of nonlinearity on the CMB lensing power spectrum reach the 10 -- 15\% level on the scales considered in this work \cite{pratten2016, takahashi2012} and might become non-negligible.\\\vfill\null

In the construction of the covariance we neglect the $\phi E$ correlation because it is confined at very large angular scales and carries little information on the parameters of interest in our analysis. For the sake of simplicity we do not include the $B$-mode power spectrum in $\hat{\mathbf{C}}_\ell$ and $\mathbf{C}_\ell$ to avoid the need to model the non-Gaussian covariance between $C_{\ell}^{BB}$ and $C_{L}^{\phi\phi}$ \cite{smith2004}. We note that more optimal formalisms to deal with the non-Gaussian correlations between CMB and lensing power spectra have been discussed in the literature \cite{Schmittfull:2013uea, peloton2017, benoit-levy2012}.
As the present analysis is intended to quantify biases on cosmological parameters estimation due to mismodeling of the lensing potential bias rather than to provide an accurate forecast of future CMB experiment constraints, the approximations adopted here are not expected to affect our conclusions at the level of accuracy considered in this work. \\

In the likelihood construction we assume a fiducial $\Lambda$CDM cosmology taken from Planck 2015 results \cite{Ade:2015xua,Adam:2016hgk} devoid of massive neutrinos, while we allow for a single neutrino to be massive in the parameter fit. 
We include angular scales $30\leq \ell \leq 3000$ and assume an observed sky fraction $f_\textrm{sky} = 40\%$ to mimic a CMB-S4-like survey with $1.4\ \mu K\textrm{-arcmin}$ white noise in polarization and a $1\ \textrm{arcmin}$ beam size in the likelihood. We summarize the values of our fiducial cosmology as well as the details of the priors adopted for the cosmological parameters sampled in our analysis in Table \ref{tab:params}. \\

\begin{table}
\centering
\caption{The cosmological parameters from Planck 2015 \cite{Ade:2015xua,Adam:2016hgk} together with their $1\sigma$ proposal scale or parameter bounds used in the cosmological parameter inference.
}
\begin{ruledtabular}
\begin{tabular}{lc} 
$\Omega_{b}h^2$&$0.02225\pm0.00016$ \\$\Omega_{c}h^2$&$0.1198\pm0.0015$\\$\tau$&$0.058\pm0.012$\\$\ln10^{10}A_{s }$&$3.094\pm0.034$\\$n_s$&$0.9645\pm0.0049$\\$100\theta_s$&$1.04077\pm 0.00032$\\$M_\nu$ [meV]&$[0,300]$\\
\end{tabular}
\end{ruledtabular}
\label{tab:params}
\end{table}

\begin{figure}[!h]
\centering
\includegraphics[width=\linewidth]{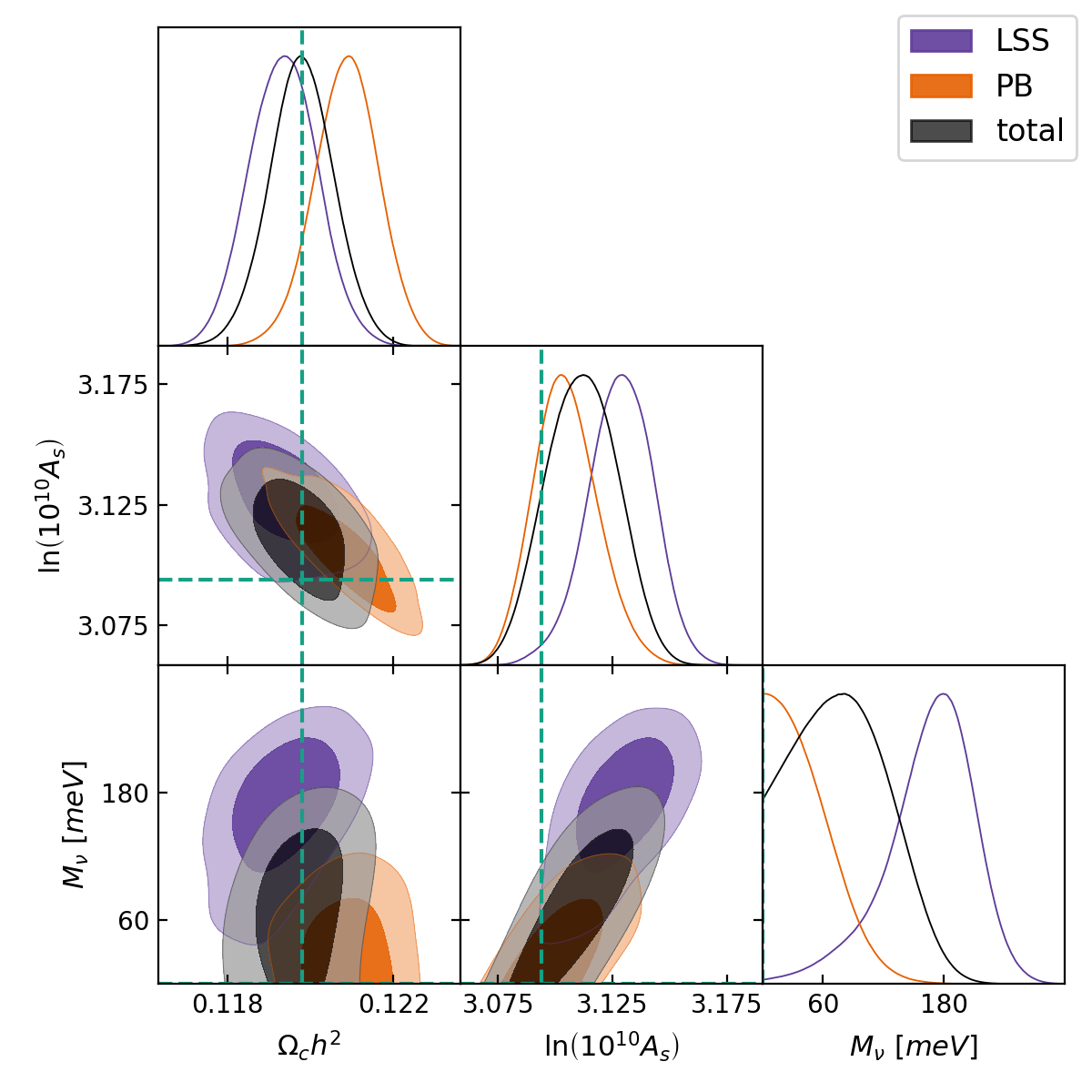}
\caption{The 2D posteriors for the cold dark matter density, $\Omega_c h^2$, the amplitude of primordial inflationary perturbations, $A_s$ and the neutrino mass, $M_\nu$, including biases from LSS nonlinearities and post-Born effect in $\hat{C}_L^{\phi\phi}$, reconstructed using the minimum variance estimator, CMB modes up to $\ell_\textrm{max}=4000$ and CMB-S4 experimental specifications.}
\label{fig:triangle40000101}
\end{figure}

We neglect the effects of the LSS non-Gaussianity and post-Born corrections on the lensed CMB TT and EE power spectra since the cumulative signal-to-noise ratio for these corrections is below the detection thresholds even for CMB-S4 sensitivity. 
In Fig.~\ref{fig:triangle40000101} we show the 2D posteriors obtained for the parameter combinations of $\Omega_c h^2$, $\ln\left(10^{10}A_{s }\right) $, and $M_\nu$ for the minimum-variance lensing estimator and CMB-S4 sensitivity. The figure shows an example of the main biases in the parameter estimation induced by different sources (LSS, PB, total) of unaccounted $N_L^{(3/2)}$ bias. Similar to what was observed in Sec.\ref{sec:n32-measurements}, the compensating effect between the LSS and PB biases observed at the level of the lensing power spectrum is also visible in the cosmological parameter estimation, where we find a cancellation of the parameter biases when both these terms are included. Each source of $N_L^{(3/2)}$ bias might considerably affect the estimation of the cosmological parameters when considered alone at the level of CMB-S4 sensitivity. Assuming we can model these biases analytically we need to include both the terms in the modeling as the inclusion of only one of the LSS or PB term would lead to an overcorrection of the effect. This is clearly visible in the case the LSS-induced $N_L^{(3/2)}$ for $A_s$ and $M_\nu$, where the large negative bias over a large range of scales in the power spectrum causes a significant false detection of a $169^{+50}_{-30}\ \textrm{meV}$ neutrino mass. The cancellation due to post-Born corrections mitigates this bias, reducing it to $83^{+40}_{-50}$, and hence still compatible with zero neutrino mass only at the 2$\sigma$ level. \\

\begin{table*}
\centering
\caption{This table shows the deviation of the best-fit from the fiducial values (bias) and 68\% confidence level ($1\sigma$) uncertainties for the cold dark matter density, $\Omega_c h^2$, the optical depth to reionization, $\tau$, the amplitude of primordial inflationary perturbations, $A_s$ and the neutrino mass $M_\nu$. A configuration with $1.4\ \mu K\textrm{-arcmin}$ white noise and $1\ \textrm{arcmin}$ beam with different CMB multipole cutoff and estimator combinations was used. We show biases using minimum-variance lensing reconstruction including CMB temperature (T+P) and using polarization only (P). Upper limits are given in terms of 95\% confidence level.}
\label{tab:totalbias}
\begin{ruledtabular}
\begin{tabular} {llcccccccc}
\multicolumn{2}{ c}{} & \multicolumn{2}{ c}{$\ell_\textrm{max}=3000$} & \multicolumn{2}{ c}{$\ell_\textrm{max}=4000$} & \multicolumn{2}{ c}{$\ell_\textrm{max}=5000$} & \multicolumn{2}{ c}{$\ell_\textrm{max}=5000$+DESI}\\
\multicolumn{2}{ c}{} & \textbf{Bias} & \textbf{$\mathbf{1\sigma}$ (stat.)} & \textbf{Bias} & \textbf{$\mathbf{1\sigma}$ (stat.)}& \textbf{Bias} & \textbf{$\mathbf{1\sigma}$ (stat.)}& \textbf{Bias} & \textbf{$\mathbf{1\sigma}$ (stat.)} \\\hline\\[-.8em]
\multirow{4}{*}{\textbf{T+P}} & $\Omega_{c}h^2 \times 10^{5}$ & $25 $ & $85$ & $14$ & $88$ & $-45 $ &$85$& $-66$ & $55$ \\
 & $\tau \times 10^{3}$ & $5$ & $9$ & $9$ & $8$ & $14$ & $9$ & $9$ & $10$ \\ 
 & $\ln\left(10^{10}A_{s }\right) \times 10^{3}$ & $11$ & $15$ & $18$ & $18$ & $27$ & $16$ & $16$ & $14$ \\ 
 & $M_{\nu}\ [\textrm{meV}] $ & $0$ &$79$& $90$ & $60$ &  $110$ & $50$&  $0$ & $55$ \\\hline\\[-.8em]
\multirow{4}{*}{\textbf{P}} & $\Omega_{c}h^2 \times 10^{5}$ & $16$ & $84$ & $26$ & $82$ & $25$ & $80$ & $-37$ & $56$ \\
 & $\tau \times 10^{3}$ & $6$ & $9$ & $7$ & $10$ & $7$ & $9$ & $8$ & $9$ \\ 
 & $\ln\left(10^{10}A_{s }\right) \times 10^{3}$ & $12$ & $16$ & $13$ & $16$ & $14$ & $15$ & $15$ & $16$ \\ 
 & $ M_{\nu}\ [\textrm{meV}] $ & $0$ & $75$ & $0$ & $84$ & $65$ & $60$ & $0$ & $44$
 \\
\end{tabular}
\end{ruledtabular}
\end{table*}

\begin{figure*}[!ht]
\centering
\includegraphics[width=.45\linewidth]{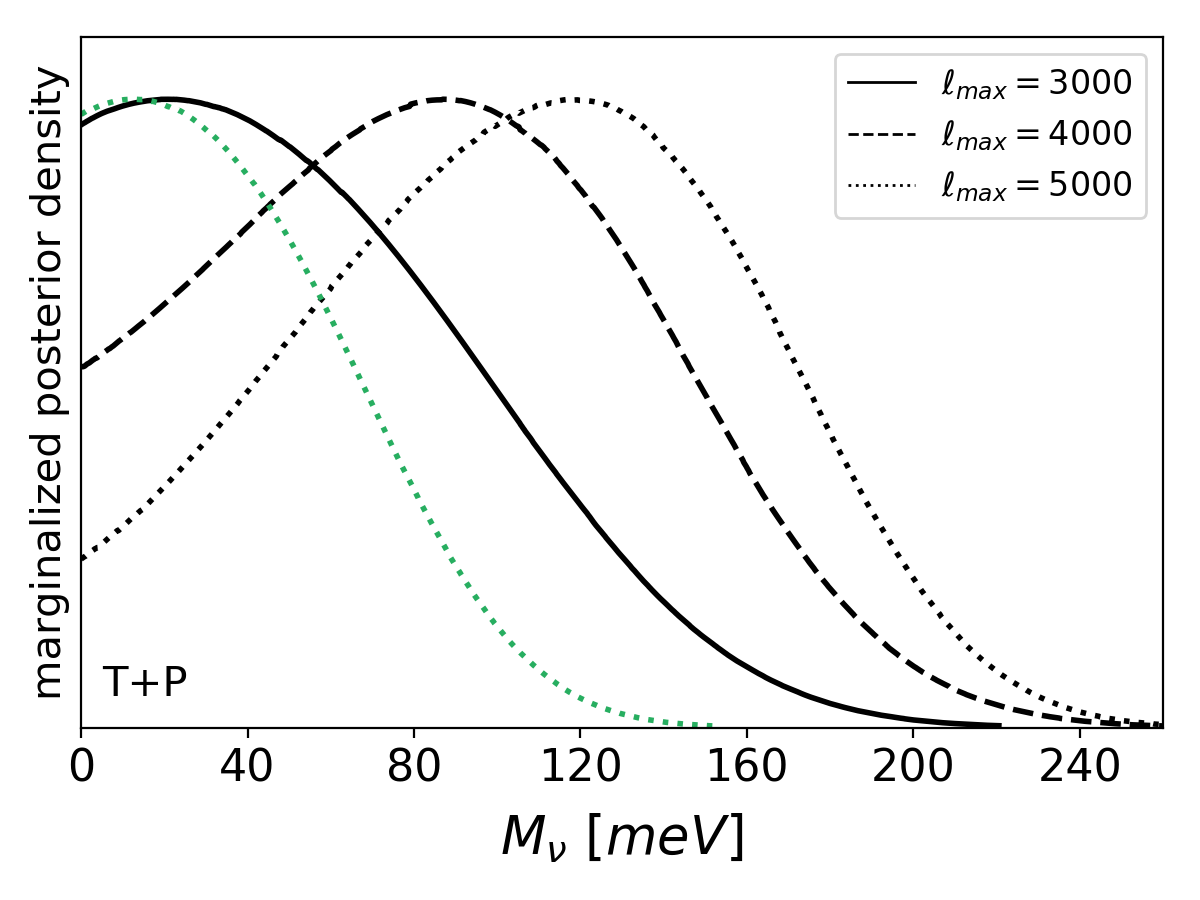}
\includegraphics[width=.45\linewidth]{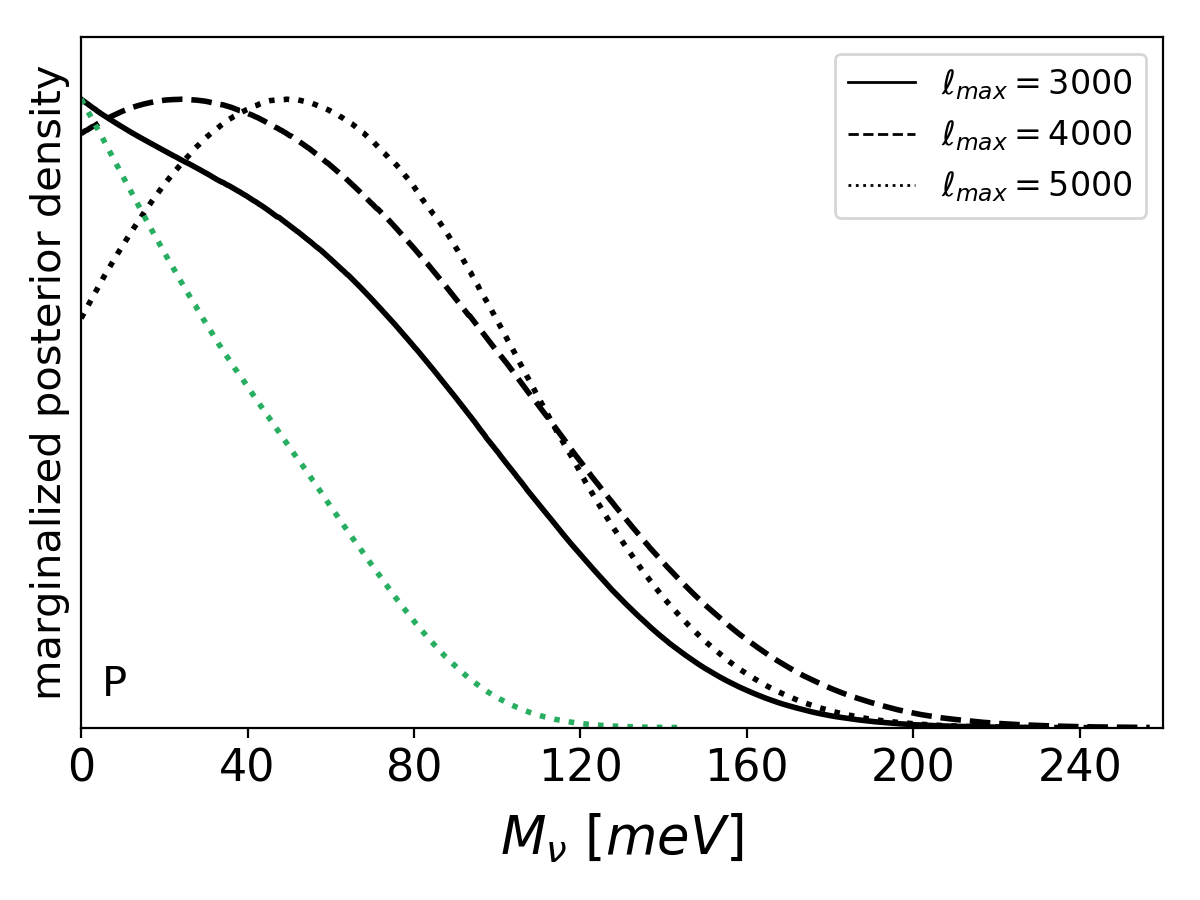}
\caption{The one-dimensional posteriors for the total neutrino mass $M_\nu$ for different CMB multipole cutoffs used in the lensing reconstruction. $\ell_\textrm{max}=3000$ case is shown as a solid line, while $\ell_\textrm{max}=4000$ and $\ell_\textrm{max}=5000$ are shown as dashed and dotted lines, respectively. The left figure shows the results obtained including all reconstruction estimators including temperature (T+P), while the right figure uses only polarization-based estimators (P). Each figure also includes the posterior after including a prior using DESI BAO data \cite{Archidiacono:2016lnv} in the sampling in green, for the most extreme case of $\ell_\textrm{max}=5000$.}
\label{fig:totalbias}
\end{figure*}

The same analysis carried out adding only the $N_L^{(3/2)}$ biases of polarization-based estimators indicates that using these reconstruction channels leads to more robust constraints on cosmological parameters, even when including the smaller angular scales in the lensing reconstruction. In Table \ref{tab:totalbias} and Fig.~\ref{fig:totalbias} we show the best-fit values and marginalized posteriors obtained including the total $N_L^{(3/2)}$ computed varying the CMB multipole cutoff used in the reconstruction for the two different cases including and excluding temperature data when forming the minimum-variance estimator. Including multipoles up to $\ell=5000$ in the reconstruction leads to a neutrino mass bias larger than $1\sigma$, even after excluding temperature data. Nevertheless, on the level of the parameter estimation we can observe that the polarization lensing estimator is more robust to these kinds of biases, which can be attributed in part to the slightly worse reconstruction lensing noise when excluding small-scale temperature data and in part to the smaller amplitude of the $N_L^{(3/2)}$ bias for polarization estimators.\\

We note, however, that the error on the total neutrino mass does not decrease significantly with decreasing noise in the CMB lensing potential power spectrum. This is due to the degeneracy of the total neutrino mass with the $A_s$ parameter and the sensitivity of the constraint on the latter (or, more precisely, on the combination $A_s e^{-2\tau}$). Since we are assuming future data from ground-based CMB-S4 instruments, which are limited to multipoles $\ell\geq 30$, we are not able to push the uncertainty on $\tau$ to the cosmic-variance limit. However, accessing the reionization bump at $\ell\leq 20$ down to cosmic-variance precision could be achieved by proposed all-sky polarized CMB surveys like CLASS \cite{Watts:2018etg}, CORE \cite{divalentino2018}, LiteBIRD \cite{Suzuki:2018cuy} or PIXIE \cite{Calabrese:2016eii}. This would provide a tighter constraint on $\tau$ \cite{Allison:2015qca}, and would lead to the expected decrease in statistical uncertainty with increasing multipole cut-off in the lensing reconstruction. Furthermore, a $N^{(3/2)}$-bias in the lensing potential estimation would bias $A_s$ and $\tau$ high. We observe that being able to include the constraining power of the reionization bump at large-scales would reduce the bias on $\tau$, and consequently significantly reduce the bias on the total neutrino mass. This would occur at the expense of a $\leq 1-\sigma$ total bias on cold-dark matter density $\Omega_m$ and a negligibly larger $\chi^2$ goodness of fit.

\section{\label{sec:conclusions} Conclusions}
In this work we investigate the properties of higher-order correlations of the CMB lensing deflection field arising from nonlinear evolution of the matter as well as post-Born corrections, modeled through numerical simulations, and their impact on the CMB lensing potential reconstruction using quadratic estimators ($N_L^{(3/2)}$  bias). We validate the numerical simulations used to model these effects comparing the expected corrections on the lensed CMB power spectrum due to both LSS nonlinearity and post-Born corrections modeled analytically, finding a good agreement. We find that both the matter nonlinearity and post-Born non-Gaussianity cause significant biases of the reconstructed CMB lensing potential power spectrum. However, when these effects are analyzed jointly, the amplitude of the total $N_L^{(3/2)}$ bias is greatly reduced both on the CMB lensing autospectrum and in the cross-correlation. This is directly related to the different shape and sign properties of the post-Born bispectrum and the matter bispectrum. The cancellation is more effective in the presence of experimental noise. Despite this fact, we find that the estimation of the  $A_\textrm{lens}$ parameter from the CMB lensing potential  could be biased by more than 3$\sigma$ for future high-sensitivity experiments like CMB-S4. \\

We further perform a MCMC analysis to evaluate the impact of the residual $N_L^{(3/2)}$ bias on the estimation of other cosmological parameters at the CMB-S4 sensitivity. We find that the best-fit value of cosmological parameters such as $M_\nu$ and $A_s$ could be biased due to the $N_L^{(3/2)}$ bias by up to 2$\sigma$, but the significance of these biases greatly depends on the type of quadratic estimator and the maximum multipole used for the lensing reconstruction. Using multipoles $\ell \leq 3000$ for the lensing reconstruction and parameter fitting would not produce any significant bias on cosmological parameters. However the inclusion of smaller angular scales in the lensing reconstruction in order to improve the sensitivity, will also bring the lensing reconstruction in a regime where the details of the cancellation of the post-Born and LSS term becomes trickier and less effective. As a consequence, the total bias due to LSS nonlinearity and post-Born effects, if unaccounted for, becomes more important. In general we find that the CMB temperature-based reconstruction channels are more prone to these biases due to their higher sensitivity to small scale lenses. In this regime, however, foreground contaminations might be the major limiting effects \cite{vanEngelen:2013rla, Osborne:2013nna, PhysRevD.97.023512}. Using only polarization-based estimators for the lensing reconstruction usually leads to cosmological constraints which are more robust to both the foreground and $N_L^{(3/2)}$ effects. The latter, in particular, is caused by a consistently more effective cancellation of LSS and post-Born effects. As an illustrative case, we perform the cosmological parameter analysis including multipoles up to $\ell=5000$. In this case we find a shift of the likelihood peak causing a detection of a nonzero neutrino mass at the 2$\sigma$ level when including all the lensing reconstruction channels. The inclusion of external data sets such as DESI BAO seems to help remove the biases, though 1$\sigma$ tensions might still remain. Nevertheless, based on the results above, we could expect inconsistencies between the inferred neutrino mass estimates from different data sets, if the $N_L^{(3/2)}$ bias is not accounted for in the parameter estimation for future, high-sensitivity/high-resolution CMB experiments. Finally, we find that the $N_L^{(3/2)}$ bias in the cross-correlation with a perfect tracer of the CMB lensing potential is reduced by a factor of roughly 2 with respect to the bias on the autospectrum, in agreement with the prediction of \cite{bohm2016}. The bias observed in cross-correlation with lower-redshift tracers might, however, be different due to the different weight that the post-Born term has for lower redshift probes, but we leave the investigation of this aspect to future work.\\

During the final stage of this work we compared our results on the $N_L^{(3/2)}$-bias with those of \cite{bohm2018}, who also performed a similar analysis using a CMB lensing field extracted from different $N$-body simulations. Despite their $N$-body simulations differing in resolution and box size, and the simulated sky area used for the lensing reconstruction being smaller than the full-sky results of our work, we find similar conclusions. This suggests that despite some quantitative conclusion of this work might still be simulation dependent and more complex physical effects are excluded from our modeling, the higher-order effects in CMB lensing should be treated carefully in future analyses in order to exploit the full scientific capacity of a CMB-S4-like observation. \\

\begin{acknowledgments}
We thank Vanessa B\"ohm and Blake Sherwin for useful discussions and the comparison of numerical results. We thank Antony Lewis for providing some of the analytical results used in the comparison with the results of this work. We thank Radek Stompor, Silvia Galli and Luca Pagano for useful discussions and comments on the manuscript. G.F. acknowledges the support of the CNES postdoctoral program. We acknowledge the use of {\sc camb}\footnote{\url{http://camb.info}} \cite{camb-1, camb-2}, {\sc class}\footnote{\url{http://github.com/lesgourg/class_public}} \cite{class}, {\sc MontePython}\footnote{\url{http://github.com/brinckmann/montepython_public}}\cite{mp-1,mp-2}, {\sc GetDist}\footnote{\url{http://github.com/cmbant/getdist}}, {\sc Lenspix}\footnote{\url{http://cosmologist.info/lenspix}} and {\sc Healpix}\footnote{\url{http://healpix.sourceforge.net}} \cite{healpix} software packages. This research used resources of the National Energy Research Scientific Computing Center (NERSC), a DOE Office of Science User Facility supported by the Office of Science of the U.S. Department of Energy under Contract No. DE-AC02-05CH11231. 
\end{acknowledgments}
\bibliography{postBorn}

\begin{thebibliography}{100}%
\makeatletter
\providecommand \@ifxundefined [1]{%
 \@ifx{#1\undefined}
}%
\providecommand \@ifnum [1]{%
 \ifnum #1\expandafter \@firstoftwo
 \else \expandafter \@secondoftwo
 \fi
}%
\providecommand \@ifx [1]{%
 \ifx #1\expandafter \@firstoftwo
 \else \expandafter \@secondoftwo
 \fi
}%
\providecommand \natexlab [1]{#1}%
\providecommand \enquote  [1]{``#1''}%
\providecommand \bibnamefont  [1]{#1}%
\providecommand \bibfnamefont [1]{#1}%
\providecommand \citenamefont [1]{#1}%
\providecommand \href@noop [0]{\@secondoftwo}%
\providecommand \href [0]{\begingroup \@sanitize@url \@href}%
\providecommand \@href[1]{\@@startlink{#1}\@@href}%
\providecommand \@@href[1]{\endgroup#1\@@endlink}%
\providecommand \@sanitize@url [0]{\catcode `\\12\catcode `\$12\catcode
  `\&12\catcode `\#12\catcode `\^12\catcode `\_12\catcode `\%12\relax}%
\providecommand \@@startlink[1]{}%
\providecommand \@@endlink[0]{}%
\providecommand \url  [0]{\begingroup\@sanitize@url \@url }%
\providecommand \@url [1]{\endgroup\@href {#1}{\urlprefix }}%
\providecommand \urlprefix  [0]{URL }%
\providecommand \Eprint [0]{\href }%
\providecommand \doibase [0]{http://dx.doi.org/}%
\providecommand \selectlanguage [0]{\@gobble}%
\providecommand \bibinfo  [0]{\@secondoftwo}%
\providecommand \bibfield  [0]{\@secondoftwo}%
\providecommand \translation [1]{[#1]}%
\providecommand \BibitemOpen [0]{}%
\providecommand \bibitemStop [0]{}%
\providecommand \bibitemNoStop [0]{.\EOS\space}%
\providecommand \EOS [0]{\spacefactor3000\relax}%
\providecommand \BibitemShut  [1]{\csname bibitem#1\endcsname}%
\let\auto@bib@innerbib\@empty
\bibitem [{\citenamefont {{Aghanim}}\ \emph {et~al.}(2008)\citenamefont
  {{Aghanim}}, \citenamefont {{Majumdar}},\ and\ \citenamefont
  {{Silk}}}]{aghanim2008}%
  \BibitemOpen
  \bibfield  {author} {\bibinfo {author} {\bibfnamefont {N.}~\bibnamefont
  {{Aghanim}}}, \bibinfo {author} {\bibfnamefont {S.}~\bibnamefont
  {{Majumdar}}}, \ and\ \bibinfo {author} {\bibfnamefont {J.}~\bibnamefont
  {{Silk}}},\ }\href {\doibase 10.1088/0034-4885/71/6/066902} {\bibfield
  {journal} {\bibinfo  {journal} {Rep. Prog. Phys.}\ }\textbf {\bibinfo
  {volume} {71}},\ \bibinfo {eid} {066902} (\bibinfo {year} {2008})},\ \Eprint
  {http://arxiv.org/abs/0711.0518} {arXiv:0711.0518} \BibitemShut {NoStop}%
\bibitem [{\citenamefont {{Sachs}}\ and\ \citenamefont {{Wolfe}}(1967)}]{isw}%
  \BibitemOpen
  \bibfield  {author} {\bibinfo {author} {\bibfnamefont {R.~K.}\ \bibnamefont
  {{Sachs}}}\ and\ \bibinfo {author} {\bibfnamefont {A.~M.}\ \bibnamefont
  {{Wolfe}}},\ }\href {\doibase 10.1086/148982} {\bibfield  {journal} {\bibinfo
   {journal} {\apj}\ }\textbf {\bibinfo {volume} {147}},\ \bibinfo {pages} {73}
  (\bibinfo {year} {1967})}\BibitemShut {NoStop}%
\bibitem [{\citenamefont {{Rees}}\ and\ \citenamefont
  {{Sciama}}(1968)}]{rees-sciama}%
  \BibitemOpen
  \bibfield  {author} {\bibinfo {author} {\bibfnamefont {M.~J.}\ \bibnamefont
  {{Rees}}}\ and\ \bibinfo {author} {\bibfnamefont {D.~W.}\ \bibnamefont
  {{Sciama}}},\ }\href {\doibase 10.1038/217511a0} {\bibfield  {journal}
  {\bibinfo  {journal} {\nat}\ }\textbf {\bibinfo {volume} {217}},\ \bibinfo
  {pages} {511} (\bibinfo {year} {1968})}\BibitemShut {NoStop}%
\bibitem [{\citenamefont {{Lewis}}\ and\ \citenamefont
  {{Challinor}}(2006)}]{lewis2006}%
  \BibitemOpen
  \bibfield  {author} {\bibinfo {author} {\bibfnamefont {A.}~\bibnamefont
  {{Lewis}}}\ and\ \bibinfo {author} {\bibfnamefont {A.}~\bibnamefont
  {{Challinor}}},\ }\href {\doibase 10.1016/j.physrep.2006.03.002} {\bibfield
  {journal} {\bibinfo  {journal} {\physrep}\ }\textbf {\bibinfo {volume}
  {429}},\ \bibinfo {pages} {1} (\bibinfo {year} {2006})},\ \Eprint
  {http://arxiv.org/abs/astro-ph/0601594} {arXiv:astro-ph/0601594} \BibitemShut
  {NoStop}%
\bibitem [{\citenamefont {{Reichardt}}\ \emph {et~al.}(2009)\citenamefont
  {{Reichardt}} \emph {et~al.}}]{reichardt2009}%
  \BibitemOpen
  \bibfield  {author} {\bibinfo {author} {\bibfnamefont {C.~L.}\ \bibnamefont
  {{Reichardt}}} \emph {et~al.},\ }\href {\doibase
  10.1088/0004-637X/694/2/1200} {\bibfield  {journal} {\bibinfo  {journal}
  {\apj}\ }\textbf {\bibinfo {volume} {694}},\ \bibinfo {pages} {1200}
  (\bibinfo {year} {2009})},\ \Eprint {http://arxiv.org/abs/0801.1491}
  {arXiv:0801.1491} \BibitemShut {NoStop}%
\bibitem [{\citenamefont {{Keisler}}\ \emph {et~al.}(2011)\citenamefont
  {{Keisler}} \emph {et~al.}}]{keisler2011}%
  \BibitemOpen
  \bibfield  {author} {\bibinfo {author} {\bibfnamefont {R.}~\bibnamefont
  {{Keisler}}} \emph {et~al.},\ }\href {\doibase 10.1088/0004-637X/743/1/28}
  {\bibfield  {journal} {\bibinfo  {journal} {\apj}\ }\textbf {\bibinfo
  {volume} {743}},\ \bibinfo {eid} {28} (\bibinfo {year} {2011})},\ \Eprint
  {http://arxiv.org/abs/1105.3182} {arXiv:1105.3182} \BibitemShut {NoStop}%
\bibitem [{\citenamefont {{Smith}}\ \emph {et~al.}(2007)\citenamefont
  {{Smith}}, \citenamefont {{Zahn}},\ and\ \citenamefont
  {{Dor{\'e}}}}]{smith2007}%
  \BibitemOpen
  \bibfield  {author} {\bibinfo {author} {\bibfnamefont {K.~M.}\ \bibnamefont
  {{Smith}}}, \bibinfo {author} {\bibfnamefont {O.}~\bibnamefont {{Zahn}}}, \
  and\ \bibinfo {author} {\bibfnamefont {O.}~\bibnamefont {{Dor{\'e}}}},\
  }\href {\doibase 10.1103/PhysRevD.76.043510} {\bibfield  {journal} {\bibinfo
  {journal} {\prd}\ }\textbf {\bibinfo {volume} {76}},\ \bibinfo {eid} {043510}
  (\bibinfo {year} {2007})},\ \Eprint {http://arxiv.org/abs/0705.3980}
  {arXiv:0705.3980} \BibitemShut {NoStop}%
\bibitem [{\citenamefont {{Hirata}}\ \emph {et~al.}(2008)\citenamefont
  {{Hirata}}, \citenamefont {{Ho}}, \citenamefont {{Padmanabhan}},
  \citenamefont {{Seljak}},\ and\ \citenamefont {{Bahcall}}}]{hirata2008}%
  \BibitemOpen
  \bibfield  {author} {\bibinfo {author} {\bibfnamefont {C.~M.}\ \bibnamefont
  {{Hirata}}}, \bibinfo {author} {\bibfnamefont {S.}~\bibnamefont {{Ho}}},
  \bibinfo {author} {\bibfnamefont {N.}~\bibnamefont {{Padmanabhan}}}, \bibinfo
  {author} {\bibfnamefont {U.}~\bibnamefont {{Seljak}}}, \ and\ \bibinfo
  {author} {\bibfnamefont {N.~A.}\ \bibnamefont {{Bahcall}}},\ }\href {\doibase
  10.1103/PhysRevD.78.043520} {\bibfield  {journal} {\bibinfo  {journal}
  {\prd}\ }\textbf {\bibinfo {volume} {78}},\ \bibinfo {eid} {043520} (\bibinfo
  {year} {2008})},\ \Eprint {http://arxiv.org/abs/0801.0644} {arXiv:0801.0644}
  \BibitemShut {NoStop}%
\bibitem [{\citenamefont {{Das}}\ \emph {et~al.}(2011)\citenamefont {{Das}}
  \emph {et~al.}}]{das2011}%
  \BibitemOpen
  \bibfield  {author} {\bibinfo {author} {\bibfnamefont {S.}~\bibnamefont
  {{Das}}} \emph {et~al.},\ }\href {\doibase 10.1103/PhysRevLett.107.021301}
  {\bibfield  {journal} {\bibinfo  {journal} {Phys. Rev. Lett.}\ }\textbf
  {\bibinfo {volume} {107}},\ \bibinfo {eid} {021301} (\bibinfo {year}
  {2011})},\ \Eprint {http://arxiv.org/abs/1103.2124} {arXiv:1103.2124}
  \BibitemShut {NoStop}%
\bibitem [{\citenamefont {{van Engelen}}\ \emph {et~al.}(2012)\citenamefont
  {{van Engelen}} \emph {et~al.}}]{vanengelen2012}%
  \BibitemOpen
  \bibfield  {author} {\bibinfo {author} {\bibfnamefont {A.}~\bibnamefont {{van
  Engelen}}} \emph {et~al.},\ }\href {\doibase 10.1088/0004-637X/756/2/142}
  {\bibfield  {journal} {\bibinfo  {journal} {\apj}\ }\textbf {\bibinfo
  {volume} {756}},\ \bibinfo {eid} {142} (\bibinfo {year} {2012})},\ \Eprint
  {http://arxiv.org/abs/1202.0546} {arXiv:1202.0546} \BibitemShut {NoStop}%
\bibitem [{\citenamefont {{Ade}}\ \emph {et~al.}(2016)\citenamefont {{Ade}}
  \emph {et~al.}}]{planck-lensing2016}%
  \BibitemOpen
  \bibfield  {author} {\bibinfo {author} {\bibfnamefont {P.~A.~R.}\
  \bibnamefont {{Ade}}} \emph {et~al.} (\bibinfo {collaboration} {Planck
  Collaboration}),\ }\href {\doibase 10.1051/0004-6361/201525941} {\bibfield
  {journal} {\bibinfo  {journal} {\aap}\ }\textbf {\bibinfo {volume} {594}},\
  \bibinfo {eid} {A15} (\bibinfo {year} {2016})},\ \Eprint
  {http://arxiv.org/abs/1502.01591} {arXiv:1502.01591} \BibitemShut {NoStop}%
\bibitem [{\citenamefont {{Ade}}\ \emph
  {et~al.}(2014{\natexlab{a}})\citenamefont {{Ade}} \emph {et~al.}}]{pb-lens}%
  \BibitemOpen
  \bibfield  {author} {\bibinfo {author} {\bibfnamefont {P.~A.~R.}\
  \bibnamefont {{Ade}}} \emph {et~al.} (\bibinfo {collaboration} {POLARBEAR
  Collaboration}),\ }\href {\doibase 10.1103/PhysRevLett.113.021301} {\bibfield
   {journal} {\bibinfo  {journal} {Phys. Rev. Lett.}\ }\textbf {\bibinfo
  {volume} {113}},\ \bibinfo {eid} {021301} (\bibinfo {year}
  {2014}{\natexlab{a}})},\ \Eprint {http://arxiv.org/abs/1312.6646}
  {arXiv:1312.6646} \BibitemShut {NoStop}%
\bibitem [{\citenamefont {{Story}}\ \emph {et~al.}(2015)\citenamefont {{Story}}
  \emph {et~al.}}]{story2015}%
  \BibitemOpen
  \bibfield  {author} {\bibinfo {author} {\bibfnamefont {K.~T.}\ \bibnamefont
  {{Story}}} \emph {et~al.},\ }\href {\doibase 10.1088/0004-637X/810/1/50}
  {\bibfield  {journal} {\bibinfo  {journal} {\apj}\ }\textbf {\bibinfo
  {volume} {810}},\ \bibinfo {eid} {50} (\bibinfo {year} {2015})},\ \Eprint
  {http://arxiv.org/abs/1412.4760} {arXiv:1412.4760} \BibitemShut {NoStop}%
\bibitem [{\citenamefont {{Sherwin}}\ \emph {et~al.}(2017)\citenamefont
  {{Sherwin}} \emph {et~al.}}]{sherwin2017}%
  \BibitemOpen
  \bibfield  {author} {\bibinfo {author} {\bibfnamefont {B.~D.}\ \bibnamefont
  {{Sherwin}}} \emph {et~al.},\ }\href {\doibase 10.1103/PhysRevD.95.123529}
  {\bibfield  {journal} {\bibinfo  {journal} {\prd}\ }\textbf {\bibinfo
  {volume} {95}},\ \bibinfo {eid} {123529} (\bibinfo {year}
  {2017})}\BibitemShut {NoStop}%
\bibitem [{\citenamefont {{Ade}}\ \emph
  {et~al.}(2014{\natexlab{b}})\citenamefont {{Ade}} \emph {et~al.}}]{pb-cib}%
  \BibitemOpen
  \bibfield  {author} {\bibinfo {author} {\bibfnamefont {P.~A.~R.}\
  \bibnamefont {{Ade}}} \emph {et~al.} (\bibinfo {collaboration} {POLARBEAR
  Collaboration}),\ }\href {\doibase 10.1103/PhysRevLett.112.131302} {\bibfield
   {journal} {\bibinfo  {journal} {Physical Review Letters}\ }\textbf {\bibinfo
  {volume} {112}},\ \bibinfo {eid} {131302} (\bibinfo {year}
  {2014}{\natexlab{b}})},\ \Eprint {http://arxiv.org/abs/1312.6645}
  {arXiv:1312.6645} \BibitemShut {NoStop}%
\bibitem [{\citenamefont {{Hanson}}\ \emph {et~al.}(2013)\citenamefont
  {{Hanson}} \emph {et~al.}}]{hanson2013}%
  \BibitemOpen
  \bibfield  {author} {\bibinfo {author} {\bibfnamefont {D.}~\bibnamefont
  {{Hanson}}} \emph {et~al.},\ }\href {\doibase 10.1103/PhysRevLett.111.141301}
  {\bibfield  {journal} {\bibinfo  {journal} {Phys. Rev. Lett.}\ }\textbf
  {\bibinfo {volume} {111}},\ \bibinfo {eid} {141301} (\bibinfo {year}
  {2013})},\ \Eprint {http://arxiv.org/abs/1307.5830} {arXiv:1307.5830}
  \BibitemShut {NoStop}%
\bibitem [{\citenamefont {{P.~A.~R.~Ade}}\ \emph {et~al.}(2014)\citenamefont
  {{P.~A.~R.~Ade}} \emph {et~al.}}]{pb-bb2014}%
  \BibitemOpen
  \bibfield  {author} {\bibinfo {author} {\bibnamefont {{P.~A.~R.~Ade}}} \emph
  {et~al.} (\bibinfo {collaboration} {POLARBEAR Collaboration}),\ }\href
  {\doibase 10.1088/0004-637X/794/2/171} {\bibfield  {journal} {\bibinfo
  {journal} {\apj}\ }\textbf {\bibinfo {volume} {794}},\ \bibinfo {eid} {171}
  (\bibinfo {year} {2014})},\ \Eprint {http://arxiv.org/abs/1403.2369}
  {arXiv:1403.2369} \BibitemShut {NoStop}%
\bibitem [{\citenamefont {{Ade}}\ \emph {et~al.}(2017)\citenamefont {{Ade}}
  \emph {et~al.}}]{pb-bb2017}%
  \BibitemOpen
  \bibfield  {author} {\bibinfo {author} {\bibfnamefont {P.~A.~R.}\
  \bibnamefont {{Ade}}} \emph {et~al.} (\bibinfo {collaboration} {POLARBEAR
  Collaboration}),\ }\href {\doibase 10.3847/1538-4357/aa8e9f} {\bibfield
  {journal} {\bibinfo  {journal} {\apj}\ }\textbf {\bibinfo {volume} {848}},\
  \bibinfo {eid} {121} (\bibinfo {year} {2017})},\ \Eprint
  {http://arxiv.org/abs/1705.02907} {arXiv:1705.02907} \BibitemShut {NoStop}%
\bibitem [{\citenamefont {{Keisler}}\ \emph {et~al.}(2015)\citenamefont
  {{Keisler}} \emph {et~al.}}]{keisler2015}%
  \BibitemOpen
  \bibfield  {author} {\bibinfo {author} {\bibfnamefont {R.}~\bibnamefont
  {{Keisler}}} \emph {et~al.},\ }\href {\doibase 10.1088/0004-637X/807/2/151}
  {\bibfield  {journal} {\bibinfo  {journal} {\apj}\ }\textbf {\bibinfo
  {volume} {807}},\ \bibinfo {eid} {151} (\bibinfo {year} {2015})},\ \Eprint
  {http://arxiv.org/abs/1503.02315} {arXiv:1503.02315} \BibitemShut {NoStop}%
\bibitem [{\citenamefont {{Abazajian}}\ \emph {et~al.}(2016)\citenamefont
  {{Abazajian}} \emph {et~al.}}]{cmbs4}%
  \BibitemOpen
  \bibfield  {author} {\bibinfo {author} {\bibfnamefont {K.~N.}\ \bibnamefont
  {{Abazajian}}} \emph {et~al.},\ }\href@noop {} {\  (\bibinfo {year}
  {2016})},\ \Eprint {http://arxiv.org/abs/1610.02743} {arXiv:1610.02743}
  \BibitemShut {NoStop}%
\bibitem [{\citenamefont {{Merkel}}\ and\ \citenamefont
  {{Sch{\"a}fer}}(2011)}]{merkel2011}%
  \BibitemOpen
  \bibfield  {author} {\bibinfo {author} {\bibfnamefont {P.~M.}\ \bibnamefont
  {{Merkel}}}\ and\ \bibinfo {author} {\bibfnamefont {B.~M.}\ \bibnamefont
  {{Sch{\"a}fer}}},\ }\href {\doibase 10.1111/j.1365-2966.2010.17739.x}
  {\bibfield  {journal} {\bibinfo  {journal} {\mnras}\ }\textbf {\bibinfo
  {volume} {411}},\ \bibinfo {pages} {1067} (\bibinfo {year} {2011})},\ \Eprint
  {http://arxiv.org/abs/1007.1408} {arXiv:1007.1408} \BibitemShut {NoStop}%
\bibitem [{\citenamefont {{Namikawa}}(2016)}]{namikawa2016}%
  \BibitemOpen
  \bibfield  {author} {\bibinfo {author} {\bibfnamefont {T.}~\bibnamefont
  {{Namikawa}}},\ }\href {\doibase 10.1103/PhysRevD.93.121301} {\bibfield
  {journal} {\bibinfo  {journal} {\prd}\ }\textbf {\bibinfo {volume} {93}},\
  \bibinfo {eid} {121301} (\bibinfo {year} {2016})},\ \Eprint
  {http://arxiv.org/abs/1604.08578} {arXiv:1604.08578} \BibitemShut {NoStop}%
\bibitem [{\citenamefont {{Pratten}}\ and\ \citenamefont
  {{Lewis}}(2016)}]{pratten2016}%
  \BibitemOpen
  \bibfield  {author} {\bibinfo {author} {\bibfnamefont {G.}~\bibnamefont
  {{Pratten}}}\ and\ \bibinfo {author} {\bibfnamefont {A.}~\bibnamefont
  {{Lewis}}},\ }\href {\doibase 10.1088/1475-7516/2016/08/047} {\bibfield
  {journal} {\bibinfo  {journal} {\jcap}\ }\textbf {\bibinfo {volume} {8}},\
  \bibinfo {eid} {047} (\bibinfo {year} {2016})},\ \Eprint
  {http://arxiv.org/abs/1605.05662} {arXiv:1605.05662} \BibitemShut {NoStop}%
\bibitem [{\citenamefont {{Carbone}}\ \emph {et~al.}(2008)\citenamefont
  {{Carbone}}, \citenamefont {{Springel}}, \citenamefont {{Baccigalupi}},
  \citenamefont {{Bartelmann}},\ and\ \citenamefont
  {{Matarrese}}}]{carbone2008}%
  \BibitemOpen
  \bibfield  {author} {\bibinfo {author} {\bibfnamefont {C.}~\bibnamefont
  {{Carbone}}}, \bibinfo {author} {\bibfnamefont {V.}~\bibnamefont
  {{Springel}}}, \bibinfo {author} {\bibfnamefont {C.}~\bibnamefont
  {{Baccigalupi}}}, \bibinfo {author} {\bibfnamefont {M.}~\bibnamefont
  {{Bartelmann}}}, \ and\ \bibinfo {author} {\bibfnamefont {S.}~\bibnamefont
  {{Matarrese}}},\ }\href {\doibase 10.1111/j.1365-2966.2008.13544.x}
  {\bibfield  {journal} {\bibinfo  {journal} {\mnras}\ }\textbf {\bibinfo
  {volume} {388}},\ \bibinfo {pages} {1618} (\bibinfo {year} {2008})},\ \Eprint
  {http://arxiv.org/abs/0711.2655} {arXiv:0711.2655} \BibitemShut {NoStop}%
\bibitem [{\citenamefont {{Carbone}}\ \emph {et~al.}(2009)\citenamefont
  {{Carbone}}, \citenamefont {{Baccigalupi}}, \citenamefont {{Bartelmann}},
  \citenamefont {{Matarrese}},\ and\ \citenamefont {{Springel}}}]{carbone2009}%
  \BibitemOpen
  \bibfield  {author} {\bibinfo {author} {\bibfnamefont {C.}~\bibnamefont
  {{Carbone}}}, \bibinfo {author} {\bibfnamefont {C.}~\bibnamefont
  {{Baccigalupi}}}, \bibinfo {author} {\bibfnamefont {M.}~\bibnamefont
  {{Bartelmann}}}, \bibinfo {author} {\bibfnamefont {S.}~\bibnamefont
  {{Matarrese}}}, \ and\ \bibinfo {author} {\bibfnamefont {V.}~\bibnamefont
  {{Springel}}},\ }\href {\doibase 10.1111/j.1365-2966.2009.14746.x} {\bibfield
   {journal} {\bibinfo  {journal} {\mnras}\ }\textbf {\bibinfo {volume}
  {396}},\ \bibinfo {pages} {668} (\bibinfo {year} {2009})},\ \Eprint
  {http://arxiv.org/abs/0810.4145} {arXiv:0810.4145} \BibitemShut {NoStop}%
\bibitem [{\citenamefont {{Hagstotz}}\ \emph {et~al.}(2015)\citenamefont
  {{Hagstotz}}, \citenamefont {{Sch{\"a}fer}},\ and\ \citenamefont
  {{Merkel}}}]{hagstotz2015}%
  \BibitemOpen
  \bibfield  {author} {\bibinfo {author} {\bibfnamefont {S.}~\bibnamefont
  {{Hagstotz}}}, \bibinfo {author} {\bibfnamefont {B.~M.}\ \bibnamefont
  {{Sch{\"a}fer}}}, \ and\ \bibinfo {author} {\bibfnamefont {P.~M.}\
  \bibnamefont {{Merkel}}},\ }\href {\doibase 10.1093/mnras/stv1977} {\bibfield
   {journal} {\bibinfo  {journal} {\mnras}\ }\textbf {\bibinfo {volume}
  {454}},\ \bibinfo {pages} {831} (\bibinfo {year} {2015})},\ \Eprint
  {http://arxiv.org/abs/1410.8452} {arXiv:1410.8452} \BibitemShut {NoStop}%
\bibitem [{\citenamefont {{Lewis}}\ and\ \citenamefont
  {{Pratten}}(2016)}]{lewis2016}%
  \BibitemOpen
  \bibfield  {author} {\bibinfo {author} {\bibfnamefont {A.}~\bibnamefont
  {{Lewis}}}\ and\ \bibinfo {author} {\bibfnamefont {G.}~\bibnamefont
  {{Pratten}}},\ }\href {\doibase 10.1088/1475-7516/2016/12/003} {\bibfield
  {journal} {\bibinfo  {journal} {\jcap}\ }\textbf {\bibinfo {volume} {12}},\
  \bibinfo {eid} {003} (\bibinfo {year} {2016})},\ \Eprint
  {http://arxiv.org/abs/1608.01263} {arXiv:1608.01263} \BibitemShut {NoStop}%
\bibitem [{\citenamefont {{Fabbian}}\ \emph {et~al.}(2018)\citenamefont
  {{Fabbian}}, \citenamefont {{Calabrese}},\ and\ \citenamefont
  {{Carbone}}}]{fabbian2018}%
  \BibitemOpen
  \bibfield  {author} {\bibinfo {author} {\bibfnamefont {G.}~\bibnamefont
  {{Fabbian}}}, \bibinfo {author} {\bibfnamefont {M.}~\bibnamefont
  {{Calabrese}}}, \ and\ \bibinfo {author} {\bibfnamefont {C.}~\bibnamefont
  {{Carbone}}},\ }\href {\doibase 10.1088/1475-7516/2018/02/050} {\bibfield
  {journal} {\bibinfo  {journal} {\jcap}\ }\textbf {\bibinfo {volume} {2}},\
  \bibinfo {eid} {050} (\bibinfo {year} {2018})},\ \Eprint
  {http://arxiv.org/abs/1702.03317} {arXiv:1702.03317} \BibitemShut {NoStop}%
\bibitem [{\citenamefont {{Cooray}}\ and\ \citenamefont
  {{Hu}}(2002)}]{cooray2002}%
  \BibitemOpen
  \bibfield  {author} {\bibinfo {author} {\bibfnamefont {A.}~\bibnamefont
  {{Cooray}}}\ and\ \bibinfo {author} {\bibfnamefont {W.}~\bibnamefont
  {{Hu}}},\ }\href {\doibase 10.1086/340892} {\bibfield  {journal} {\bibinfo
  {journal} {\apj}\ }\textbf {\bibinfo {volume} {574}},\ \bibinfo {pages} {19}
  (\bibinfo {year} {2002})},\ \Eprint {http://arxiv.org/abs/astro-ph/0202411}
  {arXiv:astro-ph/0202411} \BibitemShut {NoStop}%
\bibitem [{\citenamefont {{Shapiro}}\ and\ \citenamefont
  {{Cooray}}(2006)}]{shapiro2006}%
  \BibitemOpen
  \bibfield  {author} {\bibinfo {author} {\bibfnamefont {C.}~\bibnamefont
  {{Shapiro}}}\ and\ \bibinfo {author} {\bibfnamefont {A.}~\bibnamefont
  {{Cooray}}},\ }\href {\doibase 10.1088/1475-7516/2006/03/007} {\bibfield
  {journal} {\bibinfo  {journal} {\jcap}\ }\textbf {\bibinfo {volume} {3}},\
  \bibinfo {eid} {007} (\bibinfo {year} {2006})},\ \Eprint
  {http://arxiv.org/abs/astro-ph/0601226} {arXiv:astro-ph/0601226} \BibitemShut
  {NoStop}%
\bibitem [{\citenamefont {{Krause}}\ and\ \citenamefont
  {{Hirata}}(2010)}]{krause2010}%
  \BibitemOpen
  \bibfield  {author} {\bibinfo {author} {\bibfnamefont {E.}~\bibnamefont
  {{Krause}}}\ and\ \bibinfo {author} {\bibfnamefont {C.~M.}\ \bibnamefont
  {{Hirata}}},\ }\href {\doibase 10.1051/0004-6361/200913524} {\bibfield
  {journal} {\bibinfo  {journal} {\aap}\ }\textbf {\bibinfo {volume} {523}},\
  \bibinfo {eid} {A28} (\bibinfo {year} {2010})},\ \Eprint
  {http://arxiv.org/abs/0910.3786} {arXiv:0910.3786} \BibitemShut {NoStop}%
\bibitem [{\citenamefont {{Petri}}\ \emph {et~al.}(2017)\citenamefont
  {{Petri}}, \citenamefont {{Haiman}},\ and\ \citenamefont
  {{May}}}]{petri2017}%
  \BibitemOpen
  \bibfield  {author} {\bibinfo {author} {\bibfnamefont {A.}~\bibnamefont
  {{Petri}}}, \bibinfo {author} {\bibfnamefont {Z.}~\bibnamefont {{Haiman}}}, \
  and\ \bibinfo {author} {\bibfnamefont {M.}~\bibnamefont {{May}}},\ }\href
  {\doibase 10.1103/PhysRevD.95.123503} {\bibfield  {journal} {\bibinfo
  {journal} {\prd}\ }\textbf {\bibinfo {volume} {95}},\ \bibinfo {eid} {123503}
  (\bibinfo {year} {2017})},\ \Eprint {http://arxiv.org/abs/1612.00852}
  {arXiv:1612.00852} \BibitemShut {NoStop}%
\bibitem [{\citenamefont {{B{\"o}hm}}\ \emph {et~al.}(2016)\citenamefont
  {{B{\"o}hm}}, \citenamefont {{Schmittfull}},\ and\ \citenamefont
  {{Sherwin}}}]{bohm2016}%
  \BibitemOpen
  \bibfield  {author} {\bibinfo {author} {\bibfnamefont {V.}~\bibnamefont
  {{B{\"o}hm}}}, \bibinfo {author} {\bibfnamefont {M.}~\bibnamefont
  {{Schmittfull}}}, \ and\ \bibinfo {author} {\bibfnamefont {B.~D.}\
  \bibnamefont {{Sherwin}}},\ }\href {\doibase 10.1103/PhysRevD.94.043519}
  {\bibfield  {journal} {\bibinfo  {journal} {\prd}\ }\textbf {\bibinfo
  {volume} {94}},\ \bibinfo {eid} {043519} (\bibinfo {year} {2016})},\ \Eprint
  {http://arxiv.org/abs/1605.01392} {arXiv:1605.01392} \BibitemShut {NoStop}%
\bibitem [{\citenamefont {{Becker}}(2013)}]{becker2013}%
  \BibitemOpen
  \bibfield  {author} {\bibinfo {author} {\bibfnamefont {M.~R.}\ \bibnamefont
  {{Becker}}},\ }\href {\doibase 10.1093/mnras/stt1352} {\bibfield  {journal}
  {\bibinfo  {journal} {\mnras}\ }\textbf {\bibinfo {volume} {435}},\ \bibinfo
  {pages} {115} (\bibinfo {year} {2013})},\ \Eprint
  {http://arxiv.org/abs/arXiv:1210.3069} {arXiv:arXiv:1210.3069} \BibitemShut
  {NoStop}%
\bibitem [{\citenamefont {{Bartelmann}}\ and\ \citenamefont
  {{Schneider}}(2001)}]{bartelmann-shneider}%
  \BibitemOpen
  \bibfield  {author} {\bibinfo {author} {\bibfnamefont {M.}~\bibnamefont
  {{Bartelmann}}}\ and\ \bibinfo {author} {\bibfnamefont {P.}~\bibnamefont
  {{Schneider}}},\ }\href {\doibase 10.1016/S0370-1573(00)00082-X} {\bibfield
  {journal} {\bibinfo  {journal} {\physrep}\ }\textbf {\bibinfo {volume}
  {340}},\ \bibinfo {pages} {291} (\bibinfo {year} {2001})},\ \Eprint
  {http://arxiv.org/abs/astro-ph/9912508} {arXiv:astro-ph/9912508} \BibitemShut
  {NoStop}%
\bibitem [{\citenamefont {{Stebbins}}(1996)}]{stebbins96}%
  \BibitemOpen
  \bibfield  {author} {\bibinfo {author} {\bibfnamefont {A.}~\bibnamefont
  {{Stebbins}}},\ }\href@noop {} {\  (\bibinfo {year} {1996})},\ \Eprint
  {http://arxiv.org/abs/astro-ph/9609149} {arXiv:astro-ph/9609149} \BibitemShut
  {NoStop}%
\bibitem [{\citenamefont {{Hu}}(2000)}]{hu2000}%
  \BibitemOpen
  \bibfield  {author} {\bibinfo {author} {\bibfnamefont {W.}~\bibnamefont
  {{Hu}}},\ }\href {\doibase 10.1103/PhysRevD.62.043007} {\bibfield  {journal}
  {\bibinfo  {journal} {\prd}\ }\textbf {\bibinfo {volume} {62}},\ \bibinfo
  {eid} {043007} (\bibinfo {year} {2000})},\ \Eprint
  {http://arxiv.org/abs/astro-ph/0001303} {arXiv:astro-ph/0001303} \BibitemShut
  {NoStop}%
\bibitem [{\citenamefont {{Marozzi}}\ \emph {et~al.}(2016)\citenamefont
  {{Marozzi}}, \citenamefont {{Fanizza}}, \citenamefont {{Di Dio}},\ and\
  \citenamefont {{Durrer}}}]{marozzi2016}%
  \BibitemOpen
  \bibfield  {author} {\bibinfo {author} {\bibfnamefont {G.}~\bibnamefont
  {{Marozzi}}}, \bibinfo {author} {\bibfnamefont {G.}~\bibnamefont
  {{Fanizza}}}, \bibinfo {author} {\bibfnamefont {E.}~\bibnamefont {{Di Dio}}},
  \ and\ \bibinfo {author} {\bibfnamefont {R.}~\bibnamefont {{Durrer}}},\
  }\href {\doibase 10.1088/1475-7516/2016/09/028} {\bibfield  {journal}
  {\bibinfo  {journal} {\jcap}\ }\textbf {\bibinfo {volume} {9}},\ \bibinfo
  {eid} {028} (\bibinfo {year} {2016})},\ \Eprint
  {http://arxiv.org/abs/1605.08761} {arXiv:1605.08761} \BibitemShut {NoStop}%
\bibitem [{\citenamefont {{Hirata}}\ and\ \citenamefont
  {{Seljak}}(2003{\natexlab{a}})}]{hirata2003-curl}%
  \BibitemOpen
  \bibfield  {author} {\bibinfo {author} {\bibfnamefont {C.~M.}\ \bibnamefont
  {{Hirata}}}\ and\ \bibinfo {author} {\bibfnamefont {U.}~\bibnamefont
  {{Seljak}}},\ }\href {\doibase 10.1103/PhysRevD.67.043001} {\bibfield
  {journal} {\bibinfo  {journal} {\prd}\ }\textbf {\bibinfo {volume} {67}},\
  \bibinfo {eid} {043001} (\bibinfo {year} {2003}{\natexlab{a}})},\ \Eprint
  {http://arxiv.org/abs/astro-ph/0209489} {arXiv:astro-ph/0209489} \BibitemShut
  {NoStop}%
\bibitem [{\citenamefont {{Fabbian}}\ and\ \citenamefont
  {{Stompor}}(2013)}]{fabbian2013}%
  \BibitemOpen
  \bibfield  {author} {\bibinfo {author} {\bibfnamefont {G.}~\bibnamefont
  {{Fabbian}}}\ and\ \bibinfo {author} {\bibfnamefont {R.}~\bibnamefont
  {{Stompor}}},\ }\href {\doibase 10.1051/0004-6361/201321575} {\bibfield
  {journal} {\bibinfo  {journal} {\aap}\ }\textbf {\bibinfo {volume} {556}},\
  \bibinfo {eid} {A109} (\bibinfo {year} {2013})},\ \Eprint
  {http://arxiv.org/abs/1303.6550} {arXiv:1303.6550} \BibitemShut {NoStop}%
\bibitem [{\citenamefont {{Hu}}\ and\ \citenamefont
  {{Okamoto}}(2002)}]{hu-okamoto}%
  \BibitemOpen
  \bibfield  {author} {\bibinfo {author} {\bibfnamefont {W.}~\bibnamefont
  {{Hu}}}\ and\ \bibinfo {author} {\bibfnamefont {T.}~\bibnamefont
  {{Okamoto}}},\ }\href {\doibase 10.1086/341110} {\bibfield  {journal}
  {\bibinfo  {journal} {\apj}\ }\textbf {\bibinfo {volume} {574}},\ \bibinfo
  {pages} {566} (\bibinfo {year} {2002})},\ \Eprint
  {http://arxiv.org/abs/astro-ph/0111606} {arXiv:astro-ph/0111606} \BibitemShut
  {NoStop}%
\bibitem [{\citenamefont {{Okamoto}}\ and\ \citenamefont
  {{Hu}}(2003)}]{okamoto-hu}%
  \BibitemOpen
  \bibfield  {author} {\bibinfo {author} {\bibfnamefont {T.}~\bibnamefont
  {{Okamoto}}}\ and\ \bibinfo {author} {\bibfnamefont {W.}~\bibnamefont
  {{Hu}}},\ }\href {\doibase 10.1103/PhysRevD.67.083002} {\bibfield  {journal}
  {\bibinfo  {journal} {\prd}\ }\textbf {\bibinfo {volume} {67}},\ \bibinfo
  {eid} {083002} (\bibinfo {year} {2003})},\ \Eprint
  {http://arxiv.org/abs/astro-ph/0301031} {arXiv:astro-ph/0301031} \BibitemShut
  {NoStop}%
\bibitem [{\citenamefont {{Hirata}}\ and\ \citenamefont
  {{Seljak}}(2003{\natexlab{b}})}]{hirata2003}%
  \BibitemOpen
  \bibfield  {author} {\bibinfo {author} {\bibfnamefont {C.~M.}\ \bibnamefont
  {{Hirata}}}\ and\ \bibinfo {author} {\bibfnamefont {U.}~\bibnamefont
  {{Seljak}}},\ }\href {\doibase 10.1103/PhysRevD.68.083002} {\bibfield
  {journal} {\bibinfo  {journal} {\prd}\ }\textbf {\bibinfo {volume} {68}},\
  \bibinfo {eid} {083002} (\bibinfo {year} {2003}{\natexlab{b}})},\ \Eprint
  {http://arxiv.org/abs/astro-ph/0306354} {arXiv:astro-ph/0306354} \BibitemShut
  {NoStop}%
\bibitem [{\citenamefont {{Carron}}\ and\ \citenamefont
  {{Lewis}}(2017)}]{carron2017}%
  \BibitemOpen
  \bibfield  {author} {\bibinfo {author} {\bibfnamefont {J.}~\bibnamefont
  {{Carron}}}\ and\ \bibinfo {author} {\bibfnamefont {A.}~\bibnamefont
  {{Lewis}}},\ }\href {\doibase 10.1103/PhysRevD.96.063510} {\bibfield
  {journal} {\bibinfo  {journal} {\prd}\ }\textbf {\bibinfo {volume} {96}},\
  \bibinfo {eid} {063510} (\bibinfo {year} {2017})},\ \Eprint
  {http://arxiv.org/abs/1704.08230} {arXiv:1704.08230} \BibitemShut {NoStop}%
\bibitem [{\citenamefont {{Millea}}\ \emph {et~al.}(2017)\citenamefont
  {{Millea}}, \citenamefont {{Anderes}},\ and\ \citenamefont
  {{Wandelt}}}]{millea2017}%
  \BibitemOpen
  \bibfield  {author} {\bibinfo {author} {\bibfnamefont {M.}~\bibnamefont
  {{Millea}}}, \bibinfo {author} {\bibfnamefont {E.}~\bibnamefont {{Anderes}}},
  \ and\ \bibinfo {author} {\bibfnamefont {B.~D.}\ \bibnamefont {{Wandelt}}},\
  }\href@noop {} {\  (\bibinfo {year} {2017})},\ \Eprint
  {http://arxiv.org/abs/1708.06753} {arXiv:1708.06753} \BibitemShut {NoStop}%
\bibitem [{\citenamefont {{Hanson}}\ \emph {et~al.}(2011)\citenamefont
  {{Hanson}}, \citenamefont {{Challinor}}, \citenamefont {{Efstathiou}},\ and\
  \citenamefont {{Bielewicz}}}]{hanson2011}%
  \BibitemOpen
  \bibfield  {author} {\bibinfo {author} {\bibfnamefont {D.}~\bibnamefont
  {{Hanson}}}, \bibinfo {author} {\bibfnamefont {A.}~\bibnamefont
  {{Challinor}}}, \bibinfo {author} {\bibfnamefont {G.}~\bibnamefont
  {{Efstathiou}}}, \ and\ \bibinfo {author} {\bibfnamefont {P.}~\bibnamefont
  {{Bielewicz}}},\ }\href {\doibase 10.1103/PhysRevD.83.043005} {\bibfield
  {journal} {\bibinfo  {journal} {\prd}\ }\textbf {\bibinfo {volume} {83}},\
  \bibinfo {eid} {043005} (\bibinfo {year} {2011})},\ \Eprint
  {http://arxiv.org/abs/1008.4403} {arXiv:1008.4403} \BibitemShut {NoStop}%
\bibitem [{\citenamefont {{Peloton}}\ \emph {et~al.}(2017)\citenamefont
  {{Peloton}}, \citenamefont {{Schmittfull}}, \citenamefont {{Lewis}},
  \citenamefont {{Carron}},\ and\ \citenamefont {{Zahn}}}]{peloton2017}%
  \BibitemOpen
  \bibfield  {author} {\bibinfo {author} {\bibfnamefont {J.}~\bibnamefont
  {{Peloton}}}, \bibinfo {author} {\bibfnamefont {M.}~\bibnamefont
  {{Schmittfull}}}, \bibinfo {author} {\bibfnamefont {A.}~\bibnamefont
  {{Lewis}}}, \bibinfo {author} {\bibfnamefont {J.}~\bibnamefont {{Carron}}}, \
  and\ \bibinfo {author} {\bibfnamefont {O.}~\bibnamefont {{Zahn}}},\ }\href
  {\doibase 10.1103/PhysRevD.95.043508} {\bibfield  {journal} {\bibinfo
  {journal} {\prd}\ }\textbf {\bibinfo {volume} {95}},\ \bibinfo {eid} {043508}
  (\bibinfo {year} {2017})},\ \Eprint {http://arxiv.org/abs/1611.01446}
  {arXiv:1611.01446} \BibitemShut {NoStop}%
\bibitem [{\citenamefont {{Lewis}}\ \emph {et~al.}(2011)\citenamefont
  {{Lewis}}, \citenamefont {{Challinor}},\ and\ \citenamefont
  {{Hanson}}}]{lewis2011}%
  \BibitemOpen
  \bibfield  {author} {\bibinfo {author} {\bibfnamefont {A.}~\bibnamefont
  {{Lewis}}}, \bibinfo {author} {\bibfnamefont {A.}~\bibnamefont
  {{Challinor}}}, \ and\ \bibinfo {author} {\bibfnamefont {D.}~\bibnamefont
  {{Hanson}}},\ }\href {\doibase 10.1088/1475-7516/2011/03/018} {\bibfield
  {journal} {\bibinfo  {journal} {\jcap}\ }\textbf {\bibinfo {volume} {3}},\
  \bibinfo {eid} {018} (\bibinfo {year} {2011})},\ \Eprint
  {http://arxiv.org/abs/1101.2234} {arXiv:1101.2234} \BibitemShut {NoStop}%
\bibitem [{\citenamefont {{Kesden}}\ \emph {et~al.}(2003)\citenamefont
  {{Kesden}}, \citenamefont {{Cooray}},\ and\ \citenamefont
  {{Kamionkowski}}}]{kesden2003}%
  \BibitemOpen
  \bibfield  {author} {\bibinfo {author} {\bibfnamefont {M.}~\bibnamefont
  {{Kesden}}}, \bibinfo {author} {\bibfnamefont {A.}~\bibnamefont {{Cooray}}},
  \ and\ \bibinfo {author} {\bibfnamefont {M.}~\bibnamefont {{Kamionkowski}}},\
  }\href {\doibase 10.1103/PhysRevD.67.123507} {\bibfield  {journal} {\bibinfo
  {journal} {\prd}\ }\textbf {\bibinfo {volume} {67}},\ \bibinfo {eid} {123507}
  (\bibinfo {year} {2003})},\ \Eprint {http://arxiv.org/abs/astro-ph/0302536}
  {arXiv:astro-ph/0302536} \BibitemShut {NoStop}%
\bibitem [{\citenamefont {Namikawa}\ \emph {et~al.}(2013)\citenamefont
  {Namikawa}, \citenamefont {Hanson},\ and\ \citenamefont
  {Takahashi}}]{namikawa2013}%
  \BibitemOpen
  \bibfield  {author} {\bibinfo {author} {\bibfnamefont {T.}~\bibnamefont
  {Namikawa}}, \bibinfo {author} {\bibfnamefont {D.}~\bibnamefont {Hanson}}, \
  and\ \bibinfo {author} {\bibfnamefont {R.}~\bibnamefont {Takahashi}},\ }\href
  {\doibase 10.1093/mnras/stt195} {\bibfield  {journal} {\bibinfo  {journal}
  {Mon. Not. R. Astron. Soc.}\ }\textbf {\bibinfo {volume} {431}},\ \bibinfo
  {pages} {609} (\bibinfo {year} {2013})},\ \Eprint
  {http://arxiv.org/abs/1209.0091} {arXiv:1209.0091} \BibitemShut {NoStop}%
\bibitem [{\citenamefont {Namikawa}\ and\ \citenamefont
  {Takahashi}(2014)}]{namikawa2014}%
  \BibitemOpen
  \bibfield  {author} {\bibinfo {author} {\bibfnamefont {T.}~\bibnamefont
  {Namikawa}}\ and\ \bibinfo {author} {\bibfnamefont {R.}~\bibnamefont
  {Takahashi}},\ }\href {\doibase 10.1093/mnras/stt2290} {\bibfield  {journal}
  {\bibinfo  {journal} {Mon. Not. R. Astron. Soc.}\ }\textbf {\bibinfo {volume}
  {438}},\ \bibinfo {pages} {1507} (\bibinfo {year} {2014})},\ \Eprint
  {http://arxiv.org/abs/1310.2372} {arXiv:1310.2372} \BibitemShut {NoStop}%
\bibitem [{\citenamefont {{Anderes}}(2013)}]{anderes2013}%
  \BibitemOpen
  \bibfield  {author} {\bibinfo {author} {\bibfnamefont {E.}~\bibnamefont
  {{Anderes}}},\ }\href {\doibase 10.1103/PhysRevD.88.083517} {\bibfield
  {journal} {\bibinfo  {journal} {\prd}\ }\textbf {\bibinfo {volume} {88}},\
  \bibinfo {eid} {083517} (\bibinfo {year} {2013})},\ \Eprint
  {http://arxiv.org/abs/1301.2576} {arXiv:1301.2576} \BibitemShut {NoStop}%
\bibitem [{\citenamefont {Marozzi}\ \emph {et~al.}(2018)\citenamefont
  {Marozzi}, \citenamefont {Fanizza}, \citenamefont {Di~Dio},\ and\
  \citenamefont {Durrer}}]{marozzi2016pol}%
  \BibitemOpen
  \bibfield  {author} {\bibinfo {author} {\bibfnamefont {G.}~\bibnamefont
  {Marozzi}}, \bibinfo {author} {\bibfnamefont {G.}~\bibnamefont {Fanizza}},
  \bibinfo {author} {\bibfnamefont {E.}~\bibnamefont {Di~Dio}}, \ and\ \bibinfo
  {author} {\bibfnamefont {R.}~\bibnamefont {Durrer}},\ }\href {\doibase
  10.1103/PhysRevD.98.023535} {\bibfield  {journal} {\bibinfo  {journal} {Phys.
  Rev. D}\ }\textbf {\bibinfo {volume} {98}},\ \bibinfo {pages} {023535}
  (\bibinfo {year} {2018})},\ \Eprint {http://arxiv.org/abs/arXiv:1612.07263}
  {arXiv:arXiv:1612.07263} \BibitemShut {NoStop}%
\bibitem [{\citenamefont {{Carbone}}\ \emph {et~al.}(2016)\citenamefont
  {{Carbone}}, \citenamefont {{Petkova}},\ and\ \citenamefont
  {{Dolag}}}]{carbone2016}%
  \BibitemOpen
  \bibfield  {author} {\bibinfo {author} {\bibfnamefont {C.}~\bibnamefont
  {{Carbone}}}, \bibinfo {author} {\bibfnamefont {M.}~\bibnamefont
  {{Petkova}}}, \ and\ \bibinfo {author} {\bibfnamefont {K.}~\bibnamefont
  {{Dolag}}},\ }\href {\doibase 10.1088/1475-7516/2016/07/034} {\bibfield
  {journal} {\bibinfo  {journal} {\jcap}\ }\textbf {\bibinfo {volume} {7}},\
  \bibinfo {eid} {034} (\bibinfo {year} {2016})},\ \Eprint
  {http://arxiv.org/abs/1605.02024} {arXiv:1605.02024} \BibitemShut {NoStop}%
\bibitem [{\citenamefont {{Castorina}}\ \emph {et~al.}(2015)\citenamefont
  {{Castorina}}, \citenamefont {{Carbone}}, \citenamefont {{Bel}},
  \citenamefont {{Sefusatti}},\ and\ \citenamefont {{Dolag}}}]{castorina2015}%
  \BibitemOpen
  \bibfield  {author} {\bibinfo {author} {\bibfnamefont {E.}~\bibnamefont
  {{Castorina}}}, \bibinfo {author} {\bibfnamefont {C.}~\bibnamefont
  {{Carbone}}}, \bibinfo {author} {\bibfnamefont {J.}~\bibnamefont {{Bel}}},
  \bibinfo {author} {\bibfnamefont {E.}~\bibnamefont {{Sefusatti}}}, \ and\
  \bibinfo {author} {\bibfnamefont {K.}~\bibnamefont {{Dolag}}},\ }\href
  {\doibase 10.1088/1475-7516/2015/07/043} {\bibfield  {journal} {\bibinfo
  {journal} {\jcap}\ }\textbf {\bibinfo {volume} {7}},\ \bibinfo {eid} {043}
  (\bibinfo {year} {2015})},\ \Eprint {http://arxiv.org/abs/1505.07148}
  {arXiv:1505.07148} \BibitemShut {NoStop}%
\bibitem [{\citenamefont {{Calabrese}}\ \emph {et~al.}(2015)\citenamefont
  {{Calabrese}}, \citenamefont {{Carbone}}, \citenamefont {{Fabbian}},
  \citenamefont {{Baldi}},\ and\ \citenamefont
  {{Baccigalupi}}}]{calabrese2015}%
  \BibitemOpen
  \bibfield  {author} {\bibinfo {author} {\bibfnamefont {M.}~\bibnamefont
  {{Calabrese}}}, \bibinfo {author} {\bibfnamefont {C.}~\bibnamefont
  {{Carbone}}}, \bibinfo {author} {\bibfnamefont {G.}~\bibnamefont
  {{Fabbian}}}, \bibinfo {author} {\bibfnamefont {M.}~\bibnamefont {{Baldi}}},
  \ and\ \bibinfo {author} {\bibfnamefont {C.}~\bibnamefont {{Baccigalupi}}},\
  }\href {\doibase 10.1088/1475-7516/2015/03/049} {\bibfield  {journal}
  {\bibinfo  {journal} {\jcap}\ }\textbf {\bibinfo {volume} {3}},\ \bibinfo
  {eid} {049} (\bibinfo {year} {2015})},\ \Eprint
  {http://arxiv.org/abs/1409.7680} {arXiv:1409.7680} \BibitemShut {NoStop}%
\bibitem [{\citenamefont {{Fosalba}}\ \emph {et~al.}(2008)\citenamefont
  {{Fosalba}}, \citenamefont {{Gazta{\~n}aga}}, \citenamefont {{Castander}},\
  and\ \citenamefont {{Manera}}}]{fosalba2008}%
  \BibitemOpen
  \bibfield  {author} {\bibinfo {author} {\bibfnamefont {P.}~\bibnamefont
  {{Fosalba}}}, \bibinfo {author} {\bibfnamefont {E.}~\bibnamefont
  {{Gazta{\~n}aga}}}, \bibinfo {author} {\bibfnamefont {F.~J.}\ \bibnamefont
  {{Castander}}}, \ and\ \bibinfo {author} {\bibfnamefont {M.}~\bibnamefont
  {{Manera}}},\ }\href {\doibase 10.1111/j.1365-2966.2008.13910.x} {\bibfield
  {journal} {\bibinfo  {journal} {\mnras}\ }\textbf {\bibinfo {volume} {391}},\
  \bibinfo {pages} {435} (\bibinfo {year} {2008})},\ \Eprint
  {http://arxiv.org/abs/0711.1540} {arXiv:0711.1540} \BibitemShut {NoStop}%
\bibitem [{\citenamefont {{Jain}}\ \emph {et~al.}(2000)\citenamefont {{Jain}},
  \citenamefont {{Seljak}},\ and\ \citenamefont {{White}}}]{jain2000}%
  \BibitemOpen
  \bibfield  {author} {\bibinfo {author} {\bibfnamefont {B.}~\bibnamefont
  {{Jain}}}, \bibinfo {author} {\bibfnamefont {U.}~\bibnamefont {{Seljak}}}, \
  and\ \bibinfo {author} {\bibfnamefont {S.}~\bibnamefont {{White}}},\ }\href
  {\doibase 10.1086/308384} {\bibfield  {journal} {\bibinfo  {journal} {\apj}\
  }\textbf {\bibinfo {volume} {530}},\ \bibinfo {pages} {547} (\bibinfo {year}
  {2000})},\ \Eprint {http://arxiv.org/abs/astro-ph/9901191}
  {arXiv:astro-ph/9901191} \BibitemShut {NoStop}%
\bibitem [{\citenamefont {{Hamana}}\ and\ \citenamefont
  {{Mellier}}(2001)}]{hamana2001}%
  \BibitemOpen
  \bibfield  {author} {\bibinfo {author} {\bibfnamefont {T.}~\bibnamefont
  {{Hamana}}}\ and\ \bibinfo {author} {\bibfnamefont {Y.}~\bibnamefont
  {{Mellier}}},\ }\href {\doibase 10.1046/j.1365-8711.2001.04685.x} {\bibfield
  {journal} {\bibinfo  {journal} {\mnras}\ }\textbf {\bibinfo {volume} {327}},\
  \bibinfo {pages} {169} (\bibinfo {year} {2001})},\ \Eprint
  {http://arxiv.org/abs/astro-ph/0101333} {arXiv:astro-ph/0101333} \BibitemShut
  {NoStop}%
\bibitem [{\citenamefont {{Schneider}}\ \emph {et~al.}(1992)\citenamefont
  {{Schneider}}, \citenamefont {{Ehlers}},\ and\ \citenamefont
  {{Falco}}}]{schneider1992}%
  \BibitemOpen
  \bibfield  {author} {\bibinfo {author} {\bibfnamefont {P.}~\bibnamefont
  {{Schneider}}}, \bibinfo {author} {\bibfnamefont {J.}~\bibnamefont
  {{Ehlers}}}, \ and\ \bibinfo {author} {\bibfnamefont {E.~E.}\ \bibnamefont
  {{Falco}}},\ }\href {\doibase 10.1007/978-3-662-03758-4} {\emph {\bibinfo
  {title} {Astronomy and Astrophysics Library}}}\ (\bibinfo  {publisher}
  {Springer-Verlag},\ \bibinfo {address} {New York},\ \bibinfo {year} {1992})\
  p.\ \bibinfo {pages} {112}\BibitemShut {NoStop}%
\bibitem [{\citenamefont {{Das}}\ and\ \citenamefont {{Bode}}(2008)}]{das2008}%
  \BibitemOpen
  \bibfield  {author} {\bibinfo {author} {\bibfnamefont {S.}~\bibnamefont
  {{Das}}}\ and\ \bibinfo {author} {\bibfnamefont {P.}~\bibnamefont {{Bode}}},\
  }\href {\doibase 10.1086/589638} {\bibfield  {journal} {\bibinfo  {journal}
  {\apj}\ }\textbf {\bibinfo {volume} {682}},\ \bibinfo {eid} {1} (\bibinfo
  {year} {2008})},\ \Eprint {http://arxiv.org/abs/0711.3793} {arXiv:0711.3793}
  \BibitemShut {NoStop}%
\bibitem [{\citenamefont {{Hilbert}}\ \emph {et~al.}(2009)\citenamefont
  {{Hilbert}}, \citenamefont {{Hartlap}}, \citenamefont {{White}},\ and\
  \citenamefont {{Schneider}}}]{hilbert2009}%
  \BibitemOpen
  \bibfield  {author} {\bibinfo {author} {\bibfnamefont {S.}~\bibnamefont
  {{Hilbert}}}, \bibinfo {author} {\bibfnamefont {J.}~\bibnamefont
  {{Hartlap}}}, \bibinfo {author} {\bibfnamefont {S.~D.~M.}\ \bibnamefont
  {{White}}}, \ and\ \bibinfo {author} {\bibfnamefont {P.}~\bibnamefont
  {{Schneider}}},\ }\href {\doibase 10.1051/0004-6361/200811054} {\bibfield
  {journal} {\bibinfo  {journal} {\aap}\ }\textbf {\bibinfo {volume} {499}},\
  \bibinfo {pages} {31} (\bibinfo {year} {2009})},\ \Eprint
  {http://arxiv.org/abs/0809.5035} {arXiv:0809.5035} \BibitemShut {NoStop}%
\bibitem [{\citenamefont {{Lewis}}\ \emph {et~al.}(2017)\citenamefont
  {{Lewis}}, \citenamefont {{Hall}},\ and\ \citenamefont
  {{Challinor}}}]{lewis2017}%
  \BibitemOpen
  \bibfield  {author} {\bibinfo {author} {\bibfnamefont {A.}~\bibnamefont
  {{Lewis}}}, \bibinfo {author} {\bibfnamefont {A.}~\bibnamefont {{Hall}}}, \
  and\ \bibinfo {author} {\bibfnamefont {A.}~\bibnamefont {{Challinor}}},\
  }\href {\doibase 10.1088/1475-7516/2017/08/023} {\bibfield  {journal}
  {\bibinfo  {journal} {\jcap}\ }\textbf {\bibinfo {volume} {8}},\ \bibinfo
  {eid} {023} (\bibinfo {year} {2017})},\ \Eprint
  {http://arxiv.org/abs/1706.02673} {arXiv:1706.02673} \BibitemShut {NoStop}%
\bibitem [{\citenamefont {{Srednicki}}(1993)}]{Srednicki}%
  \BibitemOpen
  \bibfield  {author} {\bibinfo {author} {\bibfnamefont {M.}~\bibnamefont
  {{Srednicki}}},\ }\href {\doibase 10.1086/187056} {\bibfield  {journal}
  {\bibinfo  {journal} {\apjl}\ }\textbf {\bibinfo {volume} {416}},\ \bibinfo
  {pages} {L1} (\bibinfo {year} {1993})},\ \Eprint
  {http://arxiv.org/abs/astro-ph/9306012} {arXiv:astro-ph/9306012} \BibitemShut
  {NoStop}%
\bibitem [{\citenamefont {{Komatsu}}\ and\ \citenamefont
  {{Spergel}}(2001)}]{komatsu2001}%
  \BibitemOpen
  \bibfield  {author} {\bibinfo {author} {\bibfnamefont {E.}~\bibnamefont
  {{Komatsu}}}\ and\ \bibinfo {author} {\bibfnamefont {D.~N.}\ \bibnamefont
  {{Spergel}}},\ }\href {\doibase 10.1103/PhysRevD.63.063002} {\bibfield
  {journal} {\bibinfo  {journal} {\prd}\ }\textbf {\bibinfo {volume} {63}},\
  \bibinfo {eid} {063002} (\bibinfo {year} {2001})},\ \Eprint
  {http://arxiv.org/abs/astro-ph/0005036} {arXiv:astro-ph/0005036} \BibitemShut
  {NoStop}%
\bibitem [{\citenamefont {{Bernardeau}}\ \emph {et~al.}(2002)\citenamefont
  {{Bernardeau}}, \citenamefont {{Colombi}}, \citenamefont {{Gazta{\~n}aga}},\
  and\ \citenamefont {{Scoccimarro}}}]{Bernardeau}%
  \BibitemOpen
  \bibfield  {author} {\bibinfo {author} {\bibfnamefont {F.}~\bibnamefont
  {{Bernardeau}}}, \bibinfo {author} {\bibfnamefont {S.}~\bibnamefont
  {{Colombi}}}, \bibinfo {author} {\bibfnamefont {E.}~\bibnamefont
  {{Gazta{\~n}aga}}}, \ and\ \bibinfo {author} {\bibfnamefont {R.}~\bibnamefont
  {{Scoccimarro}}},\ }\href {\doibase 10.1016/S0370-1573(02)00135-7} {\bibfield
   {journal} {\bibinfo  {journal} {\physrep}\ }\textbf {\bibinfo {volume}
  {367}},\ \bibinfo {pages} {1} (\bibinfo {year} {2002})},\ \Eprint
  {http://arxiv.org/abs/astro-ph/0112551} {arXiv:astro-ph/0112551} \BibitemShut
  {NoStop}%
\bibitem [{\citenamefont {{Scoccimarro}}\ and\ \citenamefont
  {{Couchman}}(2001)}]{SC}%
  \BibitemOpen
  \bibfield  {author} {\bibinfo {author} {\bibfnamefont {R.}~\bibnamefont
  {{Scoccimarro}}}\ and\ \bibinfo {author} {\bibfnamefont {H.~M.~P.}\
  \bibnamefont {{Couchman}}},\ }\href {\doibase
  10.1046/j.1365-8711.2001.04281.x} {\bibfield  {journal} {\bibinfo  {journal}
  {\mnras}\ }\textbf {\bibinfo {volume} {325}},\ \bibinfo {pages} {1312}
  (\bibinfo {year} {2001})},\ \Eprint {http://arxiv.org/abs/astro-ph/0009427}
  {arXiv:astro-ph/0009427} \BibitemShut {NoStop}%
\bibitem [{\citenamefont {{Ruggeri}}\ \emph {et~al.}(2018)\citenamefont
  {{Ruggeri}}, \citenamefont {{Castorina}}, \citenamefont {{Carbone}},\ and\
  \citenamefont {{Sefusatti}}}]{ruggeri2018}%
  \BibitemOpen
  \bibfield  {author} {\bibinfo {author} {\bibfnamefont {R.}~\bibnamefont
  {{Ruggeri}}}, \bibinfo {author} {\bibfnamefont {E.}~\bibnamefont
  {{Castorina}}}, \bibinfo {author} {\bibfnamefont {C.}~\bibnamefont
  {{Carbone}}}, \ and\ \bibinfo {author} {\bibfnamefont {E.}~\bibnamefont
  {{Sefusatti}}},\ }\href {\doibase 10.1088/1475-7516/2018/03/003} {\bibfield
  {journal} {\bibinfo  {journal} {\jcap}\ }\textbf {\bibinfo {volume} {3}},\
  \bibinfo {eid} {003} (\bibinfo {year} {2018})},\ \Eprint
  {http://arxiv.org/abs/1712.02334} {arXiv:1712.02334} \BibitemShut {NoStop}%
\bibitem [{\citenamefont {{Takahashi}}\ \emph {et~al.}(2012)\citenamefont
  {{Takahashi}}, \citenamefont {{Sato}}, \citenamefont {{Nishimichi}},
  \citenamefont {{Taruya}},\ and\ \citenamefont {{Oguri}}}]{takahashi2012}%
  \BibitemOpen
  \bibfield  {author} {\bibinfo {author} {\bibfnamefont {R.}~\bibnamefont
  {{Takahashi}}}, \bibinfo {author} {\bibfnamefont {M.}~\bibnamefont {{Sato}}},
  \bibinfo {author} {\bibfnamefont {T.}~\bibnamefont {{Nishimichi}}}, \bibinfo
  {author} {\bibfnamefont {A.}~\bibnamefont {{Taruya}}}, \ and\ \bibinfo
  {author} {\bibfnamefont {M.}~\bibnamefont {{Oguri}}},\ }\href {\doibase
  10.1088/0004-637X/761/2/152} {\bibfield  {journal} {\bibinfo  {journal}
  {\apj}\ }\textbf {\bibinfo {volume} {761}},\ \bibinfo {eid} {152} (\bibinfo
  {year} {2012})},\ \Eprint {http://arxiv.org/abs/1208.2701} {arXiv:1208.2701}
  \BibitemShut {NoStop}%
\bibitem [{\citenamefont {{Gil-Mar{\'{\i}}n}}\ \emph
  {et~al.}(2012)\citenamefont {{Gil-Mar{\'{\i}}n}}, \citenamefont {{Wagner}},
  \citenamefont {{Fragkoudi}}, \citenamefont {{Jimenez}},\ and\ \citenamefont
  {{Verde}}}]{gilmarin2012}%
  \BibitemOpen
  \bibfield  {author} {\bibinfo {author} {\bibfnamefont {H.}~\bibnamefont
  {{Gil-Mar{\'{\i}}n}}}, \bibinfo {author} {\bibfnamefont {C.}~\bibnamefont
  {{Wagner}}}, \bibinfo {author} {\bibfnamefont {F.}~\bibnamefont
  {{Fragkoudi}}}, \bibinfo {author} {\bibfnamefont {R.}~\bibnamefont
  {{Jimenez}}}, \ and\ \bibinfo {author} {\bibfnamefont {L.}~\bibnamefont
  {{Verde}}},\ }\href {\doibase 10.1088/1475-7516/2012/02/047} {\bibfield
  {journal} {\bibinfo  {journal} {\jcap}\ }\textbf {\bibinfo {volume} {2}},\
  \bibinfo {eid} {047} (\bibinfo {year} {2012})},\ \Eprint
  {http://arxiv.org/abs/1111.4477} {arXiv:1111.4477} \BibitemShut {NoStop}%
\bibitem [{\citenamefont {{Knox}}(1995)}]{knox1995}%
  \BibitemOpen
  \bibfield  {author} {\bibinfo {author} {\bibfnamefont {L.}~\bibnamefont
  {{Knox}}},\ }\href {\doibase 10.1103/PhysRevD.52.4307} {\bibfield  {journal}
  {\bibinfo  {journal} {\prd}\ }\textbf {\bibinfo {volume} {52}},\ \bibinfo
  {pages} {4307} (\bibinfo {year} {1995})},\ \Eprint
  {http://arxiv.org/abs/astro-ph/9504054} {arXiv:astro-ph/9504054} \BibitemShut
  {NoStop}%
\bibitem [{\citenamefont {{B{\"o}hm}}\ \emph {et~al.}(2018)\citenamefont
  {{B{\"o}hm}}, \citenamefont {Sherwin}, \citenamefont {Liu}, \citenamefont
  {Hill}, \citenamefont {Schmittfull},\ and\ \citenamefont
  {Namikawa}}]{bohm2018}%
  \BibitemOpen
  \bibfield  {author} {\bibinfo {author} {\bibfnamefont {V.}~\bibnamefont
  {{B{\"o}hm}}}, \bibinfo {author} {\bibfnamefont {B.~D.}\ \bibnamefont
  {Sherwin}}, \bibinfo {author} {\bibfnamefont {J.}~\bibnamefont {Liu}},
  \bibinfo {author} {\bibfnamefont {J.~C.}\ \bibnamefont {Hill}}, \bibinfo
  {author} {\bibfnamefont {M.}~\bibnamefont {Schmittfull}}, \ and\ \bibinfo
  {author} {\bibfnamefont {T.}~\bibnamefont {Namikawa}},\ }\href@noop {} {\
  (\bibinfo {year} {2018})},\ \Eprint {http://arxiv.org/abs/1806.01157}
  {arXiv:1806.01157} \BibitemShut {NoStop}%
\bibitem [{\citenamefont {Schaan}\ \emph {et~al.}(2018)\citenamefont {Schaan},
  \citenamefont {Ferraro},\ and\ \citenamefont {Spergel}}]{schaan2018}%
  \BibitemOpen
  \bibfield  {author} {\bibinfo {author} {\bibfnamefont {E.}~\bibnamefont
  {Schaan}}, \bibinfo {author} {\bibfnamefont {S.}~\bibnamefont {Ferraro}}, \
  and\ \bibinfo {author} {\bibfnamefont {D.~N.}\ \bibnamefont {Spergel}},\
  }\href {\doibase 10.1103/PhysRevD.97.123539} {\bibfield  {journal} {\bibinfo
  {journal} {Phys. Rev. D}\ }\textbf {\bibinfo {volume} {97}},\ \bibinfo
  {pages} {123539} (\bibinfo {year} {2018})},\ \Eprint
  {http://arxiv.org/abs/1802.05706} {arXiv:1802.05706} \BibitemShut {NoStop}%
\bibitem [{\citenamefont {Castro}\ \emph {et~al.}(2018)\citenamefont {Castro},
  \citenamefont {Quartin}, \citenamefont {Giocoli}, \citenamefont {Borgani},\
  and\ \citenamefont {Dolag}}]{castro2018}%
  \BibitemOpen
  \bibfield  {author} {\bibinfo {author} {\bibfnamefont {T.}~\bibnamefont
  {Castro}}, \bibinfo {author} {\bibfnamefont {M.}~\bibnamefont {Quartin}},
  \bibinfo {author} {\bibfnamefont {C.}~\bibnamefont {Giocoli}}, \bibinfo
  {author} {\bibfnamefont {S.}~\bibnamefont {Borgani}}, \ and\ \bibinfo
  {author} {\bibfnamefont {K.}~\bibnamefont {Dolag}},\ }\href {\doibase
  10.1093/mnras/sty1117} {\bibfield  {journal} {\bibinfo  {journal} {Mon. Not.
  Roy. Astron. Soc.}\ }\textbf {\bibinfo {volume} {478}},\ \bibinfo {pages}
  {1305} (\bibinfo {year} {2018})},\ \Eprint {http://arxiv.org/abs/1711.10017}
  {arXiv:1711.10017} \BibitemShut {NoStop}%
\bibitem [{\citenamefont {{Natarajan}}\ \emph {et~al.}(2014)\citenamefont
  {{Natarajan}}, \citenamefont {{Zentner}}, \citenamefont {{Battaglia}},\ and\
  \citenamefont {{Trac}}}]{natarajan2014}%
  \BibitemOpen
  \bibfield  {author} {\bibinfo {author} {\bibfnamefont {A.}~\bibnamefont
  {{Natarajan}}}, \bibinfo {author} {\bibfnamefont {A.~R.}\ \bibnamefont
  {{Zentner}}}, \bibinfo {author} {\bibfnamefont {N.}~\bibnamefont
  {{Battaglia}}}, \ and\ \bibinfo {author} {\bibfnamefont {H.}~\bibnamefont
  {{Trac}}},\ }\href {\doibase 10.1103/PhysRevD.90.063516} {\bibfield
  {journal} {\bibinfo  {journal} {\prd}\ }\textbf {\bibinfo {volume} {90}},\
  \bibinfo {eid} {063516} (\bibinfo {year} {2014})},\ \Eprint
  {http://arxiv.org/abs/1405.6205} {arXiv:1405.6205} \BibitemShut {NoStop}%
\bibitem [{\citenamefont {{Merkel}}\ and\ \citenamefont
  {{Sch{\"a}fer}}(2017)}]{merkel2017}%
  \BibitemOpen
  \bibfield  {author} {\bibinfo {author} {\bibfnamefont {P.~M.}\ \bibnamefont
  {{Merkel}}}\ and\ \bibinfo {author} {\bibfnamefont {B.~M.}\ \bibnamefont
  {{Sch{\"a}fer}}},\ }\href {\doibase 10.1093/mnras/stx1664} {\bibfield
  {journal} {\bibinfo  {journal} {\mnras}\ }\textbf {\bibinfo {volume} {471}},\
  \bibinfo {pages} {2431} (\bibinfo {year} {2017})},\ \Eprint
  {http://arxiv.org/abs/1709.04444} {arXiv:1709.04444} \BibitemShut {NoStop}%
\bibitem [{\citenamefont {{Larsen}}\ and\ \citenamefont
  {{Challinor}}(2016)}]{larsen2016}%
  \BibitemOpen
  \bibfield  {author} {\bibinfo {author} {\bibfnamefont {P.}~\bibnamefont
  {{Larsen}}}\ and\ \bibinfo {author} {\bibfnamefont {A.}~\bibnamefont
  {{Challinor}}},\ }\href {\doibase 10.1093/mnras/stw1645} {\bibfield
  {journal} {\bibinfo  {journal} {\mnras}\ }\textbf {\bibinfo {volume} {461}},\
  \bibinfo {pages} {4343} (\bibinfo {year} {2016})},\ \Eprint
  {http://arxiv.org/abs/1510.02617} {arXiv:1510.02617} \BibitemShut {NoStop}%
\bibitem [{\citenamefont {{Troxel}}\ and\ \citenamefont
  {{Ishak}}(2014)}]{troxel2014}%
  \BibitemOpen
  \bibfield  {author} {\bibinfo {author} {\bibfnamefont {M.~A.}\ \bibnamefont
  {{Troxel}}}\ and\ \bibinfo {author} {\bibfnamefont {M.}~\bibnamefont
  {{Ishak}}},\ }\href {\doibase 10.1103/PhysRevD.89.063528} {\bibfield
  {journal} {\bibinfo  {journal} {\prd}\ }\textbf {\bibinfo {volume} {89}},\
  \bibinfo {eid} {063528} (\bibinfo {year} {2014})},\ \Eprint
  {http://arxiv.org/abs/1401.7051} {arXiv:1401.7051} \BibitemShut {NoStop}%
\bibitem [{\citenamefont {Brinckmann}\ and\ \citenamefont
  {Lesgourgues}(2018{\natexlab{a}})}]{Brinckmann:2018cvx}%
  \BibitemOpen
  \bibfield  {author} {\bibinfo {author} {\bibfnamefont {T.}~\bibnamefont
  {Brinckmann}}\ and\ \bibinfo {author} {\bibfnamefont {J.}~\bibnamefont
  {Lesgourgues}},\ }\href@noop {} {\  (\bibinfo {year} {2018}{\natexlab{a}})},\
  \Eprint {http://arxiv.org/abs/1804.07261} {arXiv:1804.07261} \BibitemShut
  {NoStop}%
\bibitem [{\citenamefont {Audren}\ \emph {et~al.}(2013)\citenamefont {Audren},
  \citenamefont {Lesgourgues}, \citenamefont {Benabed},\ and\ \citenamefont
  {Prunet}}]{Audren:2012wb}%
  \BibitemOpen
  \bibfield  {author} {\bibinfo {author} {\bibfnamefont {B.}~\bibnamefont
  {Audren}}, \bibinfo {author} {\bibfnamefont {J.}~\bibnamefont {Lesgourgues}},
  \bibinfo {author} {\bibfnamefont {K.}~\bibnamefont {Benabed}}, \ and\
  \bibinfo {author} {\bibfnamefont {S.}~\bibnamefont {Prunet}},\ }\href
  {\doibase 10.1088/1475-7516/2013/02/001} {\bibfield  {journal} {\bibinfo
  {journal} {\jcap}\ }\textbf {\bibinfo {volume} {1302}},\ \bibinfo {eid} {001}
  (\bibinfo {year} {2013})},\ \Eprint {http://arxiv.org/abs/1210.7183}
  {arXiv:1210.7183} \BibitemShut {NoStop}%
\bibitem [{\citenamefont {{Di Valentino}}\ \emph {et~al.}(2018)\citenamefont
  {{Di Valentino}}, \citenamefont {{Brinckmann}}, \citenamefont {{Gerbino}},
  \citenamefont {{Poulin}}, \citenamefont {{Bouchet}}, \citenamefont
  {{Lesgourgues}}, \citenamefont {{Melchiorri}}, \citenamefont {{Chluba}},
  \citenamefont {{Clesse}}, \citenamefont {{Delabrouille}} \emph
  {et~al.}}]{divalentino2018}%
  \BibitemOpen
  \bibfield  {author} {\bibinfo {author} {\bibfnamefont {E.}~\bibnamefont {{Di
  Valentino}}}, \bibinfo {author} {\bibfnamefont {T.}~\bibnamefont
  {{Brinckmann}}}, \bibinfo {author} {\bibfnamefont {M.}~\bibnamefont
  {{Gerbino}}}, \bibinfo {author} {\bibfnamefont {V.}~\bibnamefont {{Poulin}}},
  \bibinfo {author} {\bibfnamefont {F.~R.}\ \bibnamefont {{Bouchet}}}, \bibinfo
  {author} {\bibfnamefont {J.}~\bibnamefont {{Lesgourgues}}}, \bibinfo {author}
  {\bibfnamefont {A.}~\bibnamefont {{Melchiorri}}}, \bibinfo {author}
  {\bibfnamefont {J.}~\bibnamefont {{Chluba}}}, \bibinfo {author}
  {\bibfnamefont {S.}~\bibnamefont {{Clesse}}}, \bibinfo {author}
  {\bibfnamefont {J.}~\bibnamefont {{Delabrouille}}},  \emph {et~al.},\ }\href
  {\doibase 10.1088/1475-7516/2018/04/017} {\bibfield  {journal} {\bibinfo
  {journal} {\jcap}\ }\textbf {\bibinfo {volume} {4}},\ \bibinfo {eid} {017}
  (\bibinfo {year} {2018})},\ \Eprint {http://arxiv.org/abs/1612.00021}
  {arXiv:1612.00021} \BibitemShut {NoStop}%
\bibitem [{\citenamefont {{Smith}}\ \emph {et~al.}(2004)\citenamefont
  {{Smith}}, \citenamefont {{Hu}},\ and\ \citenamefont
  {{Kaplinghat}}}]{smith2004}%
  \BibitemOpen
  \bibfield  {author} {\bibinfo {author} {\bibfnamefont {K.~M.}\ \bibnamefont
  {{Smith}}}, \bibinfo {author} {\bibfnamefont {W.}~\bibnamefont {{Hu}}}, \
  and\ \bibinfo {author} {\bibfnamefont {M.}~\bibnamefont {{Kaplinghat}}},\
  }\href {\doibase 10.1103/PhysRevD.70.043002} {\bibfield  {journal} {\bibinfo
  {journal} {\prd}\ }\textbf {\bibinfo {volume} {70}},\ \bibinfo {eid} {043002}
  (\bibinfo {year} {2004})},\ \Eprint {http://arxiv.org/abs/astro-ph/0402442}
  {astro-ph/0402442} \BibitemShut {NoStop}%
\bibitem [{\citenamefont {Schmittfull}\ \emph {et~al.}(2013)\citenamefont
  {Schmittfull}, \citenamefont {Challinor}, \citenamefont {Hanson},\ and\
  \citenamefont {Lewis}}]{Schmittfull:2013uea}%
  \BibitemOpen
  \bibfield  {author} {\bibinfo {author} {\bibfnamefont {M.~M.}\ \bibnamefont
  {Schmittfull}}, \bibinfo {author} {\bibfnamefont {A.}~\bibnamefont
  {Challinor}}, \bibinfo {author} {\bibfnamefont {D.}~\bibnamefont {Hanson}}, \
  and\ \bibinfo {author} {\bibfnamefont {A.}~\bibnamefont {Lewis}},\ }\href
  {\doibase 10.1103/PhysRevD.88.063012} {\bibfield  {journal} {\bibinfo
  {journal} {Phys. Rev. D}\ }\textbf {\bibinfo {volume} {88}},\ \bibinfo
  {pages} {063012} (\bibinfo {year} {2013})},\ \Eprint
  {http://arxiv.org/abs/1308.0286} {arXiv:1308.0286} \BibitemShut {NoStop}%
\bibitem [{\citenamefont {{Benoit-L{\'e}vy}}\ \emph {et~al.}(2012)\citenamefont
  {{Benoit-L{\'e}vy}}, \citenamefont {{Smith}},\ and\ \citenamefont
  {{Hu}}}]{benoit-levy2012}%
  \BibitemOpen
  \bibfield  {author} {\bibinfo {author} {\bibfnamefont {A.}~\bibnamefont
  {{Benoit-L{\'e}vy}}}, \bibinfo {author} {\bibfnamefont {K.~M.}\ \bibnamefont
  {{Smith}}}, \ and\ \bibinfo {author} {\bibfnamefont {W.}~\bibnamefont
  {{Hu}}},\ }\href {\doibase 10.1103/PhysRevD.86.123008} {\bibfield  {journal}
  {\bibinfo  {journal} {\prd}\ }\textbf {\bibinfo {volume} {86}},\ \bibinfo
  {eid} {123008} (\bibinfo {year} {2012})},\ \Eprint
  {http://arxiv.org/abs/1205.0474} {arXiv:1205.0474} \BibitemShut {NoStop}%
\bibitem [{\citenamefont {Ade}\ \emph {et~al.}(2016)\citenamefont {Ade} \emph
  {et~al.}}]{Ade:2015xua}%
  \BibitemOpen
  \bibfield  {author} {\bibinfo {author} {\bibfnamefont {P.~A.~R.}\
  \bibnamefont {Ade}} \emph {et~al.} (\bibinfo {collaboration} {Planck
  Collaboration}),\ }\href {\doibase 10.1051/0004-6361/201525830} {\bibfield
  {journal} {\bibinfo  {journal} {Astron. Astrophys.}\ }\textbf {\bibinfo
  {volume} {594}},\ \bibinfo {pages} {A13} (\bibinfo {year} {2016})},\ \Eprint
  {http://arxiv.org/abs/1502.01589} {arXiv:1502.01589} \BibitemShut {NoStop}%
\bibitem [{\citenamefont {Adam}\ \emph {et~al.}(2016)\citenamefont {Adam} \emph
  {et~al.}}]{Adam:2016hgk}%
  \BibitemOpen
  \bibfield  {author} {\bibinfo {author} {\bibfnamefont {R.}~\bibnamefont
  {Adam}} \emph {et~al.} (\bibinfo {collaboration} {Planck}),\ }\href {\doibase
  10.1051/0004-6361/201628897} {\bibfield  {journal} {\bibinfo  {journal}
  {Astron. Astrophys.}\ }\textbf {\bibinfo {volume} {596}},\ \bibinfo {pages}
  {A108} (\bibinfo {year} {2016})},\ \Eprint {http://arxiv.org/abs/1605.03507}
  {arXiv:1605.03507} \BibitemShut {NoStop}%
\bibitem [{\citenamefont {Archidiacono}\ \emph {et~al.}(2017)\citenamefont
  {Archidiacono}, \citenamefont {Brinckmann}, \citenamefont {Lesgourgues},\
  and\ \citenamefont {Poulin}}]{Archidiacono:2016lnv}%
  \BibitemOpen
  \bibfield  {author} {\bibinfo {author} {\bibfnamefont {M.}~\bibnamefont
  {Archidiacono}}, \bibinfo {author} {\bibfnamefont {T.}~\bibnamefont
  {Brinckmann}}, \bibinfo {author} {\bibfnamefont {J.}~\bibnamefont
  {Lesgourgues}}, \ and\ \bibinfo {author} {\bibfnamefont {V.}~\bibnamefont
  {Poulin}},\ }\href {\doibase 10.1088/1475-7516/2017/02/052} {\bibfield
  {journal} {\bibinfo  {journal} {\jcap}\ }\textbf {\bibinfo {volume} {1702}},\
  \bibinfo {pages} {052} (\bibinfo {year} {2017})},\ \Eprint
  {http://arxiv.org/abs/1610.09852} {arXiv:1610.09852} \BibitemShut {NoStop}%
\bibitem [{\citenamefont {Watts}\ \emph {et~al.}(2018)\citenamefont {Watts}
  \emph {et~al.}}]{Watts:2018etg}%
  \BibitemOpen
  \bibfield  {author} {\bibinfo {author} {\bibfnamefont {D.~J.}\ \bibnamefont
  {Watts}} \emph {et~al.},\ }\href {\doibase 10.3847/1538-4357/aad283}
  {\bibfield  {journal} {\bibinfo  {journal} {Astrophys. J.}\ }\textbf
  {\bibinfo {volume} {863}},\ \bibinfo {pages} {121} (\bibinfo {year}
  {2018})},\ \Eprint {http://arxiv.org/abs/1801.01481} {arXiv:1801.01481
  [astro-ph.CO]} \BibitemShut {NoStop}%
\bibitem [{\citenamefont {Suzuki}\ \emph {et~al.}(2018)\citenamefont {Suzuki}
  \emph {et~al.}}]{Suzuki:2018cuy}%
  \BibitemOpen
  \bibfield  {author} {\bibinfo {author} {\bibfnamefont {A.}~\bibnamefont
  {Suzuki}} \emph {et~al.}\ }(\bibinfo {year} {2018})\ \Eprint
  {http://arxiv.org/abs/1801.06987} {arXiv:1801.06987} \BibitemShut {NoStop}%
\bibitem [{\citenamefont {Calabrese}\ \emph {et~al.}(2017)\citenamefont
  {Calabrese}, \citenamefont {Alonso},\ and\ \citenamefont
  {Dunkley}}]{Calabrese:2016eii}%
  \BibitemOpen
  \bibfield  {author} {\bibinfo {author} {\bibfnamefont {E.}~\bibnamefont
  {Calabrese}}, \bibinfo {author} {\bibfnamefont {D.}~\bibnamefont {Alonso}}, \
  and\ \bibinfo {author} {\bibfnamefont {J.}~\bibnamefont {Dunkley}},\ }\href
  {\doibase 10.1103/PhysRevD.95.063504} {\bibfield  {journal} {\bibinfo
  {journal} {Phys. Rev. D}\ }\textbf {\bibinfo {volume} {95}},\ \bibinfo
  {pages} {063504} (\bibinfo {year} {2017})},\ \Eprint
  {http://arxiv.org/abs/1611.10269} {arXiv:1611.10269} \BibitemShut {NoStop}%
\bibitem [{\citenamefont {Allison}\ \emph {et~al.}(2015)\citenamefont
  {Allison}, \citenamefont {Caucal}, \citenamefont {Calabrese}, \citenamefont
  {Dunkley},\ and\ \citenamefont {Louis}}]{Allison:2015qca}%
  \BibitemOpen
  \bibfield  {author} {\bibinfo {author} {\bibfnamefont {R.}~\bibnamefont
  {Allison}}, \bibinfo {author} {\bibfnamefont {P.}~\bibnamefont {Caucal}},
  \bibinfo {author} {\bibfnamefont {E.}~\bibnamefont {Calabrese}}, \bibinfo
  {author} {\bibfnamefont {J.}~\bibnamefont {Dunkley}}, \ and\ \bibinfo
  {author} {\bibfnamefont {T.}~\bibnamefont {Louis}},\ }\href {\doibase
  10.1103/PhysRevD.92.123535} {\bibfield  {journal} {\bibinfo  {journal} {Phys.
  Rev. D}\ }\textbf {\bibinfo {volume} {92}},\ \bibinfo {pages} {123535}
  (\bibinfo {year} {2015})},\ \Eprint {http://arxiv.org/abs/1509.07471}
  {arXiv:1509.07471} \BibitemShut {NoStop}%
\bibitem [{\citenamefont {van Engelen}\ \emph {et~al.}(2014)\citenamefont {van
  Engelen}, \citenamefont {Bhattacharya}, \citenamefont {Sehgal}, \citenamefont
  {Holder}, \citenamefont {Zahn},\ and\ \citenamefont
  {Nagai}}]{vanEngelen:2013rla}%
  \BibitemOpen
  \bibfield  {author} {\bibinfo {author} {\bibfnamefont {A.}~\bibnamefont {van
  Engelen}}, \bibinfo {author} {\bibfnamefont {S.}~\bibnamefont
  {Bhattacharya}}, \bibinfo {author} {\bibfnamefont {N.}~\bibnamefont
  {Sehgal}}, \bibinfo {author} {\bibfnamefont {G.~P.}\ \bibnamefont {Holder}},
  \bibinfo {author} {\bibfnamefont {O.}~\bibnamefont {Zahn}}, \ and\ \bibinfo
  {author} {\bibfnamefont {D.}~\bibnamefont {Nagai}},\ }\href {\doibase
  10.1088/0004-637X/786/1/13} {\bibfield  {journal} {\bibinfo  {journal}
  {Astrophys. J.}\ }\textbf {\bibinfo {volume} {786}},\ \bibinfo {pages} {13}
  (\bibinfo {year} {2014})},\ \Eprint {http://arxiv.org/abs/1310.7023}
  {arXiv:1310.7023} \BibitemShut {NoStop}%
\bibitem [{\citenamefont {{Osborne}}\ \emph {et~al.}(2014)\citenamefont
  {{Osborne}}, \citenamefont {{Hanson}},\ and\ \citenamefont
  {{Dor{\'e}}}}]{Osborne:2013nna}%
  \BibitemOpen
  \bibfield  {author} {\bibinfo {author} {\bibfnamefont {S.~J.}\ \bibnamefont
  {{Osborne}}}, \bibinfo {author} {\bibfnamefont {D.}~\bibnamefont {{Hanson}}},
  \ and\ \bibinfo {author} {\bibfnamefont {O.}~\bibnamefont {{Dor{\'e}}}},\
  }\href {\doibase 10.1088/1475-7516/2014/03/024} {\bibfield  {journal}
  {\bibinfo  {journal} {\jcap}\ }\textbf {\bibinfo {volume} {3}},\ \bibinfo
  {eid} {024} (\bibinfo {year} {2014})},\ \Eprint
  {http://arxiv.org/abs/1310.7547} {arXiv:1310.7547} \BibitemShut {NoStop}%
\bibitem [{\citenamefont {Ferraro}\ and\ \citenamefont
  {Hill}(2018)}]{PhysRevD.97.023512}%
  \BibitemOpen
  \bibfield  {author} {\bibinfo {author} {\bibfnamefont {S.}~\bibnamefont
  {Ferraro}}\ and\ \bibinfo {author} {\bibfnamefont {J.~C.}\ \bibnamefont
  {Hill}},\ }\href {\doibase 10.1103/PhysRevD.97.023512} {\bibfield  {journal}
  {\bibinfo  {journal} {Phys. Rev. D}\ }\textbf {\bibinfo {volume} {97}},\
  \bibinfo {pages} {023512} (\bibinfo {year} {2018})},\ \Eprint
  {http://arxiv.org/abs/1705.06751} {arXiv:1705.06751} \BibitemShut {NoStop}%
\bibitem [{\citenamefont {Lewis}\ \emph {et~al.}(2000)\citenamefont {Lewis},
  \citenamefont {Challinor},\ and\ \citenamefont {Lasenby}}]{camb-1}%
  \BibitemOpen
  \bibfield  {author} {\bibinfo {author} {\bibfnamefont {A.}~\bibnamefont
  {Lewis}}, \bibinfo {author} {\bibfnamefont {A.}~\bibnamefont {Challinor}}, \
  and\ \bibinfo {author} {\bibfnamefont {A.}~\bibnamefont {Lasenby}},\ }\href
  {\doibase 10.1086/309179} {\bibfield  {journal} {\bibinfo  {journal}
  {Astrophys. J.}\ }\textbf {\bibinfo {volume} {538}},\ \bibinfo {pages} {473}
  (\bibinfo {year} {2000})},\ \Eprint {http://arxiv.org/abs/astro-ph/9911177}
  {arXiv:astro-ph/9911177} \BibitemShut {NoStop}%
\bibitem [{\citenamefont {Howlett}\ \emph {et~al.}(2012)\citenamefont
  {Howlett}, \citenamefont {Lewis}, \citenamefont {Hall},\ and\ \citenamefont
  {Challinor}}]{camb-2}%
  \BibitemOpen
  \bibfield  {author} {\bibinfo {author} {\bibfnamefont {C.}~\bibnamefont
  {Howlett}}, \bibinfo {author} {\bibfnamefont {A.}~\bibnamefont {Lewis}},
  \bibinfo {author} {\bibfnamefont {A.}~\bibnamefont {Hall}}, \ and\ \bibinfo
  {author} {\bibfnamefont {A.}~\bibnamefont {Challinor}},\ }\href {\doibase
  10.1088/1475-7516/2012/04/027} {\bibfield  {journal} {\bibinfo  {journal}
  {\jcap}\ }\textbf {\bibinfo {volume} {1204}},\ \bibinfo {eid} {027} (\bibinfo
  {year} {2012})},\ \Eprint {http://arxiv.org/abs/1201.3654} {arXiv:1201.3654}
  \BibitemShut {NoStop}%
\bibitem [{\citenamefont {{Blas}}\ \emph {et~al.}(2011)\citenamefont {{Blas}},
  \citenamefont {{Lesgourgues}},\ and\ \citenamefont {{Tram}}}]{class}%
  \BibitemOpen
  \bibfield  {author} {\bibinfo {author} {\bibfnamefont {D.}~\bibnamefont
  {{Blas}}}, \bibinfo {author} {\bibfnamefont {J.}~\bibnamefont
  {{Lesgourgues}}}, \ and\ \bibinfo {author} {\bibfnamefont {T.}~\bibnamefont
  {{Tram}}},\ }\href {\doibase 10.1088/1475-7516/2011/07/034} {\bibfield
  {journal} {\bibinfo  {journal} {\jcap}\ }\textbf {\bibinfo {volume} {7}},\
  \bibinfo {eid} {034} (\bibinfo {year} {2011})},\ \Eprint
  {http://arxiv.org/abs/1104.2933} {arXiv:1104.2933} \BibitemShut {NoStop}%
\bibitem [{\citenamefont {Brinckmann}\ and\ \citenamefont
  {Lesgourgues}(2018{\natexlab{b}})}]{mp-1}%
  \BibitemOpen
  \bibfield  {author} {\bibinfo {author} {\bibfnamefont {T.}~\bibnamefont
  {Brinckmann}}\ and\ \bibinfo {author} {\bibfnamefont {J.}~\bibnamefont
  {Lesgourgues}},\ }\href@noop {} {\  (\bibinfo {year} {2018}{\natexlab{b}})},\
  \Eprint {http://arxiv.org/abs/1804.07261} {arXiv:1804.07261} \BibitemShut
  {NoStop}%
\bibitem [{\citenamefont {{Audren}}\ \emph {et~al.}(2013)\citenamefont
  {{Audren}}, \citenamefont {{Lesgourgues}}, \citenamefont {{Benabed}},\ and\
  \citenamefont {{Prunet}}}]{mp-2}%
  \BibitemOpen
  \bibfield  {author} {\bibinfo {author} {\bibfnamefont {B.}~\bibnamefont
  {{Audren}}}, \bibinfo {author} {\bibfnamefont {J.}~\bibnamefont
  {{Lesgourgues}}}, \bibinfo {author} {\bibfnamefont {K.}~\bibnamefont
  {{Benabed}}}, \ and\ \bibinfo {author} {\bibfnamefont {S.}~\bibnamefont
  {{Prunet}}},\ }\href {\doibase 10.1088/1475-7516/2013/02/001} {\bibfield
  {journal} {\bibinfo  {journal} {\jcap}\ }\textbf {\bibinfo {volume} {2}},\
  \bibinfo {eid} {001} (\bibinfo {year} {2013})},\ \Eprint
  {http://arxiv.org/abs/1210.7183} {arXiv:1210.7183} \BibitemShut {NoStop}%
\bibitem [{\citenamefont {{G{\'o}rski}}\ \emph {et~al.}(2005)\citenamefont
  {{G{\'o}rski}}, \citenamefont {{Hivon}}, \citenamefont {{Banday}},
  \citenamefont {{Wandelt}}, \citenamefont {{Hansen}}, \citenamefont
  {{Reinecke}},\ and\ \citenamefont {{Bartelmann}}}]{healpix}%
  \BibitemOpen
  \bibfield  {author} {\bibinfo {author} {\bibfnamefont {K.~M.}\ \bibnamefont
  {{G{\'o}rski}}}, \bibinfo {author} {\bibfnamefont {E.}~\bibnamefont
  {{Hivon}}}, \bibinfo {author} {\bibfnamefont {A.~J.}\ \bibnamefont
  {{Banday}}}, \bibinfo {author} {\bibfnamefont {B.~D.}\ \bibnamefont
  {{Wandelt}}}, \bibinfo {author} {\bibfnamefont {F.~K.}\ \bibnamefont
  {{Hansen}}}, \bibinfo {author} {\bibfnamefont {M.}~\bibnamefont
  {{Reinecke}}}, \ and\ \bibinfo {author} {\bibfnamefont {M.}~\bibnamefont
  {{Bartelmann}}},\ }\href {\doibase 10.1086/427976} {\bibfield  {journal}
  {\bibinfo  {journal} {\apj}\ }\textbf {\bibinfo {volume} {622}},\ \bibinfo
  {pages} {759} (\bibinfo {year} {2005})},\ \Eprint
  {http://arxiv.org/abs/astro-ph/0409513} {arXiv:astro-ph/0409513} \BibitemShut
  {NoStop}%
\end{thebibliography}%
\end{document}